\documentclass[a4paper,10pt,notitlepage]{article}
\usepackage{amsfonts,amssymb,amsmath,amsthm,hyperref,colonequals}
\usepackage{caption}
\usepackage{cite}
\usepackage{tabularx}
\usepackage{graphicx}
\usepackage{shuffle}
\usepackage[nice]{nicefrac} 
\usepackage{url}
\theoremstyle{definition}
\newcommand{\beqa}{\begin{eqnarray}}
\newcommand{\eeqa}{\end{eqnarray}}
\newcommand{\beq}{\begin{equation}}
\newcommand{\eeq}{\end{equation}}

\newcommand{\ft}{\mathfrak{t}}
\newcommand{\fq}{\mathfrak{q}}
\newcommand{\fC}{\mathfrak{C}}
\newcommand{\fQ}{\mathbf{\mathcal{Q}}}
\newcommand{\fZ}{\boldsymbol{\mathfrak{Z}}}
\newcommand{\fA}{\mathsf{A}}
\newcommand{\fR}{\mathsf{R}}
\newcommand{\fw}{\mathsf{w}}
\newcommand{\calR}{\mathcal{R}}

\newcommand{\calO}{\mathcal{O}}
\newcommand{\calZ}{\mathcal{Z}}
\newcommand{\calM}{\mathcal{M}}\newcommand{\calN}{\mathcal{N}}

\newcommand{\tA}{\textsf{A}}

\newcommand{\ba}{\textbf{a}}\newcommand{\bA}{\textbf{A}}

\def\balpha{\boldsymbol{\alpha}}
\newcommand{\form}[2]{\ensuremath{\left( #1, #2\right)}}

\newcommand{\Up}{\Upsilon_q}
\newcommand{\tN}{\textsf{N}^{\beta}}
\newcommand{\ttN}{\textsf{N}}
\begin{document}
%%%%%%%%%%%%%%%%%%%%%%%%%%%%%%%%%%%%%%%%%%%%%%%%%%%%%%%%%%%%%%%%%%%%

%%%%%%%%%%%%%%%%%%%%%% TITLE PAGE %%%%%%%%%%%%%%%%%%%%%%%%%%%%%%%%%%%
\thispagestyle{empty}
\setcounter{page}{0}
\begin{flushright}\footnotesize
\texttt{DESY 14-160}\\
\texttt{HU-Mathematik-19-2014}\\
\texttt{HU-EP-14/33}\\
\vspace{0.5cm}
\end{flushright}
\setcounter{footnote}{0}

\begin{center}
{\huge{
\textbf{Toda 3-Point Functions \vspace{0.5cm}\\ From Topological Strings
}
}}
\vspace{15mm}

{\sc 
Vladimir Mitev$^{a}$,   Elli Pomoni$^{b,c}$ }\\[5mm]

{\it $^a$ Institut f\"ur Mathematik und Institut f\"ur Physik,\\ Humboldt-Universit\"at zu Berlin\\
IRIS Haus, Zum Gro{\ss}en Windkanal 6,  12489 Berlin, Germany
}\\[5mm]

{\it $^b$ DESY Hamburg, Theory Group, \\
Notkestrasse 85, D--22607 Hamburg, Germany
}\\[5mm]

{\it $^c$ Physics Division, National Technical University of Athens,\\
15780 Zografou Campus, Athens, Greece
}\\[5mm]

\texttt{mitev@math.hu-berlin.de}\\
\texttt{elli.pomoni@desy.de}\\[25mm]

\textbf{Abstract}\\[2mm]
\end{center}
We consider the long-standing problem of obtaining the 3-point functions of Toda CFT. Our main tools are topological strings and the AGT-W relation between gauge theories and 2D CFTs. In \cite{Bao:2013pwa} we computed the partition function of 5D $T_N$ theories on $S^4\times S^1$ and suggested that they should be interpreted as the three-point structure constants of $q$-deformed Toda. In this paper, we provide the exact AGT-W dictionary for this relation and rewrite the 5D $T_N$ partition function in a form that makes taking the 4D limit possible. Thus, we obtain a prescription for the computation of the partition function of the 4D $T_N$ theories on $S^4$, or equivalently the undeformed 3-point Toda structure constants.  Our formula, has the correct symmetry properties, the zeros that it should and, for $N=2$, gives the known answer for Liouville CFT.

%%%%%%%%%%%%%%%%%%%%%%%%% END OF TITLE PAGE %%%%%%%%%%%%%%%%%%%%%%%%%%%%%

\newpage
\setcounter{page}{1}

%%%%%%%%%%%%%%%%%%%%%%%%%%%%%%%%%%%%%%%%%%%%%%%%%%%%%%%%%%%%%%%%%%%%
%%%%%%%%%%%%%%%%%%%%%%%%%%%%%%%%%%%%%%%%%%%%%%%%%%%%%%%%%%%%%%%%%%%%

\tableofcontents
\addtolength{\baselineskip}{5pt}

%%%%%%%%%%%%%%%%%%%%%%%%%%%%%%%%%%%%%%
\section{Introduction}
%%%%%%%%%%%%%%%%%%%%%%%%%%%%%%%%%%%%%%
%\input{Introduction}

The AGT(-W) correspondence \cite{Alday:2009aq,Wyllard:2009hg,Gaiotto:2009ma} is a relationship between, on one side, the 2D Liouville (Toda)  CFT on a Riemann surface of genus $g$ with $n$ punctures and, on the other side,  the 4D $\mathcal{N}=2$ SU(2) (SU$(N)$) quiver gauge theories obtained by compactifying the 6D (2,0) SCFT on that same surface . The correlation functions of the 2D Toda \textbf{W}$_N$ conformal field theories are obtained from by the partition functions of the corresponding 4D $\mathcal{N}=2$ gauge theories as
\begin{align}
\label{eq:AGTWrelation}
\calZ^{S^4}=\int [da] \Big|\calZ_{\textrm{Nek}}^{\textrm{4D}}(a,m,\epsilon_{1,2})\Big|^2 
\propto 
\langle V_{\boldsymbol{\alpha}_1}(z_1)\cdots V_{\boldsymbol{\alpha}_n}(z_n)
\rangle_{\textrm{Toda}}
\, .
\end{align}
The conformal blocks of the 2D CFTs are given by the appropriate instanton partition functions, while the three point structure constants  should be obtained by the $S^4$ partition functions of the $T_N$ superconformal theories. These  partition functions were until recently \cite{Bao:2013pwa, Hayashi:2013qwa} unknown, with the sole exception of the \textbf{W}$_2$ case, {\it i.e.} the Liouville case, whose three point structure constants are given by the famous DOZZ formula \cite{Dorn:1994xn,Zamolodchikov:1995aa}.
 The  AGT(-W) relation \eqref{eq:AGTWrelation} holds
 after the mass parameters $m$ of the gauge theory, the UV coupling constants and the vacuum expectation values $a$  of the scalars in the vector multiplet (the Coulomb moduli) are appropriately identified with, respectively, the external momenta $\balpha$ of the primary fields, the moduli $z_i$ of the 2D surface ({\it i.e.} the sewing parameters) and the internal momenta over which we integrate. Finally, the IR regulators of the gauge theory, which are given by the Omega deformation parameters $\epsilon_{1,2}$, are identified with the Toda dimensionless coupling constant via $b=\epsilon_1=\epsilon_2^{-1}$.
 The AGT conjecture, {\it i.e.} the $N=2$ case, was recently proven in \cite{Fateev:2009aw,Mironov:2009qn,Hadasz:2010xp,Alba:2010qc, Mironov:2010qe, Mironov:2010pi}, while a lot of evidence and even proofs for specific cases exist \cite{Fateev:2011hq,Kanno:2013vi,Mironov:2013oaa} in support of the AGT-W correspondence for $N>2$.

Similarly, there exists a 5D version of the AGT(-W) relation\footnote{Originally  suggested in \cite{Schiappa:2009cc}.} \cite{Awata:2009ur,Awata:2010yy} (see also \cite{Mironov:2011dk,Itoyama:2013mca,Bao:2011rc,Nieri:2013yra,Bao:2013pwa,Nieri:2013vba,Aganagic:2013tta,Aganagic:2014oia,Taki:2014fva}) which relates the 5D Nekrasov partition functions on $S^4 \times S^1$ to correlation functions of $q$-deformed Liouville (Toda) field theory:
\begin{align}
\calZ^{S^4\times S^1}=\int [da] \Big|
\calZ_{\textrm{Nek}}^{\textrm{5D}}(a,m,\beta,\epsilon_{1,2})\Big|^2
\propto  \langle
V_{\boldsymbol{\alpha}_1}(z_1)\cdots V_{\boldsymbol{\alpha}_n}(z_n)
\rangle_{q\textrm{-Toda}} \, ,
\end{align}
where $\beta= - \log q$ is the circumference  of the $ S^1$.
Importantly, the integral of the norm squared of 5D Nekrasov partition function is  the 5D superconformal index $\calZ^{S^4\times S^1}$, 
which as discussed recently in \cite{Iqbal:2012xm} 
 can be computed using the topological string partition function
\beq
\calZ^{S^4\times S^1}= \int [d a] \, |\calZ_{\textrm{Nek}}^{\textrm{5D}} (a)|^2\propto \int [d a] \, |\calZ_{\textrm{top}} (a)|^2 \, .
\eeq

From both the 4D and the 5D AGT-W relations  a very important element is missing:  the three point functions of  the \textbf{W}$_N$ Toda CFT.
Computing the three point functions of  the \textbf{W}$_N$ Toda CFT has been a long standing  unsolved problem.
 From the the CFT side, the state of the art is due to Fateev and Litvinov, who in
\cite{Fateev:2005gs,Fateev:2007ab,Fateev:2008bm}, were able to compute the 3-point functions of Toda primaries for the special case in which one of the fields is semi-degenerate, using \cite{Dotsenko:1984ad}.
On the gauge theory side, the 3-point functions correspond to the partition functions of the $T_N$  theories, but since these theories lack any known Lagrangian description, the usual methods of computing the partition functions are not applicable.

In \cite{Bao:2013pwa} we computed the partition functions of the 5D $T_N$ theories on $S^4\times S^1$ by using the web diagram  provided by \cite{Benini:2009gi} and by employing the refined topological vertex formalism of \cite{Awata:2005fa,Iqbal:2007ii}. We further argued that these partition functions  should give the three point functions of $q$-deformed Toda, which was also proposed earlier in \cite{Kozcaz:2010af}. Our results were checked by computing the 5D superconformal  index, {\it i.e.} the partition function on $S^4\times S^1$, using the prescription in \cite{Iqbal:2012xm} and comparing it to the result obtained via localization in \cite{Kim:2012gu}. The same partition functions were also obtained in \cite{Hayashi:2013qwa} and the two computations agree. More comparisons with the superconformal index were given in the recent work \cite{Hayashi:2014wfa}.

In this paper we show how to, in principle\footnote{The specification ``in principle'' refers to the fact that there is still a missing ingredient which  is to perform the sums in \eqref{eq:ZTNsum}. This work will appear in a separate \cite{Misha} publication, where we will show that some of the sums can be computed.}, take the 4D limit, thus obtaining the  4D $T_N$ partition functions.  Through the AGT(-W) relation, they are identified with the usual, undeformed Toda three point functions.  Our formula has the correct symmetry properties, zeros and reproduces the known answer for the Liouville CFT. Furthermore, we carefully study the 5D AGT-W dictionary. For that, it was very important to examine the known $q$-Liouville case \cite{Kozcaz:2010af, Nieri:2013yra} for which for the first time we were able to write the formula with the complete factors, thanks to the exact definition of the functions $\Up$, see appendix \ref{app:qUpsilon}.

Our method of attacking the problem of solving Toda, even though indirect, is very powerful for the following reasons.
For 2D CFTs with only Virasoro symmetry the multipoint correlation functions of  Virasoro descendants can be obtained from the ones  containing only Virasoro primary fields \cite{Belavin:1984vu}. On the other hand, for the \textbf{W}$_N$ Toda CFTs with $N>2$ complete knowledge of the correlation functions of \textbf{W}$_N$ primary fields is not enough to obtain the correlation functions of descendents.
Fully solving Toda means being able  to construct
the complete set of correlation functions \textit{both} of primaries and descendants.   Obtaining the three point functions with descendants is very naturally done using topological strings and is work in progress \cite{FutoshiMasato}.

Since this article relates two somewhat disjointed fields, each used to its own notations, we wish to include a reader's guide to the other sections. 
We begin in section \ref{sec:AGT} with a presentation of the parametrizations and the precise relations between the partition functions of section \ref{sec:TNpartitionfunction} and the correlators of section \ref{sec:Toda3ptfunctions}. In the following section \ref{sec:Toda3ptfunctions}, we review shortly the Toda CFT, introduce the associated notation and make some observations regarding the symmetries of the correlation functions that to our knowledge are not available in the literature. We finish section \ref{sec:Toda3ptfunctions} by a discussion of the pole structures and the $q$-deformations of the correlation functions.
In section \ref{sec:TNpartitionfunction},  we give a short review of the derivation of the partition functions of the $T_N$ theories, rewrite them using the functions $\Up$ that in our opinion are the appropriate tools to use in this context. We then discuss their 4D limit.    In sections \ref{sec:W2} and \ref{sec:W3} we illustrate our claims for the two simplest cases with $N=2$ and $N=3$.
The reader can find a collection of useful formulas, notations and parametrizations in the in appendices.
Finally, the exact definition of the functions $\Up$ is given in appendix \ref{app:qUpsilon} together with a discussion of their properties.

%%%%%%%%%%%%%%%%%%%%%%%%%%%%%%%%%%%%%%%%%%%%%%%%%%%%
\section{The AGT dictionary}
\label{sec:AGT}
%%%%%%%%%%%%%%%%%%%%%%%%%%%%%%%%%%%%%%%%%%%%%%%%%%%%

The main goal of this section is to provide the dictionary needed to relate the topological string amplitudes of section \ref{sec:TNpartitionfunction} to the Toda CFT correlation functions of section \ref{sec:Toda3ptfunctions}. First, we review the parameters of the  Omega deformation. The circumference of the 5D circle is $\beta>0$ and the $\Omega$ background parameters are $\epsilon_1$ and $\epsilon_2$ from which we derive
\beq
\label{eq:deftq}
\fq\colonequals e^{-\beta \epsilon_1},\qquad \ft\colonequals e^{\beta \epsilon_2}.
\eeq
Furthermore, we need to also define\footnote{The  combinations  $\beta \,  \epsilon_{i}$ are dimensionless, but not $\beta$ or $\epsilon_{i}$ separately. In this paper we rescale them by the  dimensionful constant $\sqrt{\epsilon_1\epsilon_2}$ while keeping their product $\beta \,  \epsilon_{i}$ fixed so that each one of them $\beta$ and $\epsilon_{i}$ are separately dimensionless.
}
\beq
q\colonequals e^{-\beta},\qquad x\colonequals\sqrt{\frac{\fq}{\ft}}=q^{\frac{\epsilon_+}{2}},\qquad y\colonequals \sqrt{\fq\ft}=q^{\frac{\epsilon_-}{2}},
\eeq
with $\epsilon_{\pm}\colonequals\epsilon_{1}\pm \epsilon_2$, and
$q$ the $q$-deformation parameter. The combinations $x$ and $y$ are the natural variables, fugacities of the 5D superconformal index.
 When we need to relate the topological string partition functions to the Toda CFT correlators, the $\Omega$ background parameters need to be specialized as 
\beq
\label{eq:bcase}
\epsilon_1=b, \qquad  \epsilon_2=b^{-1},
\eeq
which implies in particular that $|q|<1$, $|\fq|<1$, $|\ft|>1$ and $|x|<1$ since we take $b$ to be positive. 

\begin{figure}[ht]
 \centering
  \includegraphics[height=4cm]{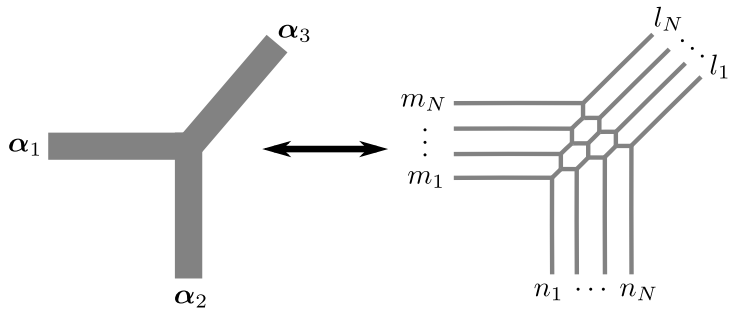}
  \caption{\it This figure depicts the identification of the $\balpha$ weights appearing on the Toda CFT side with the position of the flavor branes on the $T_N$ side, here drawn for the case $N=4$.}
  \label{fig:3point}
\end{figure}

On the Toda CFT side, see section \ref{sec:Toda3ptfunctions}, one uses the weights $\balpha_i$ parametrized by \eqref{eq:paramalphaN} to label the primary fields, while on the $T_N$ theory side, one uses the positions of the exterior branes, see section \ref{sec:TNpartitionfunction} and appendix \ref{app:notation}, as parameters. The rough relationship is illustrated in figure~\ref{fig:3point}
and the precise identifications are
\beq
\label{eq:identificationparameters}
\begin{split}
m_i&=\form{\balpha_1-\fQ}{h_i}=N\sum_{j=i}^{N-1}\alpha_1^j-\sum_{j=1}^{N-1}j\alpha_1^j-\frac{N+1-2i}{2}Q,\\
n_i&=-\form{\balpha_2-\fQ}{h_i}=-N\sum_{j=i}^{N-1}\alpha_2^j+\sum_{j=1}^{N-1}j\alpha_2^j+\frac{N+1-2i}{2}Q,\\
l_i&=\form{\balpha_3-\fQ}{h_i}=N\sum_{j=i}^{N-1}\alpha_3^j-\sum_{j=1}^{N-1}j\alpha_3^j-\frac{N+1-2i}{2}Q,
\end{split}
\eeq
where $h_i$ are the weights of the fundamental representation of SU$(N)$.
In appendix \ref{app:sln} the reader can find all the group theory conventions.
 In particular, for $N=2$, we have
\beq
\label{eq:paramcorrespondenceN2}
m_1=-m_2=\alpha_1^1-\frac{Q}{2},\qquad n_1=-n_2=-\alpha_2^1+\frac{Q}{2},\qquad l_1=-l_2=\alpha_3^1+\frac{Q}{2},
\eeq
while for $N=3$ we have
\beq
\label{eq:paramcorrespondenceN3}
m_1=2\alpha_1^1+\alpha_1^2-Q,\qquad m_2=-\alpha_1^1+\alpha_1^2,\qquad m_3=-\alpha_1^1-2\alpha_1^2+Q,
\eeq
with similar expressions for the $n_i$ and $l_i$.

Having set up the parametrization, we are ready to present our full claim. For that it is important to stress that
from the Toda CFT 3-point structure constants $C$, see \eqref{eq:3pointcorrelator}, we can extract the \textit{Weyl-invariant structure constants} $\fC$ as
\beq
C(\balpha_1,\balpha_2,\balpha_3)=\left(\left[\pi \mu \gamma(b^2)b^{2-2b^2}\right]^{\frac{\form{2\fQ}{\rho}}{b}}\prod_{i=1}^3Y(\balpha_i)\right)\times \fC(\balpha_1,\balpha_2,\balpha_3),
\eeq
with the functions $Y(\alpha)$ defined in \eqref{eq:deffunctionY} encoding all the information about the Weyl transformation.
All the details needed are introduced in section \ref{sec:Toda3ptfunctions}.
We claim that the exact AGT-W dictionary relates the Weyl-invariant structure constants  $\fC$ to the 4D $T_N$ partition function on $S^4$ ($\calZ_{N}^{S^4}$) as
\beq
\label{eq:mainequality}
\fC(\balpha_1,\balpha_2,\balpha_3)=\text{const}\times\calZ_{N}^{S^4}
\eeq
where the constant part can be a function of $N$ and of the Omega deformation parameters but {\it cannot} be a function of the masses. The partition function on $S^4$ itself is obtained from the partition function on $S^4\times S^1$, also called the 5D \textit{superconformal index}, by taking the appropriate limit when the circumference $\beta$ of the $S^1$ goes to zero:
\beq
\label{eq:def4DlimitcalZ}
\calZ_{N}^{S^4}=\text{const}\times \lim_{\beta\rightarrow 0} \beta^{-\frac{\chi_N}{\epsilon_1\epsilon_2}}\calZ_{N}^{S^4\times S^1}.
\eeq
The partition function $\calZ_{N}^{S^4\times S^1}$ is contained in \eqref{eq:defindex} and the power $\chi_N$ of the divergence in \eqref{eq:chiNgeneral}.
Moreover, as far as the 5D AGT-W dictionary is concerned, we need \eqref{eq:bcase} to set $b=\epsilon_1=\epsilon_2^{-1}$ and obtain 
\begin{align}
\label{eq:main5Dequality}
&\fC_q(\balpha_1,\balpha_2,\balpha_3)=\frac{C_q(\balpha_1,\balpha_2,\balpha_3)}{J_q(\balpha_1,\balpha_2,\balpha_3)}=\text{const}\times \frac{C_q(\balpha_1,\balpha_2,\balpha_3)}{\prod_{j=1}^3Y_q(\balpha_j)}=\text{const}\times (1-q)^{-\chi_N} \calZ_{N}^{S^4\times S^1} \, ,\nonumber\\
&\prod_{j=1}^3Y_q(\balpha_j) =\text{const}\times \left[\big(1-q^b\big)^{2b^{-1}}\big(1-q^{b^{-1}}\big)^{2b}\right]^{-\sum_{k=1}^3\form{\balpha_k}{\rho}}\big(1-q\big)^{\chi_N}|\calZ_{N}^{\text{dec}} |^2   \, ,
\end{align}
where again the constant parts can only depend on $N$ and of the Omega deformation parameters but {\it cannot} be functions of the parameters that define the theory, {\it i.e.} the masses. The $\fC_q$ are the $q$-deformed   Weyl-invariant structure constants \eqref{eq:qdeformedfC}, $J_q$ the $q$-deformation of the Weyl-covariant part of the structure constants  \eqref{eq:qdefdivisor}
 and $\calZ_{N}^{\text{dec}}$ the partition function of some extra degrees of freedom \eqref{eq:defnonfullspincontent} that are included in the topological string calculation but then decouple from the 5D theory. In \cite{Bao:2013pwa} we refer to them as {\it non-full spin content}. Note that the second line of \eqref{eq:main5Dequality} is the same as equation \eqref{eq:defnonfullspincontentv2}, where the constant factor is explicitly written.
 
 Finally, putting \eqref{eq:def4DlimitcalZ} and \eqref{eq:main5Dequality} together, we obtain the final identification
 \beq
 \label{eq:3pointfunctionsastopologicalstrings}
 \boxed{
 C(\balpha_1,\balpha_2,\balpha_3)=\,\text{const}\times \left(\pi \mu\gamma(b^2)b^{2-2b^2}\right)^{\frac{\form{2\fQ-\sum_{i=1}^3\balpha_i}{\rho}}{b}} \lim_{\beta\rightarrow 0}\frac{\left|\calZ^{\text{dec}}_N\right|^2\calZ_N^{S^4\times S^1}}{\beta^{2Q\sum_{i=1}^3\form{\balpha_i}{\rho}}}}
 \eeq
where the limit is well defined up to an overall divergent term that only depends on $\beta$ and $b$. The above equation gives the complete relationship between the Toda 3-point structure constants and the partition functions of the $T_N$ theories.

%%%%%%%%%%%%%%%%%%%%%%%%%%%%%%%%%%%%%%%%%%%%%%%%%%%%%%%
\section{Toda 3-point functions}
\label{sec:Toda3ptfunctions}
%%%%%%%%%%%%%%%%%%%%%%%%%%%%%%%%%%%%%%%%%%%%%%%%%%%%%%%

We begin this section with a review of known facts about Toda 3-point functions of three primaries that we will need in later sections.
We follow \cite{Fateev:2005gs,Fateev:2007ab,Fateev:2008bm} whenever possible. We then discuss the symmetry enhancement of the Weyl invariant part of the 3-point functions as well as it's pole structure. We conclude the section with a generalization of these facts for the $q$-deformed Toda.

%%%%%%%%%%%%%%%%%%%%%%%%%%%%%%%%%%%%%%%%%%%%%%%%%%%%%%%
\subsection{Review}
\label{subsec:review}
%%%%%%%%%%%%%%%%%%%%%%%%%%%%%%%%%%%%%%%%%%%%%%%%%%%%%%%

The Lagrangian of the Toda CFT theory is given by
\beq
\label{eq:Lagrangian}
L=\frac{1}{8\pi}\form{\partial_{\nu}\varphi}{\partial^{\nu}\varphi}+\mu\sum_{k=1}^{N-1}e^{b\form{e_k}{\varphi}},
\eeq
where $\varphi\colonequals \sum_{i=1}^{N-1}\varphi_i\omega_i$ and $e_k$, respectively $\omega_k$ are the simple roots, respectively fundamental weights of SU$(N)$. We have collected all useful definitions and notations in in appendix \ref{app:sln} for the convenience of the reader. The parameter $\mu$ is called the \textit{cosmological constant}. The theory defined by \eqref{eq:Lagrangian} is invariant under the exchange $b\leftrightarrow b^{-1}$, which sends the cosmological constant to its dual $\tilde{\mu}$, defined in such a way that
\beq
\left(\pi\tilde{\mu}\gamma(b^{-2})\right)^b\stackrel{!}{=}\left(\pi\mu\gamma(b^{2})\right)^{\frac{1}{b}}\Longrightarrow 
\tilde{\mu}=\frac{\left(\pi\mu\gamma(b^2)\right)^{\nicefrac{1}{b^2}}}{\pi\gamma(\nicefrac{1}{b^2})}.
\eeq
The Toda CFT has a \textbf{W}$_N$ higher spin chiral symmetry generated by the spin $k$ fields $W_2\equiv T$, $W_3, \ldots, W_N$. The  fields that are primary under $W_N$ are denoted by $V_{\balpha}$, are labeled by a weight of SU$(N)$, {\it i.e.} an $(N-1)$-component vector $\balpha$ and are given explicitly by 
\beq
\label{eq:defprimaryfield}
V_{\balpha}\colonequals e^{\form{\balpha}{\varphi}}.
\eeq
 For the sake of avoiding some fractions, we shall parametrize the weights $\balpha$ of the fields $V_{\balpha_i}$ entering the correlation functions as follows
\beq
\label{eq:paramalphaN}
\balpha_i=N\sum_{j=1}^{N-1}\alpha_i^j\omega_j.
\eeq
The central charge of the Toda CFT and the  conformal dimension of the primary fields are
\beq 
c=N-1+12\form{\fQ}{\fQ}=(N-1)\left(1+N(N+1)Q^2\right),\qquad  \Delta(\alpha)=\frac{\form{2\fQ-\balpha}{\balpha}}{2},
\eeq
where $\fQ\colonequals Q\rho=(b+b^{-1})\rho$ with  the Weyl vector $\rho$ defined in \eqref{eq:defWeylvector}.
The conformal dimension, as well as the eigenvalues of all the other higher spin currents $W_k$ are invariant under the affine\footnote{One should not confuse the affine Weyl tranformation, \textit{i.e.} Weyl reflections accompanied by two translations, with Weyl reflections belonging to the Weyl group of the affine Lie algebra.} Weyl transformations \eqref{eq:defaffineWeyltransformations} of the weights $\balpha_i$. Furthermore, the primary fields of Toda CFT  transform under an affine Weyl transformations $\balpha\rightarrow \fw\circ \balpha$ as follows
\beq
\label{eq:reflectionprimaryfield}
V_{\fw\circ \balpha} = \fR^{\fw}(\balpha) V_{\balpha}
\eeq
with the reflection amplitude $\fR$ given by the expression
\beq
\label{eq:defreflectionamplitude}
\fR^{\fw}(\balpha)\colonequals \frac{\fA(\balpha)}{\fA(\fw\circ\balpha)} \, .
\eeq
Here, as in \cite{Fateev:2008bm}, we define the function
\beq
\fA(\balpha)\colonequals \left(\pi \mu\gamma(b^2)\right)^{\frac{\form{\balpha-\fQ}{\rho}}{b}}\prod_{e>0}\Gamma\left(1-b\form{\balpha-\fQ}{e}\right)\Gamma\left(-b^{-1}\form{\balpha-\fQ}{e}\right).
\eeq

The 2-point correlation functions of primary fields are fixed by conformal invariance and by the normalization \eqref{eq:defprimaryfield}. They read
\beq
\label{eq:twopointfunctions}
\left\langle V_{\balpha_1}(z_1,\bar{z}_1)V_{\balpha_2}(z_2,\bar{z}_2)\right\rangle=\frac{(2\pi)^{N-1}\delta(\balpha_1+\balpha_2-2\fQ)+\text{Weyl-reflections}}{|z_1-z_2|^{4\Delta(\balpha_1)}},
\eeq
where ``Weyl-reflections'' stands for additional $\delta$-contributions that come from the field identifications \eqref{eq:reflectionprimaryfield}. 

In this article, we shall be mostly interested in the three point functions of primary fields. Their coordinate dependence is fixed by conformal symmetry up to an overall coefficient $C(\balpha_1,\balpha_2,\balpha_3)$ called the 3-point structure constants as
\beq
\label{eq:3pointcorrelator}
\left\langle V_{\balpha_1}(z_1,\bar{z}_1)V_{\balpha_2}(z_2,\bar{z}_2)V_{\balpha_3}(z_3,\bar{z}_3)\right\rangle=\frac{C(\balpha_1,\balpha_2,\balpha_3)}{|z_{12}|^{2(\Delta_1+\Delta_2-\Delta_3)}|z_{13}|^{2(\Delta_1+\Delta_3-\Delta_2)}|z_{23}|^{2(\Delta_2+\Delta_3-\Delta_1)}},
\eeq
where $z_{ij}\colonequals z_i-z_j$. 

Due to the property \eqref{eq:reflectionprimaryfield}, the 3-point structure constants are not invariant under affine Weyl reflections of the weights $\alpha_i$, but are instead \textit{covariant} and transform like the primaries themselves. As \cite{Fateev:2008bm}, we will find it advantageous to talk about the Weyl invariant part of the 3-point structure constants.  For that purpose, it is useful to define the functions\footnote{For the Liouville case, these functions are also introduced by AGT \cite{Alday:2009aq} and labeled by $f(\alpha)$.} $Y$ as 
\beq
\label{eq:deffunctionY}
\begin{split}
Y(\balpha)&\colonequals \left[\pi \mu \gamma(b^2)b^{2-2b^2}\right]^{-\frac{\form{\balpha}{\rho}}{b}}\prod_{e>0}\Upsilon\left(\form{\fQ-\balpha}{e}\right)\\
&=\left[\pi \mu \gamma(b^2)b^{2-2b^2}\right]^{-\frac{N}{2b}\sum_{j=1}^{N-1}\alpha^jj(N-j)}\prod_{k=1}^{N-1}\prod_{i=1}^{N-k}\Upsilon\left(kQ-N(\alpha^{i}+\cdots +\alpha^{i+k-1})\right),
\end{split}
\eeq
where the product in the first line goes over all $\frac{N(N-1)}{2}$ positive roots of SU$(N)$.
These functions obeys  the  same reflection property as the primary fields, {\it i.e.}
\beq
\label{eq:reflectionY}
Y(\fw\circ\balpha) = \fR^{\fw}(\balpha) Y(\balpha) \, .
\eeq
The transformation property \eqref{eq:reflectionY} under affine Weyl transformation can be easily derived for reflections on the simple roots $e_i$ by noting that  for any function $f$
\beq
\prod_{e>0}f(\form{\fQ-\balpha}{e})\mapsto \prod_{e>0}f(\form{\fQ-\balpha}{e-e_j\form{e_j}{e}})=\prod_{\substack{e>0\\e\neq e_j}}f(\form{\fQ-\balpha}{e})\times f(-\form{\fQ-\balpha}{e_j}) \,  ,
\eeq
where the transformation acts as $w_i\circ\balpha= \balpha-\form{\balpha-\fQ}{e_i}e_i$. After that one uses $\Upsilon(-x)=\Upsilon(x+Q)$ as well as equation \eqref{eq:extrashiftsUpsilon4D} to show \eqref{eq:reflectionY}. As a final remark on $Y(\balpha)$, we observe that this function is zero if $\balpha$ is a multiple of a fundamental weight and in particular it has a zero of order $\frac{(N-1)(N-2)}{2}$ if we set $\balpha=\kappa \omega_{1}$ or $\balpha=\kappa \omega_{N-1}$. 

Now, we can introduce the Weyl invariant part of the structure constants
\beq
\label{eq:deffC}
\fC(\balpha_1,\balpha_2,\balpha_3)\colonequals\frac{C(\balpha_1,\balpha_2,\balpha_3)}{ \left[\pi \mu \gamma(b^2)b^{2-2b^2}\right]^{\frac{\form{2\fQ}{\rho}}{b}}\prod_{i=1}^3Y(\balpha_i)}.
\eeq
by dividing out the piece that transforms non-trivially under Weyl transformations.
The function $\fC$ of the weights $\balpha$ is independent of the cosmological constant $\mu$ and is invariant under affine Weyl reflections in the $\balpha$. Anticipating a bit, we will show in the later sections that the Weyl invariant part of the 3-point structure constants has a higher symmetry than the naive  affine Weyl symmetry of SU$(N)^3$. In particular, for $N=2$ it is invariant under the SU(4) Weyl group, while for $N=3$ it is invariant under the $E_6$ Weyl group.

While the general formula for the 3-point structure constants of Toda CFT is not known, they have been computed in special cases. The formula for the structure constants of \textbf{W}$_N$  for the \textit{degenerate case} in which one of the three weights becomes proportional to the first or the last fundamental weight, {\it i.e.} $\balpha_3=\varkappa\omega_1$ or $\balpha_3=\varkappa \omega_{N-1}$ is known from \cite{Fateev:2005gs} and reads\footnote{In \cite{Fateev:2008bm} a more general formula was derived for $N=3$ for the case of semi-degenerate fields $\balpha_3=\varkappa \omega_2-mb\omega_1$ with $m$ integer. We will not need it here.}
\beq
\begin{split}
\label{eq:FLTodacorr}
C(\balpha_1,\balpha_2,\varkappa \omega_{N-1})=&\left(\pi\mu\gamma(b^2)b^{2-2b^2}\right)^{\frac{\form{2\fQ-\sum_{i=1}^3\balpha_i}{\rho}}{b}}\times \\&\times\frac{\Upsilon'(0)^{N-1}\Upsilon(\varkappa)\prod_{e>0}\Upsilon(\form{\fQ-\balpha_1}{e})\Upsilon(\form{\fQ-\balpha_2}{e})}{\prod_{i,j=1}^N\Upsilon(\frac{\varkappa}{N}+\form{\balpha_1-\fQ}{h_i}+\form{\balpha_2-\fQ}{h_j})}.
\end{split}
\eeq
We remark that in the limit in which the degenerate field becomes the identity, {\it i.e.} $\varkappa\rightarrow 0$
one can show that the 3-point structure constants \eqref{eq:FLTodacorr} converge to \eqref{eq:twopointfunctions}.

In the $N=2$ case, the degeneration doesn't matter since there is only one fundamental weight anyway and \eqref{eq:FLTodacorr} reduces to (we set $\varkappa=2\alpha_3$) the famous DOZZ formula\footnote{For $N=2$ we set $\balpha_i=2\alpha_i\omega_1$, {\it i.e.} we omit the unnecessary second index and set $\alpha_i^1\equiv \alpha_i$.} 
\beq
\label{eq:DOZZ}
C(\balpha_1,\balpha_2,\balpha_3)=\left(\pi\mu\gamma(b^2)b^{2-2b^2}\right)^{\frac{Q-\sum_{i=1}^3\alpha_i}{b}}\frac{\Upsilon'(0)\prod_{i=1}^3\Upsilon(2\alpha_i)}{\Upsilon(\sum_{i=1}^3\alpha_i-Q)\prod_{j=1}^3\Upsilon(\sum_{i=1}^3\alpha_i-2\alpha_j)},
\eeq
which was conjecture by \cite{Dorn:1994xn,Zamolodchikov:1995aa} and derived by \cite{Teschner:1995yf,Teschner:2001rv}.

%%%%%%%%%%%%%%%%%%%%%%%%%%%%%%%%%%%
\subsection{Enhanced symmetry of the Weyl invariant part}
%%%%%%%%%%%%%%%%%%%%%%%%%%%%%%%%%%%
\label{subsec:enchancedsymmetry}

In this subsection, we shall make a couple of observations on the symmetries of the Weyl invariant part of the structure constants that to our knowledge are not found in the literature.

In the Liouville case ($N=2$) the Weyl invariant piece of the structure constants \eqref{eq:deffC} take the form 
\beq
\label{eq:deffCT2}
\fC(\balpha_1,\balpha_2,\balpha_3)=\frac{\Upsilon'(0)}{\Upsilon(\alpha_1+\alpha_2+\alpha_3-Q)\Upsilon(\alpha_1+\alpha_2-\alpha_3)\Upsilon(\alpha_1-\alpha_2+\alpha_3)
\Upsilon(-\alpha_1+\alpha_2+\alpha_3)}.
\eeq
At this point,   we use \eqref{eq:paramcorrespondenceN2} and replace the $\alpha_i$ by the $m_1$, $n_1$ and $l_1$ as 
\beq
\alpha_1=m_1+\frac{Q}{2}\, ,\qquad \alpha_2=-n_1+\frac{Q}{2}\, ,\qquad \alpha_3=l_1+\frac{Q}{2} \, .
\eeq
Setting then 
\beq
\label{eq:mass2u}
m_1=\frac{u_1+u_3}{2}
\, , \quad
n_1=\frac{u_2+u_3}{2}
\, , \quad
l_1=\frac{u_1+u_2}{2}
\eeq
and using the symmetries of the $\Upsilon$ functions leads to the following compact expression for the Weyl invariant structure constants of the Liouville CFT
\beq
\label{eq:weylinvariantT2structureconstants}
\fC(\balpha_1,\balpha_2,\balpha_3)=\frac{\Upsilon'(0)}{\prod_{i=1}^4\Upsilon(u_i+\frac{Q}{2})},\quad \text{ where }\quad  \sum_{i=1}^4u_i=0.
\eeq
We observe that the above is invariant under the SU$(4)$ Weyl group that acts as the permutation group $S^4$ on the variables $u_i$. We have thus uncovered the presence of a ``hidden'' symmetry group.

\begin{figure}[ht]
 \centering
  \includegraphics[height=1.5cm]{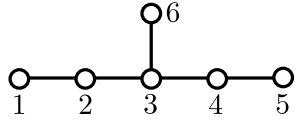}
  \caption{\it The figure shows the $E_6$ Dynkin diagram together with our labeling of the simple roots. }
\label{fig:E6Dynkin}
\end{figure}

The $N=3$ case is considerably more involved. For reasons that will become apparent shortly, we will label by an index $j=1,\,2,\,3$ the weights $h_i^{(j)}$ of  the three different SU$(3)$s that appear, \textit{i.e.} each $\balpha_j$ lives in its own copy of the SU$(3)$ weight space labeled by $j$. 
Using \cite{Fateev:2008bm}, we know that $\fC$ is invariant not only under SU(3) affine Weyl reflections of the $\balpha_j$'s, but also under the 27 new transformations
\beq
\label{eq:newtransformationFL}
\balpha_1\rightarrow \balpha_1-\varsigma_{ijk}h_i^{(1)},\qquad  
\balpha_2\rightarrow \balpha_2-\varsigma_{ijk}h_j^{(2)},\qquad 
\balpha_3\rightarrow \balpha_3-\varsigma_{ijk}h_k^{(3)},
\eeq
where $i$, $j$ and $k$ are fixed and we have defined
\beq
\varsigma_{ijk}\colonequals \form{\balpha_1-\fQ}{h_i^{(1)}}+\form{\balpha_2-\fQ}{h_j^{(2)}}+\form{\balpha_3-\fQ}{h_k^{(3)}}.
\eeq
We can now make the following set of observations. 
First, the affine SU$(3)$ Weyl transformations in the $\balpha_i$ become the usual SU$(3)$  Weyl reflections when expressed in the variables $m_i$, $n_i$ and $l_i$ defined via \eqref{eq:identificationparameters}, {\it i.e.} they act as the $S^3$ permutations. 
Using the parametrization \eqref{eq:identificationparameters}, we then observe that 
\beq
\varsigma_{ijk}=m_i-n_j+l_k, \quad \text{ where }\quad \sum_{i=1}^3m_i= \sum_{i=1}^3n_i=\sum_{i=1}^3l_i=0.
\eeq
Therefore, the transformation \eqref{eq:newtransformationFL} for a given choice of $i$, $j$ and $k$ acts of the variables $m_a$, $n_b$ and $l_c$ as
\beq
\label{eq:E6Weyltransformations}
\begin{split}
m_a&\rightarrow m_a-(m_i-n_j+l_k)\left(\delta_{ai}-\frac{1}{3}\right),\\
n_b&\rightarrow n_b+(m_i-n_j+l_k)\left(\delta_{bj}-\frac{1}{3}\right),\\
l_c&\rightarrow l_c-(m_i-n_j+l_k)\left(\delta_{ck}-\frac{1}{3}\right),
\end{split}
\eeq
where no sum over $i$, $j$, $k$ is to be taken.
We now want to interpret the new transformations \eqref{eq:newtransformationFL} as being the result of the (non-affine) action of the  Weyl group of $E_6$. Since the Weyl group is generated by the Weyl reflections associated to the simple roots, we only need to consider those. We have 9 weights $h_i^{(j)}$ subject to the three constraints $\sum_{i=1}^3h_i^{(j)}=0$ and we can build the $E_6$ root system from them as 
\begin{align}
&e^{E_6}_1=h_1^{(1)}-h_2^{(1)},&   &e^{E_6}_2=h_2^{(1)}-h_3^{(1)},& &e^{E_6}_3=h_3^{(1)}+h_3^{(2)}+h_1^{(3)},&\nonumber\\
&e^{E_6}_4=-h_1^{(3)}+h_2^{(3)},&   &e^{E_6}_5=-h_2^{(3)}+h_3^{(3)},& &e^{E_6}_6=h_2^{(2)}-h_3^{(2)},&
\end{align}
where we refer to figure~\ref{fig:E6Dynkin} for the numbering of the $E_6$ simple roots.
We observe that $\form{e^{E_6}_i}{e^{E_6}_j}$ is the Cartan matrix of $E_6$, if we require $\form{h_a^{(k)}}{h_b^{(l)}}=0$ if $k\neq l$. Therefore, we have constructed the $E_6$ root system within the space spanned by the $h_i^{(j)}$. Furthermore, we can obtain all the variables $m_i$, $n_i$ and $l_i$ by taking the scalar products $\form{\sum_{i=1}^3(\balpha_i-\fQ)}{e^{E_6}_j}$, where each $\balpha_k-\fQ$ is expressed only through the $h_i^{(k)}$. We find that Weyl reflections for the simple roots $e^{E_6}_i$ with $i\neq 3$ correspond to permutations of the $m$'s, $n$'s and $l$'s among themselves. However, the Weyl reflection corresponding to $e^{E_6}_3$ transforms the variables as
\begin{align}
&m_1\rightarrow m_1+\lambda,& &m_2\rightarrow m_2+\lambda,& &m_3\rightarrow m_3-2\lambda,&\nonumber\\
&n_1\rightarrow n_1-\lambda,& &n_2\rightarrow n_2-\lambda,& &n_3\rightarrow n_3+2\lambda,&\\
&l_1\rightarrow l_1-2\lambda,& &l_2\rightarrow l_2+\lambda,& &l_3\rightarrow l_3+\lambda,&
\end{align}
where $3\lambda=m_3-n_3+l_1$. We easily see that this transformation corresponds to \eqref{eq:newtransformationFL} for $i=3$, $j=3$ and $k=1$. The transformations corresponding to the other choices of $i$, $j$ and $k$ can be obtained by acting with some other $e^{E_6}_l$ first. Hence, the Weyl transformations of the three SU$(3)$ can be combined with \eqref{eq:newtransformationFL} to generate the Weyl group of the entire $E_6$.

For the cases $N\geq 4$ the full enhanced symmetry of the Weyl invariant structure constants is not completely known. We shall argue in the conclusions that the enhanced symmetry should contain $E_7$ in the case $N=4$ and $E_8$ for $N=6$. 

%%%%%%%%%%%%%%%%%%%%%%%%%%%%%%%%%%%%
\subsection{Pole structure of the Weyl invariant part}
%%%%%%%%%%%%%%%%%%%%%%%%%%%%%%%%%%%%

We see from \eqref{eq:weylinvariantT2structureconstants}, that the poles for the $N=2$ Liouville case are all captured by the expression
\beq
\label{eq:polestructureLiouville}
\left[\prod_{i=1}^4\Upsilon(u_i+\frac{Q}{2})\right]^{-1} =  \left[\prod_{h \in \mathbf{4} \oplus \overline{\mathbf{4}}} \textbf{G}\left(\frac{Q}{2} + \form{\sum_{i=1}^3(\balpha_i-\fQ)}{h}\right) \right]^{-1}  \, ,
\eeq
where we used the function $\textbf{G}(x)=\frac{1}{\Gamma_b(x)}$ with $\Upsilon(x)=\textbf{G}(x)\textbf{G}(Q-x)$ introduced in \cite{Fateev:2008bm}, see \eqref{eq:propertiesG}. The weights $h$ are SU(4) weights and in the fundamental representation $\mathbf{4}$ they are\footnote{Note that in order to get the suitably normalized scalar product for SU$(4)$, we need to define $\form{\balpha_1}{\balpha_2}^{\text{SU}(4)}\colonequals \nicefrac{\form{\balpha_1}{\balpha_2}^{\text{SU}(2)^3}}{2}$, {\it i.e.} we compute the scalar products as before and divide the answer by two.}
\begin{align}
&h^{\text{SU}(4)}_1=h_1^{(1)}-h_1^{(2)}+h_1^{(3)},& &h^{\text{SU}(4)}_2=-h_1^{(1)}+h_1^{(2)}+h_1^{(3)},&\nonumber\\
&h^{\text{SU}(4)}_3=h_1^{(1)}+h_1^{(2)}-h_1^{(3)},& &h^{\text{SU}(4)}_4=-h_1^{(1)}-h_1^{(2)}-h_1^{(3)},&
\end{align}
with the weights of $\overline{\mathbf{4}}$ being the negatives of the above.

Moving on to the case with $N=3$,  it was argued in \cite{Fateev:2008bm} that the pole structure of the full correlation function $C(\balpha_1,\balpha_2,\balpha_3)$ is given by 
\beq
\label{eq:polestructureToda3}
C(\balpha_1,\balpha_2,\balpha_3) =\mathfrak{F} \left[\prod_{i_1,i_2,i_3=1}^3\fZ\left(\sum_{k=1}^3(\balpha_k-\fQ,h_{i_k}^{(k)})\right)\right]^{-1
}=\mathfrak{F} \left[\prod_{i,j,k=1}^3\fZ(m_i-n_j+l_k)\right]^{-1},
\eeq
where $\mathfrak{F} $ is some unknown entire function and the function $\fZ$ is defined in \eqref{eq:deffrakZ}, 
$\fZ(x)\colonequals \textbf{G}(Q+x)\textbf{G}(Q-x)$.
Using the $E_6$ Weyl symmetry of $\fC$, it follows that the poles of $\fC$ are contained in 
\beq
\label{eq:polesofW3}
\mathfrak{C}(\balpha_1,\balpha_2,\balpha_3) \sim \left[\fZ(0)^3\prod_{i,j,k=1}^3\fZ(m_i-n_j+l_k)\prod_{i<j=1}^3\fZ(m_i-m_j)\fZ(n_i-n_j)\fZ(l_i-l_j)\right]^{-1},
\eeq
where $\fZ(0)^3$ is just convenient normalization.
We recognize in this expression the weights of the $78$-dimensional adjoint representation of $E_6$ expressed using the weights of SU$(3)^3\subset E_6$,
\beq
\label{eq:polestructureW3Weyl}
\mathfrak{C}(\balpha_1,\balpha_2,\balpha_3) \sim \frac{1}{\prod_{h \in \textbf{78}}  \textbf{G}\left(Q + \form{\sum_{i=1}^3(\balpha_i-\fQ)}{h}\right)}  \, .
\eeq
The additional poles introduced in \eqref{eq:polesofW3} are completely canceled by the Weyl covariant part  in the formula  \eqref{eq:deffC} relating them to the 3-point structure constants, because
\beq
\label{eq:nonfullspingame}
\prod_{k=1}^3Y(\balpha_k)\propto\frac{\prod_{i<j=1}^3\fZ(m_i-m_j)\fZ(n_i-n_j)\fZ(l_i-l_j)}{\prod_{k=1}^3\prod_{e>0}\form{\fQ-\balpha_k}{e}\Gamma\big(b\form{\fQ-\balpha_k}{e}\big)\Gamma\big(b^{-1}\form{\fQ-\balpha_k}{e}\big)}
\eeq
where we have used \eqref{eq:deffunctionY} and \eqref{eq:deffrakZ}. The proportionality factor in \eqref{eq:nonfullspingame} depends only on $\mu$ and $b$ and has no zeroes or poles while the additional factors of $\Gamma$ in the denominator of \eqref{eq:nonfullspingame} lead only to more zeroes of $\prod_{k=1}^3Y(\balpha_k)$. Thus, multiplying \eqref{eq:polesofW3} with the Weyl covariant part, see \eqref{eq:deffC}, in order to get the full 3-point structure constants will cancel the extra poles that we introduced.

 Finally, it is compelling to conjecture that for any $N$ the poles of the Weyl invariant structure constants should behave as
\beq
\label{eq:polestructureWNWeyl}
\mathfrak{C}(\balpha_1,\balpha_2,\balpha_3) \sim \frac{1}{\prod_{h \in \textbf{R}}  \textbf{G}\left(\frac{N-1}{2}Q + \form{\sum_{i=1}^3(\balpha_i-\fQ)}{h}\right)}  \, 
\eeq
for  an appropriate representation $\textbf{R}$ of SU$(N)^3$.

%%%%%%%%%%%%%%%%%%%%%%%%%%%%%%%%%%%%%%%%%%%%%%%%%%%%
\subsection{The \texorpdfstring{$q$}{q}-deformed Toda field theory}
\label{subsec:qdefToda}
%%%%%%%%%%%%%%%%%%%%%%%%%%%%%%%%%%%%%%%%%%%%%%%%%%%%

One of our goals in this paper is to show how to use the topological string formalism to solve the Toda field theory.   This will require a careful study of the $q$-deformed Toda correlation functions which topological strings naturally provide and to then learn how to take the $q\rightarrow 1$ limit. For this purpose here we generalize some of the formulas that we discussed in the previous sections. An incomplete list of references includes\cite{Awata:2009ur,Awata:2010yy,Mironov:2011dk,Itoyama:2013mca,Bao:2011rc,Nieri:2013yra,Bao:2013pwa,Nieri:2013vba,Aganagic:2013tta,Aganagic:2014oia,Taki:2014fva} .  This section goes hand in hand with  appendix \ref{app:qUpsilon}, where we define the $q$-deformed version of the $\Upsilon$ functions and discuss in detail its symmetry properties as well as its zeros. To our knowledge these formulas do not exist in the literature.

We begin by stressing some defining properties that all the $q$-deformed formulas must have:
\begin{itemize}
\item 
They must reproduce the exact undeformed formula in the $q\rightarrow 1$ limit. With no further prefactor, unless stated otherwise. That will be the case of the $C_q$ \eqref{Cq2C}.
\item 
For the N=2 case, they must give the known answers, insofar they are available \cite{Nieri:2013yra}.
\item
They must have exactly the same symmetries and transformation properties as the undeformed ones under
the (affine) Weyl, as well as the enhanced symmetry group.
\item
They must have their poles and zeros in the same place with the undeformed ones. To be more precise, the $q$-deformed functions have more zeroes/poles, specifically a whole tower of zeroes/poles for each zero/pole of the undeformed function as discussed in  \eqref{eq:zeroesofqUpsilon}.
The tower is generated by beginning with the undeformed  zero/pole and translating it by
$m\frac{2\pi i }{\log q} = - m \frac{2\pi i }{\beta}$, where $m$ is a positive integer.
 \end{itemize}
We moreover want to stress  
 that the $q$-deformed version of Toda field theory does not have a known
 Lagrangian description. Everything is defined algebraically in analogy to the usual case via a deformation of the \textbf{W}$_N$ algebra. Since no Lagrangian description is known for the $q$-deformed Toda field theory, we can compute everything up to overall factors that in the $q\rightarrow 1$ limit give  the cosmological constant. Thus, we define the 5D correlation functions  up to the  $\pi \mu \gamma(b^2)$ term, since they together form  the $b\rightarrow b^{-1}$ invariant combination.
 Explicitly, we have for the $q$-deformed 3-point structure constants
\beq
\label{Cq2C}
C_q(\balpha_1,\balpha_2,\balpha_3)\stackrel{q\rightarrow 1}{\longrightarrow}\left(\pi \mu\gamma(b^2)\right)^{-\frac{Q-\sum_i\alpha_i}{b}}C(\balpha_1,\balpha_2,\balpha_3) \, .
\eeq
Obviously, after the $q\rightarrow 1$ limit is taken and the undeformed answer is obtained, it is clear how one can put back the appropriate $\pi \mu \gamma(b^2)$ factors for a given correlation function, thus obtaining the full result with all the factors.

 As we already said in section \ref{subsec:review} the Weyl invariant part  $\fC$  is independent of the cosmological constant and thus it's $q$-deformed version should  be straightforward. However, the Weyl covariant part with which we need to multiply in order to obtain the full $C_q$ will converge to its undeformed version, up to an $\pi \mu \gamma(b^2)$ factor.
 In particular, the $q$-deformed version of the functions $Y$ defined in \eqref{eq:deffunctionY} is
\beq
\label{eq:deffunctionYqdef}
Y_q(\balpha)\colonequals \left[\frac{\big(1-q^b\big)^{2b^{-1}}\big(1-q^{b^{-1}}\big)^{2b}}{(1-q)^{2Q}}\right]^{-\form{\balpha}{\rho}}\prod_{e>0}\Up\left(\form{\fQ-\balpha}{e}\right),
\eeq
where the functions $\Up$ are introduced in \eqref{eq:defUp}. Using \eqref{eq:specialshiftidentityqUpsilon}, we find that this function behaves under affine Weyl transformations as  
\beq
Y_q(\fw\circ \balpha)=\fR_q^{\fw}(\balpha) Y_q( \balpha)
\eeq
with the $q$-deformed version of the reflection amplitude 
\beq
\label{eq:qreflection}
\fR_q^{\fw}(\balpha)\colonequals \frac{\fA_q(\balpha)}{\fA_q(\fw\circ\balpha)}
\eeq 
being composed out of
\beq
\fA_q(\balpha)\colonequals \prod_{e>0}\Gamma_{q^{b^{-1}}}\left(1-b\form{\balpha-\fQ}{e}\right)\Gamma_{q^b}\left(-b^{-1}\form{\balpha-\fQ}{e}\right).
\eeq
Note that also the $q$-deformed version of the reflection amplitude in the $q\rightarrow 1$ limit gives $\fR^{\fw}$ up to an overall $\pi \mu \gamma(b^2)$ factor.
The $q$-deformed factor that we need to divide by in order to get the Weyl invariant structure constants is
\beq
\label{eq:qdefdivisor}
J_q(\balpha_1,\balpha_2,\balpha_3)=\left[\frac{\big(1-q^b\big)^2\big(1-q^{b^{-1}}\big)^{2b^2}}{(1-q)^{2(1+b^2)}}\right]^{\frac{\form{2\fQ}{\rho}}{b}}\prod_{i=1} ^3Y_q(\balpha_i)=\text{const}\times  \prod_{i=1} ^3Y_q(\balpha_i),
\eeq
so that like in \eqref{eq:deffC}
\beq
\label{eq:qdeformedfC}
\fC_q(\balpha_1,\balpha_2,\balpha_3)\colonequals \frac{C_q(\balpha_1,\balpha_2,\balpha_3)}{J_q(\balpha_1,\balpha_2,\balpha_3)}.
\eeq

The $q$-deformation version of the Fateev and Litvinov formula \eqref{eq:FLTodacorr} for the 3-point correlation functions with one degenerate insertion reads
\beq
\begin{split}
\label{eq:qdefFLTodacorr}
C_q(\balpha_1,\balpha_2,\varkappa \omega_{N-1})=&\left(\frac{\big(1-q^b\big)^2\big(1-q^{b^{-1}}\big)^{2b^2}}{(1-q)^{2(1+b^2)}}\right)^{\frac{\form{2\fQ-\sum_{i=1}^3\balpha_i}{\rho}}{b}}\times \\&\times\frac{\Up'(0)^{N-1}\Up(\varkappa)\prod_{e>0}\Up(\form{\fQ-\balpha_1}{e})\Up(\form{\fQ-\balpha_2}{e})}{\prod_{i,j=1}^N\Up(\frac{\varkappa}{N}+\form{\balpha_1-\fQ}{h_i}+\form{\balpha_2-\fQ}{h_j})}.
\end{split}
\eeq
This formula to our knowledge does not appear anywhere else in the literature. We write it down as the unique formula that has the properties mentioned at the beginning of the section. First, it has its poles and zeros in the correct positions, see \eqref{eq:zeroesofUpsilon}. Second, it has the correct covariance properties under the affine Weyl symmetries of the non-degenerate fields \eqref{eq:qreflection}. Finally, for the $N= 2$ case, \eqref{eq:qdefFLTodacorr} reduces to the $q$-deformation of the DOZZ formula (up to  the $\mu$ dependence) 
\begin{multline}
\label{eq:qdefDOZZ}
C_q(\balpha_1,\balpha_2,\balpha_3)=\left(\frac{\big(1-q^b\big)^2\big(1-q^{b^{-1}}\big)^{2b^2}}{(1-q)^{2(1+b^2)}}\right)^{\frac{Q-\sum_{i=1}^3\alpha_i}{b}}\\\times \frac{\Up'(0)\prod_{i=1}^3\Up(2\alpha_i)}{\Up(\sum_{i=1}^3\alpha_i-Q)\prod_{j=1}^3\Up(\sum_{i=1}^3\alpha_i-2\alpha_j)},
\end{multline}
derived in \cite{Kozcaz:2010af}.
From it we can extract the $q$-deformed version of the Weyl invariant part using equation \eqref{eq:qdeformedfC},
\beq
\label{eq:qweylinvariantT2structureconstants}
\fC_q(\balpha_1,\balpha_2,\balpha_3)= \frac{\Up'(0)}{\Up(\sum_{i=1}^3\alpha_i-Q)\prod_{j=1}^3\Up(\sum_{i=1}^3\alpha_i-2\alpha_j)}
 =\frac{\Up'(0)}{\prod_{i=1}^4\Up(u_i+\frac{Q}{2})} \, ,
\eeq
which immediately gives the he correct undeformed $ \fC$
\beq
\fC_q (\balpha_1,\balpha_2,\balpha_3)  \quad  \stackrel{q\rightarrow 1}{\longrightarrow}    \quad \fC(\balpha_1,\balpha_2,\balpha_3)
\eeq
 as it is in equations \eqref{eq:deffCT2} and  \eqref{eq:weylinvariantT2structureconstants} with no further factors.

%%%%%%%%%%%%%%%%%%%%%%%%%%%%%%%%%%%%%%%%%%%%%%%%%%%%
\section{The \texorpdfstring{$T_{N}$}{TN} partition function from topological strings}
\label{sec:TNpartitionfunction}
%%%%%%%%%%%%%%%%%%%%%%%%%%%%%%%%%%%%%%%%%%%%%%%%%%%%
In this section we introduce the formula for the 5D  $T_{N}$ partition functions that we computed in  \cite{Bao:2013pwa} and we discuss how they can be brought to a form that allows us to take the 4D limit ($\beta \rightarrow 0$) in order to obtain  the  $T_{N}$ partition functions  on $S^4$.
Since the parametrization is crucial, we begin by carefully discussing it  and the way it is read  off from the web diagrams.
Some details of the computations are presented in appendix \ref{app:TNpartitionfunction}.

The $T_N$ theories are isolated strongly coupled fixed points that one can discover by taking the strong coupling limit of the $SU(N)^{N-2}$ or of the $U(N-1)\times U(N-2)\times \cdots\times U(1)$  linear quivers.
The calculation of the $T_N$ partition function is not possible using any purely field theoretic method currently known, because the $T_N$ theories have no known Lagrangian description. The only applicable method is string theory and in particular 5-brane webs \cite{Aharony:1997ju,Aharony:1997bh} from which the answer is derived using topological strings.

%%%%%%%%%%%%%%%%%%%%%%%%%%%%%%%%%
\subsection{The 5-brane webs}

A very short review of 5-brane webs is in order.
First, 5D $\mathcal{N}=1$  gauge theories can be embedded in string theory  by using type IIB
$(p, q)$ 5-brane webs \cite{Aharony:1997ju,Aharony:1997bh}. 
All the information needed to describe the low energy effective theory on the Coulomb branch is encoded in  the web diagrams, through which the 5D SW curves  can be easily derived \cite{Aharony:1997ju,Aharony:1997bh,Brandhuber:1997ua,Bao:2011rc}.
Furthermore, 5D  $\mathcal{N}=1$  gauge theories can also be realized using geometric engineering \cite{Katz:1996fh,Katz:1997eq}, in particular M-theory compactified on Calabi-Yau
threefolds. This alternative description provides an efficient way of computing the Nekrasov partition functions of the gauge theories by computing the partition functions of topological strings living on these backgrounds. Recent reviews on the subject can be found in \cite{Assi:2014exa,Kashani-Poor:2014lxa}.
In particular, the dual  to the Calabi-Yau toric diagram is exactly equal to the web diagram of the type IIB
$(p, q)$ 5-brane systems \cite{Leung:1997tw}.

The SW curves  and the Nekrasov partition functions are parametrized by the Coulomb moduli $a$ as well as the UV  masses $m$ and coupling constants $\tau$ of the gauge theory. These parameters are encoded in the web diagrams as follows.
On the one hand, deformations of the webs that do not change the asymptotic form of the 5-branes correspond to the Coulomb moduli $a$ and their number is the  number of {\it faces} of the web diagram.
On the other hand, deformations of the webs that do change the asymptotic form of the 5-branes correspond to parameters that define the theory, namely masses and coupling constants and they are equal to the number of external branes minus three.
Note that at each vertex there is a no-force condition (D5/NS5 ($p,q$) charge conservation) that serves to preserve 8 supersymmetries.

Having said all the above, we can now return to the  $T_{N}$  theories. The first step towards being able to calculate the  $T_{N}$ partition functions  was taken by Benini,
Benvenuti and Tachikawa, who  gave in \cite{Benini:2009gi} the web diagrams of the 5D $T_{N}$ theories. Subsequently, in \cite{Bao:2013pwa} we tested their proposal by deriving the corresponding SW curves and Nekrasov
partition functions.
Most importantly, we were able to cross-check our results for the partition functions against the 5D
superconformal index that was recently calculated in \cite{Kim:2012gu}. 
For similar work see also \cite{Hayashi:2013qwa,Hayashi:2014wfa}.

We now turn to the parametrization of the  $T_N$ web diagrams. The general parametrization in contained in the appendix, see figure~\ref{fig:TNparam} and here we just give a short introduction. We have one parameter $a_i^{(j)}$ for each face, or hexagon, of the diagram, that will also appear as $\tilde{A}_{i}^{(j)}= e^{-\beta a_i^{(j)} }$. They can be thought of as  Coulomb moduli that will be integrated over and are called {\it breathing modes}.
The number of faces in the web diagram of the $T_N$ theory is  $\frac{(N-1)(N-2)}{2}$.
  In addition, we have $3N$ parameters  $m_i$, $n_i$, $l_i$  labeling the positions of the exterior flavor branes for the branes on the, respectively, left, lower and upper right side of the diagram. From them, we define the fugacities
\beq
\tilde{M}_i\colonequals e^{-\beta m_i}, \qquad \tilde{N}_i\colonequals e^{-\beta n_i}, \qquad \tilde{L}_i\colonequals e^{-\beta l_i},
\eeq
that are subject to the relation 
\beq
\label{eq:externalparametersconstraintsfirst}
\prod_{k=1}^N\tilde{M}_k=\prod_{k=1}^N\tilde{N}_k=\prod_{k=1}^N\tilde{L}_k=1\Longleftrightarrow \sum_{k=1}^Nm_k=\sum_{k=1}^Nn_k=\sum_{k=1}^Nl_k=0.
\eeq
From the mass parameters, we  also define
the ``boundary'' Coulomb parameters. They are the $\tilde{A}_{i}^{(j)}$ with $i+j=N$,  with $i=0$ or with $j=0$ and are given as functions of the positions of the flavor branes in \eqref{eq:borderAdefMNL}.

In the dual, geometric engineering description, the parameters above correspond to  the  K\"ahler parameters of the Calabi-Yau threefold.
On the web diagram, the  K\"ahler parameters correspond to the horizontal, the diagonal and the vertical lines and are labeled by $Q_{n;i}^{(j)}$, $Q_{m;i}^{(j)}$ and $Q_{l;i}^{(j)}$ respectively. They are derived quantities through the equations \eqref{eq:PQR} and are useful because they are the ones that enter in the computation of the partition function via the topological vertex.

\begin{figure}[ht]
 \centering
  \includegraphics[height=6cm]{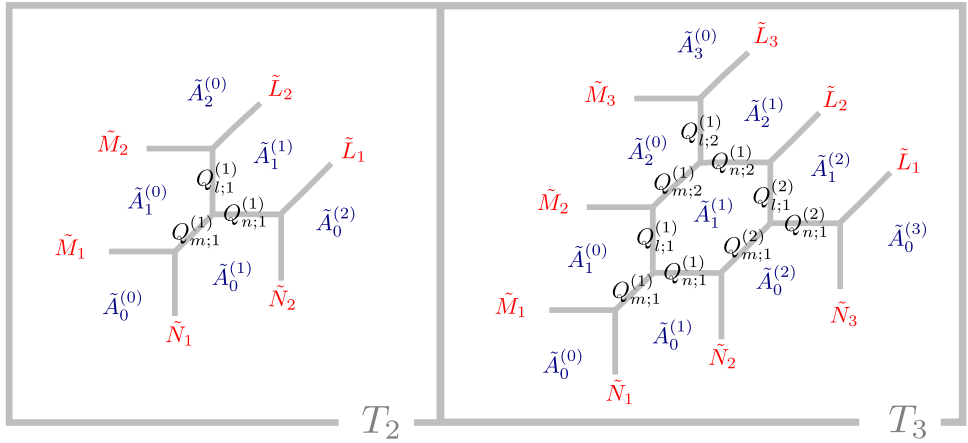}
  \caption{\it The parametrization and K\"ahler parameters of the $T_2$ and $T_3$ junctions. The external ``mass'' parameters are shown in red, the ``face'' moduli in blue and the ``edge'' ones in black.}
\label{fig:T23parm}
\end{figure}
 
   In order to familiarize the reader with the parametrization, we shall illustrate   the simplest cases $N=2$ and $N=3$ with some examples. The parametrization in those cases is contained in figure~\ref{fig:T23parm}.  For $N=2$, we see that we have no Coulomb moduli and the  K\"ahler parameters obey the relation
\beq
Q_{m;1}^{(1)} Q_{l;1}^{(1)} = \frac{\tilde{M}_1}{\tilde{M}_{2}},
\qquad
Q_{m;1}^{(1)} Q_{n;1}^{(1)} = \frac{\tilde{N}_1}{\tilde{N}_{2}},
\qquad
Q_{n;1}^{(1)} Q_{l;1}^{(1)} = \frac{\tilde{L}_1}{\tilde{L}_{2}}.
\eeq
Using \eqref{eq:externalparametersconstraintsfirst}, we find $Q_{m;1}^{(1)}=\frac{\tilde{M}_1\tilde{N}_1}{\tilde{L}_1}$, $Q_{n;1}^{(1)}=\frac{\tilde{M}_1\tilde{L}_1}{\tilde{N}_1}$  and $Q_{l;1}^{(1)}=\frac{\tilde{N}_1\tilde{L}_1}{\tilde{M}_1}$. For $N=3$ we have seven independent parameters: one Coulomb modulus $\tA\equiv \tilde{A}_{1}^{(1)}$  and $3\times (3-1)$ independent brane positions. A straightforward computation gives the nine K\"ahler parameters of the web diagram as 
\begin{align}
&Q_{m;1}^{(1)}=\tA^{-1}\tilde{M}_1\tilde{N}_1,& &Q_{m;2}^{(1)}=\tA\tilde{M}_2\tilde{L}_3,&
&Q_{m;1}^{(2)}=\tA\tilde{N}_2\tilde{L}_1^{-1},&\nonumber\\
&Q_{n;1}^{(1)}=\tA\tilde{M}_1^{-1}\tilde{N}_2^{-1},& &Q_{n;2}^{(1)}=\tA\tilde{M}_3\tilde{L}_2,&
&Q_{n;1}^{(2)}=\tA^{-1}\tilde{N}_3^{-1}\tilde{L}_1,&\\
&Q_{l;1}^{(1)}=\tA\tilde{M}_2^{-1}\tilde{N}_1^{-1},& &Q_{l;2}^{(1)}=\tA^{-1}\tilde{M}_3^{-1}\tilde{L}_3^{-1},&
&Q_{l;1}^{(2)}=\tA\tilde{N}_3\tilde{L}_2^{-1}.&\nonumber
\end{align}
It is easy to check that the above solutions obey the set of equations \eqref{eq:quotientMNL} relating them to the brane position parameters and that they furthermore satisfy the two constraints coming from matching the height and widths of the hexagon of figure~\ref{fig:T23parm}
\beq
Q_{m;1}^{(2)}Q_{n;1}^{(1)}=Q_{m;2}^{(1)}Q_{n;2}^{(1)},\qquad Q_{m;2}^{(1)}Q_{l;1}^{(1)}=Q_{m;1}^{(2)}Q_{l;1}^{(2)}.
\eeq

%%%%%%%%%%%%%%%%%%%%%%%%%%%%%%%%
\subsection{The topological vertex computation}
%%%%%%%%%%%%%%%%%%%%%%%%%%%%%%%%

Now that we have gained some understanding of the parametrization of the $T_N$ web diagram, we would like to compute its refined topological string amplitude. For this, we use the refined topological vertex, choose the preferred direction to be the horizontal one and cut the toric diagram diagonally into sub-diagrams called \emph{strips}. The calculation was carried out in \cite{Bao:2013pwa}, here we just reproduce the results for the reader's convenience.
\begin{figure}[ht]
 \centering
  \includegraphics[height=5.6cm]{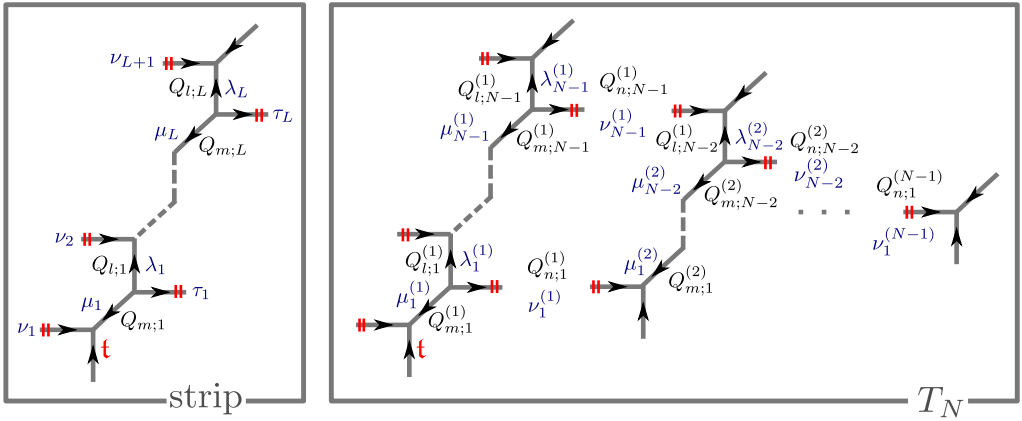}
  \caption{\it The left part of the figure shows the strip diagram, while the right one depicts the dissection of the $T_N$ diagram into $N$ strips. The partitions associated with the horizontal, diagonal and vertical lines are $\nu_{i}^{(j)}$, $\mu_{i}^{(j)}$ and $\lambda_{i}^{(j)}$ with $j=1,\ldots, N-1$, $i=1,\ldots, N-j$ respectively. The K\"ahler parameters of the horizontal, diagonal and vertical lines are $Q_{n;i}^{(j)}$, $Q_{m;i}^{(j)}$, $Q_{l;i}^{(j)}$ respectively with the same range of indices. }
\label{fig:strip}
\end{figure}
We consider the strip diagram of arbitrary length $L\geq 0$, drawn on the left in figure~\ref{fig:strip}. The corresponding partition function depends on the external horizontal partitions $\boldsymbol{\nu}=(\nu_1,\ldots, \nu_{L+1})$, $\boldsymbol{\tau}=(\tau_1,\ldots, \tau_{L})$ as well as the parameters $\boldsymbol{Q}_m=(Q_{m;1},\ldots, Q_{m;L})$ and $\boldsymbol{Q}_l=(Q_{l;1},\ldots, Q_{l;L})$. It takes the form
\begin{equation}
\label{eq:strippartitionfunction1}
\calZ_{\boldsymbol{\nu}\boldsymbol{\tau}}^{\text{strip}}(\boldsymbol{Q}_m, \boldsymbol{Q}_l;\ft,\fq)= \sum_{\boldsymbol{\lambda},\boldsymbol{\mu}}\prod_{i=1}^L(-Q_{m;i})^{|\mu_i|}(-Q_{l;i})^{|\lambda_i|}\prod_{j=1}^{L+1}C_{\mu_j^t\lambda_{j-1}^t\nu_j^t}(\fq,\ft)\prod_{k=1}^LC_{\mu_k\lambda_k\tau_k}(\ft,\fq),
\end{equation}
where $\mu_{L+1}=\lambda_0=\emptyset$. We refer to \cite{Iqbal:2007ii} for a definition of the topological vertex $C_{\lambda\mu\nu}$. The full topological string partition function is then given by 
\beq
\label{eq:toppartitionfunction}
\calZ_{N}^{\text{top}}=\sum_{\boldsymbol{\nu}}\prod_{r=1}^{N}\Big(-\boldsymbol{Q}_n^{(r)}\Big)^{|\boldsymbol{\nu}^{(r)}|}\calZ^{\text{strip}}_{\boldsymbol{\nu}^{(r-1)}\boldsymbol{\nu}^{(r)}}(\boldsymbol{Q}_m^{(r)}, \boldsymbol{Q}_l^{(r)};\ft,\fq).
\eeq
The strip partition function \eqref{eq:strippartitionfunction1} was computed in  \cite{Bao:2013pwa}. In appendix \ref{app:TNpartitionfunction}, we show that it is useful to redefine the strip slightly, {\it i.e.} to ``cut'' the $T_N$ junction in a different way by moving  some factors  from one strip to its neighbors. These redefinitions do not change the full topological string partition function of the $T_N$ junction. The technical details are left to appendix \ref{app:TNpartitionfunction}. Combining everything, we obtain
\beq
\label{eq:finalexpressionforZTN}
\calZ_N^{\text{top}}=\calZ_N^{\text{pert}}\calZ_N^{\text{inst}},
\eeq
where we have defined the ``perturbative'' partition function
\begin{align}
\label{eq:ZTNperturbative}
\calZ_N^{\text{pert}}&\colonequals \prod_{r=1}^{N-1}\prod_{i\leq j=1}^{N-r}\frac{\calM\Big(\frac{\tilde{A}_i^{(r-1)}\tilde{A}_j^{(r-1)}}{\tilde{A}_{i-1}^{(r-1)}\tilde{A}_{j+1}^{(r-1)}}\Big)}{\calM\Big(\sqrt{\frac{\ft}{\fq}}\frac{\tilde{A}_i^{(r-1)}\tilde{A}_{j-1}^{(r)}}{\tilde{A}_{i-1}^{(r-1)}\tilde{A}_{j}^{(r)}}\Big)\calM\Big(\sqrt{\frac{\ft}{\fq}}\frac{\tilde{A}_i^{(r)}\tilde{A}_j^{(r-1)}}{\tilde{A}_{i-1}^{(r)}\tilde{A}_{j+1}^{(r-1)}}\Big)}\prod_{i\leq j=1}^{N-r-1}\calM\Big(\frac{\ft}{\fq}\frac{\tilde{A}_i^{(r)}\tilde{A}_{j}^{(r)}}{\tilde{A}_{i-1}^{(r)}\tilde{A}_{j+1}^{(r)}}\Big),
\end{align}
and the ``instanton'' one
\begin{align}
\label{eq:ZTNsum}
\calZ_N^{\text{inst}}&\colonequals \sum_{\boldsymbol{\nu}}\prod_{r=1}^N\prod_{i=1}^{N-r}\left(\frac{\tilde{N}_r \tilde{L}_{N-r}}{\tilde{N}_{r+1}\tilde{L}_{N-r+1}}\right)^{\frac{|\nu_i^{(r)}|}{2}}\nonumber\\&\times \prod_{r=1}^N\prod_{i\leq j=1}^{N-r}\left[\frac{\tN_{\nu_i^{(r-1)}\nu_j^{(r)}}\Big(a_i^{(r-1)}+a_{j-1}^{(r)}-a_{i-1}^{(r-1)}-a_j^{(r)}-\nicefrac{\epsilon_+}{2}\Big)}{\tN_{\nu_i^{(r-1)}\nu_{j+1}^{(r-1)}}\Big(a_i^{(r-1)}+a_j^{(r-1)}-a_{i-1}^{(r-1)}-a_{j+1}^{(r-1)}\Big)}\right.\\&\times\left. \frac{\tN_{\nu_i^{(r)}\nu_{j+1}^{(r-1)}}\Big(a_i^{(r)}+a_j^{(r-1)}-a_{i-1}^{(r)}-a_{j+1}^{(r-1)}-\nicefrac{\epsilon_+}{2}\Big)}{\tN_{\nu_i^{(r)}\nu_j^{(r)}}\Big(a_i^{(r)}+a_{j-1}^{(r)}-a_{i-1}^{(r)}-a_{j}^{(r)}-\epsilon_+\Big)}\right],\nonumber
\end{align}
where the $a_{i}^{(j)}$ are defined via $\tilde{A}_i^{(j)}=e^{-\beta a_{j}^{(r)}}$. We put the words  ``perturbative'' and ``instanton'' inside quotation marks because for the $T_N$ there is not really a notion of  instanton expansion. There is no coupling constant, since there is no gauge group. We recall that the boundary $a_{i}^{(j)}$ are related to the masses via \eqref{eq:borderAdefMNL}. In writing \eqref{eq:ZTNperturbative} and \eqref{eq:ZTNsum} we have introduced the notation\footnote{We often drop the explicit dependence of these functions on the parameters $\ft$ and $\fq$.}
\beq
\begin{split}
\calM(u;\ft,\fq)&\equiv \calM(u)=\prod_{i,j=1}^{\infty}(1-u\ft^{-i}\fq^j),\\
\tN_{\lambda\mu}(m;\ft,\fq)&\equiv \tN_{\lambda\mu}(m)=\prod_{(i,j)\in \lambda}2\sinh\frac{\beta}{2}\left[m+\epsilon_1(\lambda_i-j+1)+\epsilon_2(i-\mu^t_j)\right]\\&\times \prod_{(i,j)\in \mu}2\sinh\frac{\beta}{2}\left[m+\epsilon_1(j-\mu_i)+\epsilon_2(\lambda^t_j-i+1)\right].
\end{split}
\eeq
We refer to appendix \ref{app:qUpsilon}, respectively \ref{app:finiteproduct} for more details concerning $\calM$, respectively $\tN_{\lambda\mu}$.

As in \cite{Bao:2013pwa}, we define the non-full spin content (also called U(1) factor in \cite{Hayashi:2013qwa})
\beq
\label{eq:defnonfullspincontent}
\calZ_{N}^{\text{dec}}\colonequals \prod_{i<j=1}^N\calM(\tilde{M}_i\tilde{M}_j^{-1})
\calM(\nicefrac{\ft}{\fq}\tilde{N}_i\tilde{N}_j^{-1})\calM(\tilde{L}_i\tilde{L}_j^{-1}).
\eeq
We remark that for $b=\epsilon_1=-\epsilon_2$, we can write 
\beq
\begin{split}
\label{eq:defnonfullspincontentv2}
\left|\calZ_{N}^{\text{dec}}\right|^2=&\Lambda^{\frac{3N(N-1)}{2}}(1-q)^{\frac{N(N-1)(2N^2-5)}{8}Q^2}\times\\&\times \prod_{k=1}^3\left(1-q\right)^{N\form{\balpha_k}{\balpha_k-2\fQ}}\left(\big(1-q^b\big)^{2b^{-1}}\big(1-q^{b^{-1}}\big)^{2b}\right)^{\form{\balpha_k}{\rho}}Y_q(\balpha_k)
\end{split}
\eeq
where we have used \eqref{eq:identificationparameters}, the identity \eqref{eq:SUNscalarproducts3} and the $q$-deformed function \eqref{eq:deffunctionYqdef}. Thus, up to some ambiguities, there is a clear identification of the decoupled part $\left|\calZ_{N}^{\text{dec}}\right|^2$ with the Weyl covariant part \eqref{eq:qdefdivisor} of the correlation functions, see \eqref{eq:main5Dequality}.

The contributions \eqref{eq:defnonfullspincontent} \textit{decouple} from the gauge theory and need to be removed in order to obtain the $S^4\times S^1$ partition function. In particular, using \eqref{eq:AfunctionMNL}, we find for the quotient
\beq
\label{eq:ZpertminusZtop}
\begin{split}
\frac{\calZ_{N}^{\text{pert}}}{\calZ_{N}^{\text{dec}}}=\prod_{r=1}^{N-1}\frac{\prod_{i\leq j=1}^{N-r-1}\calM\Big(\frac{\tilde{A}_i^{(r)}\tilde{A}_j^{(r)}}{\tilde{A}_{i-1}^{(r)}\tilde{A}_{j+1}^{(r)}}\Big)\calM\Big(\frac{\ft}{\fq}\frac{\tilde{A}_i^{(r)}\tilde{A}_{j}^{(r)}}{\tilde{A}_{i-1}^{(r)}\tilde{A}_{j+1}^{(r)}}\Big)}{\prod_{i\leq j=1}^{N-r}\calM\Big(\sqrt{\frac{\ft}{\fq}}\frac{\tilde{A}_i^{(r-1)}\tilde{A}_{j-1}^{(r)}}{\tilde{A}_{i-1}^{(r-1)}\tilde{A}_{j}^{(r)}}\Big)\calM\Big(\sqrt{\frac{\ft}{\fq}}\frac{\tilde{A}_i^{(r)}\tilde{A}_j^{(r-1)}}{\tilde{A}_{i-1}^{(r)}\tilde{A}_{j+1}^{(r-1)}}\Big)}\\
\times \left[\prod_{i<j=1}^{N}\calM\left(\frac{\ft}{\fq}\frac{\tilde{A}_0^{(i)}\tilde{A}_0^{(j-1)}}{\tilde{A}_0^{(i-1)}\tilde{A}_0^{(j)}}\right)\calM\left(\frac{\tilde{A}_i^{(N-i)}\tilde{A}_{j-1}^{(N-j+1)}}{\tilde{A}_{i-1}^{(N-i+1)}\tilde{A}_{j}^{(N-j)}}\right)\right]^{-1}.
\end{split}
\eeq
We now want to compute the norm squared of the above expression and write it in a way that would make the 4D limit more accessible. First, from the definition \eqref{eq:defUp} of the $q$-deformed $\Upsilon$ function, we see that
\beq
\label{eq:Mnormsquared}
|\calM(e^{-\beta x};\ft,\fq)|^2=|\calM(q^{-\frac{\epsilon_+}{2}};\ft,\fq)|^2(1-q)^{\frac{1}{\epsilon_1\epsilon_2}\left(x+\frac{\epsilon_+}{2}\right)^2} \Up(-x|\epsilon_1,\epsilon_2).
\eeq
Here and elsewhere, we shall use the notation
\beq
\label{eq:normsquared}
|f(u_1,\ldots, u_r;\ft,\fq)|^2\colonequals f(u_1,\ldots, u_r;\ft,\fq)f(u_1^{-1},\ldots, u_r^{-1};\ft^{-1},\fq^{-1}). 
\eeq
For the remainder of the section we shall write $\Up(x)$ instead of $\Up(x|\epsilon_1,\epsilon_2)$.  Since it will appear often, it is convenient to define
\beq
\Lambda\colonequals  |\calM(q^{-\frac{\epsilon_+}{2}};\ft,\fq)|^2.
\eeq
Furthermore, we need to carefully define the norm squared of the refined McMahon function in order to avoid a trivial zero. We follow \cite{Iqbal:2012xm} and define
\beq
\label{eq:McMahonUp}
\begin{split}
|M(\ft,\fq)|^2&\colonequals \lim_{u\rightarrow 1}\frac{|\calM(u;\ft,\fq)|^2}{1-u^{-1}}=|\calM(\fq^{-1};\ft,\fq)|^2=(1-q)^{\frac{\left(\epsilon_1-\epsilon_2\right)^2}{4\epsilon_1\epsilon_2}}\Lambda\Up(\epsilon_1).
\end{split}
\eeq
The advantage of using the functions $\Up$ is the fact that they have a well defined 4D limit $\beta\rightarrow 0$ or $q\rightarrow 1$, while the $\calM$ do not. We can apply this to the norm squared of \eqref{eq:ZpertminusZtop} with the result
\beq
\label{eq:normsquareofZpertZdec}
\begin{split}
&\left|\frac{\calZ_{N}^{\text{pert}}}{\calZ_{N}^{\text{dec}}}\right|^2=\Lambda^{-2N(N-1)}(1-q)^{\frac{\chi'_N}{\epsilon_1\epsilon_2}}\prod_{r=1}^{N-1}\left[\frac{ \prod_{i\leq j=1}^{N-r-1}\Up\left(a_{i-1}^{(r)}+a_{j+1}^{(r)}-a_i^{(r)} - a_j^{(r)}\right)}{\prod_{i\leq j=1}^{N-r}\Up\left(\frac{\epsilon_+}{2}+a_{i-1}^{(r-1)}+a_{j}^{(r)}-a_i^{(r-1)} - a_{j-1}^{(r)}\right)}\right.\\
&\times \left.\frac{\prod_{i\leq j=1}^{N-r-1}\Up\left(a_{i}^{(r)}+a_{j}^{(r)}-a_{i-1}^{(r)} - a_{j+1}^{(r)}\right)}{\prod_{i\leq j=1}^{N-r}\Up\left(\frac{\epsilon_+}{2}+a_{i}^{(r)}+a_{j}^{(r-1)}-a_{i-1}^{(r)} - a_{j+1}^{(r-1)}\right)}\right]\left[\prod_{i<j=1}^N\Up\left(n_i-n_j\right)\Up\left(l_j-l_i\right)\right]^{-1},
\end{split}
\eeq
with the exponent
\beq
\begin{split}
\chi'_N=&\sum_{r=1}^{N-1}\Big[\sum_{i\leq j=1}^{N-r}\left(a_i^{(r-1)}+a_j^{(r-1)}-a_{i-1}^{(r-1)}-a_{j+1}^{(r-1)}+\frac{\epsilon_+}{2}\right)^2\\&-\left(a_i^{(r-1)}+a_{j-1}^{(r)}-a_{i-1}^{(r-1)}-a_{j}^{(r)}\right)^2-\left(a_i^{(r)}+a_j^{(r-1)}-a_{i-1}^{(r)}-a_{j+1}^{(r-1)}\right)^2\\&+\sum_{i\leq j=1}^{N-r-1}\left(a_i^{(r)}+a_j^{(r)}-a_{i-1}^{(r)}-a_{j+1}^{(r)}-\frac{\epsilon_+}{2}\right)^2\Big]\\&-\sum_{i<j=1}^{N}\left[\left(m_i-m_j+\frac{\epsilon_+}{2}\right)^2+\left(n_j-n_i+\frac{\epsilon_+}{2}\right)^2+\left(l_i-l_j+\frac{\epsilon_+}{2}\right)^2 \right],
\end{split}
\eeq
that miraculously depends only on the boundary parameters
\beq
\label{eq:chiprimeN}
\begin{split}
\chi'_N&=-(N-1)\sum_{i=1}^Nm_i^2-\sum_{i<j=1}^{N}\left[\left(n_j-n_i+\frac{\epsilon_+}{2}\right)^2+\left(l_i-l_j+\frac{\epsilon_+}{2}\right)^2 \right]\\&-\frac{1}{N}\sum_{i<j=1}^N\left[(n_j-n_i)^2+(l_i-l_j)^2\right]-2\sum_{i=1}^Nn_il_{N+1-i}+\frac{N(N-1)(N-2)}{12}\epsilon_+^2.
\end{split}
\eeq
Now we have all the ingredients in order to compute the partition function on $S^4\times S^1$. First, we should remember that we need \cite{Iqbal:2012xm,Bao:2013pwa,Hayashi:2013qwa} to add a copy $\left|M(\ft,\fq)\right|^2$  of the norm squared of the refined McMahon function for each one of the $\frac{(N-1)(N-2)}{2}$ faces of the diagram and integrate over all the Coulomb moduli.  Then, the partition function on  $S^4\times S^1$ for the $T_N$ superconformal theory reads 
\beq
\label{eq:defindex}
\calZ_N^{S^4\times S^1}\colonequals \int_{-\frac{i\pi}{\beta}}^{\frac{i\pi}{\beta}}\prod_{k=1}^{N-2}\prod_{l=1}^{N-1-k}\frac{\beta da_k^{(l)}}{2\pi i}\left|M(\ft,\fq)\right|^{(N-1)(N-2)} \left|\frac{\calZ_N^{\text{pert}}}{\calZ_{N}^{\text{dec}}}\right|^2\left|\calZ_N^{\text{inst}}\right|^2,
\eeq
where we need to plug in  \eqref{eq:normsquareofZpertZdec} for the perturbative part $\left|\nicefrac{\calZ_N^{\text{pert}}}{\calZ_{N}^{\text{dec}}}\right|^2$, while we use \eqref{eq:ZTNsum} for the instanton part.  
The integrals over the $a_k^{(l)}$ originate as contour integrals $\oint \frac{d\tilde{A}_k^{(l)}}{2\pi i \tilde{A}_k^{(l)}}$ after the substitution $\tilde{A}_k^{(l)}=e^{-\beta a_k^{(l)}}$. 
Observe that there are $\frac{(N-1)(N-2)}{2}$ integrals to be done which is equal to the number of faces of the web diagram and that  in the simplest $T_2$ case
 no integrals have to be done. Furthermore, in order to compute the final expression for the partition function, we still need to perform $\frac{N(N-1)}{2}$ sums over the partitions $\nu_i^{(j)}$. This can unfortunately for now only be done exactly in the $N=2$ case. Finally, the derivation of \eqref{eq:finalexpressionforZTN} depended strongly on a choice of a preferred direction for the refined topological vertex. It is conjectured \cite{Iqbal:2007ii, Awata:2009yc}, under a principle called \textit{slicing invariance}, that the final answer will not depend on the choice of the preferred direction. We can make three different choices of preferred direction for the $T_N$ web diagram and in section \ref{sec:W3}, we shall do it for $T_3$. In the Toda field theory interpretation, each choice puts one of the primary fields on a special footing.

%%%%%%%%%%%%%%%%%%%%%%%%%%%%%%%%%%%%%%%%%%%%%%%%%%%%
\subsection{The 4D limit}
\label{subsec:TN4D}
%%%%%%%%%%%%%%%%%%%%%%%%%%%%%%%%%%%%%%%%%%%%%%%%%%%%

Naively, taking the 4D limit requires simply taking $\beta\rightarrow 0$. However, as we show in the previous subsection for the perturbative part for $N>2$ and for the full partition function for $N=2$ (see also section \ref{sec:W2}), in the limit most quantities diverge, but thankfully only with an overall factor of  $(1-q)$ raised to the appropriate power.
We conjecture that this will also be the case for the full partition function for every $N$, even after the instantons are accounted for.
Our conjecture is supported by symmetry arguments, a careful study of the $N=2$ case and from the lessons  we extracted form section \ref{subsec:qdefToda}, in particular equation \eqref{eq:qdefFLTodacorr}.
What is more, it is supported by \cite{Iqbal:2007ii}, where it was conjectured that the refined topological string partition function read off using the refined topological vertex from any web diagram should always at the end be possible to be written as a product of   $\mathcal{M}$'s \footnote{One might worry that the product would be infinite, but our symmetry argument that  $\chi_N$ should be given by  the quadratic Casimir suggests that  cancellations will always happen so that the degree of divergence  $\chi_N$ is finite!}.
Thus, we  define the partition function of the $T_N$ theory on $S^4$ to be
\beq
\label{eq:limit}
\calZ_{N}^{S^4}=\text{const}\times \lim_{\beta\rightarrow 0} \left(\beta^{-\frac{\chi_N}{\epsilon_1\epsilon_2}}\,\calZ_{N}^{S^4\times S^1}\right) \, ,
\eeq
where by definition the power $\chi_N$ is taken so that the limit is convergent. The constant factor cannot depend on the parameters of the theory, {\it i.e.} the masses, though it can, and in the cases checked does, depend on the Omega background parameters.

In what follows we want to use symmetries and the known limits for the partition function to argue that  the exponent $\chi_N$ of $\beta$ is given in terms of the quadratic Casimir of SU$(N)^3$ 
\beq
\label{eq:chiNgeneral}
\chi_N=-\sum_{i<j=1}^N\left[(m_i-m_j)^2+(n_j-n_i)^2+(l_i-l_j)^2\right]=-N\sum_{i=1}^3 \form{\balpha_i-\fQ}{\balpha_i-\fQ} \, .
\eeq
First, for the $N=2$ case, we can explicitly calculate the exponent and we find
\beq
\label{eq:chi2}
\chi_2=-\sum_{i<j=1}^2\left[(m_i-m_j)^2+(n_j-n_i)^2+(l_i-l_j)^2\right]=-2\sum_{i=1}^3 \form{\balpha_i-\fQ}{\balpha_i-\fQ}
\eeq
where we have made use of formulas \eqref{eq:identificationparameters} and \eqref{eq:SUNscalarproducts3}. 
Moreover,  for the perturbative part \eqref{eq:chiprimeN} we can also explicitly calculate $\chi'_N$ and we find that it is quadratic in the masses. 
What is more, we know the answer for the case with one degenerate insertion \eqref{eq:qdefFLTodacorr}, it is 
 expressed in terms of $\Up$-functions, which when combined with \eqref{eq:Mnormsquared} tells us that the power $\chi_N$ is a quadratic function in the masses.
 Furthermore,  both $\calZ_{N}^{S^4}$ and $\calZ_{N}^{S^4\times S^1}$ are invariant under (affine) Weyl reflections of SU$(N)^3$
  and since the constant term is independent of the parameters, the power $\chi_N$ has to be Weyl invariant as well. Therefore, we have to have\footnote{Usually the eigenvalue of the quadratic Casimir is written $\form{\balpha}{\balpha+2\rho}$, where $\rho$ is the Weyl vector. After a rescaling of the weight $\balpha$, this is the same as \eqref{eq:ansatzchi}.}
\beq
\label{eq:ansatzchi}
\chi_N=c_1\sum_{i=1}^3\form{\balpha_i}{\balpha_i-2\fQ}+c_2,
\eeq
where the $c_i$ are constants that symmetry cannot fix. The second constant $c_2 = -3N\form{\fQ}{\fQ}$ in \eqref{eq:chi2} can in any case be reabsorbed in the constant prefactor of \eqref{eq:limit} as it does not depend on the masses.
For $c_1$ we compare with \eqref{eq:qdefFLTodacorr}. When the $l$-parameters are degenerate\footnote{In that case $l_i=\frac{N-i}{N}\varkappa-\frac{N+1-2i}{2}Q$ for $i<N$ an $l_N=-\frac{N-1}{N}\varkappa +\frac{N-1}{2}Q$, implying $\varkappa=l_{N-1}-l_N+Q$.} 
\beq
\label{eq:testFL}
C_q(\balpha_1,\balpha_2,\varkappa \omega_{N-1})=\text{const}\times\left|\frac{\calM(\frac{\tilde{L}_N}{\tilde{L}_{N-1}})
\prod_{i<j=1}^N\calM\left(\frac{\tilde{M}_i}{\tilde{M}_j}\right)\calM\left(\frac{\ft}{\fq}\frac{\tilde{N}_i}{\tilde{N}_j}\right)}{\prod_{i,j=1}^N\calM\left(\tilde{M}_i^{-1}\tilde{N}_j\tilde{L}_1^{-1}\left(\frac{\ft}{\fq}\right)^{\frac{N-1}{2}}\right)}\right|^2\, .
\eeq
Since the $l$-part is degenerate and some zeroes from the non-full spin content have canceled some poles from the index in order to obtain \eqref{eq:testFL}, we don't expect to get the correct $l$-dependence in $\chi_N$. Thus, if we ignore $l$, subtract the remaining non-full spin content in the numerator for the $m$ and $n$ parts and compute the power of the $\beta$ divergence using \eqref{eq:Mnormsquared}, we obtain
\beq
-\sum_{i,j=1}^N\left(m_i-n_j+\frac{\epsilon_+}{2}\right)^2=-\sum_{i<j=1}^N\left[(m_i-m_j)^2+(n_i-n_j)^2\right]+\text{const} \, 
\eeq
which sets $c_1=-N$ and supports our claim \eqref{eq:chiNgeneral}.

We would like to conclude this section by stressing that even though in the present paper we do not show how to do the sums, we know that their outcome will be a product of functions $\mathcal{M}$, exactly as in \eqref{eq:testFL}, but of course for the general non-degenerate case with more  $\mathcal{M}$s.
That was already conjectured in  \cite{Iqbal:2007ii} for any topological partition function coming from  a toric diagram,  see \cite{Iqbal:2012mt} for a more recent discussion. This statement is just the refinement of the Gopakumar-Vafa formula \cite{Gopakumar:1998ii,Gopakumar:1998jq}. This is fully in agreement with our claim that the power $\chi_N$ has to be at most quadratic in the masses.

 %%%%%%%%%%%%%%%%%%%%%%%%%%%%%%%%%%%%%%%%%%%%%%%%%%%%
\section{Liouville from topological strings}
\label{sec:W2}
%%%%%%%%%%%%%%%%%%%%%%%%%%%%%%%%%%%%%%%%%%%%%%%%%%%%

In this section we show in detail how one can start from the partition function of $T_2$ that we computed in section \ref{sec:TNpartitionfunction} and derive the known Liouville 3-point function.
This exercise allows us to draw experience and learn some tricks that we shall be able to use for $N>2$, fix our conversions and
test the dictionary we presented in section \ref{sec:AGT}.

For $N=2$ there are no Coulomb moduli. The perturbative part \eqref{eq:ZTNperturbative} is
\beq
\calZ_2^{\text{pert}}=\frac{\calM\left(\left(\tilde{A}_1^{(0)}\right)^2\right)}{\calM\big(\sqrt{\frac{\ft}{\fq}}\frac{\tilde{A}_0^{(1)}\tilde{A}_1^{(0)}}{\tilde{A}_1^{(1)}}\big)\calM\big(\sqrt{\frac{\ft}{\fq}}\frac{\tilde{A}_1^{(0)}\tilde{A}_1^{(1)}}{\tilde{A}_0^{(1)}}\big)}
\eeq
while the instanton one \eqref{eq:ZTNsum} reads
\beq
\calZ_2^{\text{inst}}=\sum_{\nu}\left(\frac{\tilde{L}_1\tilde{N}_1}{\tilde{L}_2\tilde{N}_2}\right)^{\frac{|\nu|}{2}}\frac{\tN_{\nu\emptyset}(a_1^{(0)}+a_1^{(1)}-a_0^{(1)}-\frac{\epsilon_+}{2})\tN_{\emptyset\nu}(a_0^{(1)}+a_1^{(0)}-a_1^{(1)}-\frac{\epsilon_+}{2})}{\tN_{\nu\nu}(0)}
\eeq
so that \eqref{eq:finalexpressionforZTN}  becomes after replacing the $\tilde{A}$'s with the mass parameters via \eqref{eq:borderAdefMNL}
\beq
\begin{split}
\calZ_2^{\text{top}}=&\frac{\calM(\tilde{M}_1^2)}{\calM\big(\sqrt{\frac{\ft}{\fq}}\frac{\tilde{L}_1\tilde{M}_1}{\tilde{N}_1}\big)\calM\big(\sqrt{\frac{\ft}{\fq}}\frac{\tilde{N}_1\tilde{M}_1}{\tilde{L}_1}\big)}\\&\times\sum_{\nu}\left(\tilde{L}_1\tilde{N}_1\right)^{|\nu|}\frac{\tN_{\nu\emptyset}(l_1+m_1-n_1-\frac{\epsilon_+}{2})\tN_{\emptyset\nu}(n_1+m_1-l_1-\frac{\epsilon_+}{2})}{\tN_{\nu\nu}(0)}.
\end{split}
\eeq
We can use the identity of equation \eqref{eq:Masatostrick2} to perform the sum over partitions and get
\beqa
\label{eq:toppartfunctT2nonfullspin}
\calZ_2^{\text{top}}
&=&\frac{\calM(\tilde{M}_1^2)\calM(\tilde{N}_1^2\frac{\ft}{\fq})\calM(\tilde{L}_1^2) }{\calM(\frac{\tilde{M}_1\tilde{L}_1}{\tilde{N}_1}\sqrt{\frac{\ft}{\fq}})\calM(\frac{\tilde{N}_1\tilde{L}_1}{\tilde{M}_1}\sqrt{\frac{\ft}{\fq}})\calM(\frac{\tilde{M}_1\tilde{N}_1}{\tilde{L}_1}\sqrt{\frac{\ft}{\fq}})\calM(\tilde{M}_1\tilde{N}_1\tilde{L}_1\sqrt{\frac{\ft}{\fq}})}.
\eeqa
Setting $b=\epsilon_1=\epsilon_2^{-1}$ and using the definition \eqref{eq:defUp} of the $\Up$ functions as well as the parametrization \eqref{eq:identificationparameters}, we get from \eqref{eq:toppartfunctT2nonfullspin} the following expression for $|\calZ_2^{\text{top}}|^2$
\beq
|\calZ_2^{\text{top}}|^2
=\Lambda^{-1}(1-q)^{2Q(\sum_{i=1}^3\alpha_i-Q)-\frac{Q^2}{4}}\frac{\prod_{i=1}^3\Up(2\alpha_i)}{\Up(\sum_k\alpha_k-Q)\prod_{i=1}^3\Up(\sum_k\alpha_k-2\alpha_i)}
\eeq
where we have used the symmetry $\Up(x)=\Up(Q-x)$. Up to an infinite constant prefactor, the same formula was obtained in equation (3.73) of \cite{Nieri:2013yra}  as well as equation (5.10) of \cite{Kozcaz:2010af}. We can use the expression for the derivative \eqref{eq:derivativeofUp} found in the appendix as well as the McMahon function \eqref{eq:McMahonUp} to combine some factors into $\Up'(0)$ leading to
\beq
\left|\calZ_2^{\text{top}}\right|^2=
\frac{(1-q)^{2Q(\sum_{i=1}^3\alpha_i-Q)}}{\beta|M(\ft,\fq)|^2}\frac{\Up'(0)\prod_{i=1}^3\Up(2\alpha_i)}{\Up(\sum_k\alpha_k-Q)\prod_{i=1}^3\Up(\sum_k\alpha_k-2\alpha_i)}.
\eeq
This result is almost the $q$-deformed structure constants. In fact, we see by looking at \eqref{eq:qdefDOZZ} that
\beq
\label{eq:identificationZtopqDOZZ}
C_q(\balpha_1,\balpha_2,\balpha_3)=\left[\beta|M(\ft,\fq)|^2\left(\big(1-q^b\big)^{2b^{-1}}\big(1-q^{b^{-1}}\big)^{2b}\right)^{Q-\sum_i\alpha_i}\right]\left|\calZ_2^{\text{top}}\right|^2,
\eeq
which is the $q$-deformed version of \eqref{eq:3pointfunctionsastopologicalstrings} for $T_2$.

Already in \cite{Bao:2013pwa} we computed the superconformal index for the $T_2$ theory. It is obtained from $|\calZ_2^{\text{top}}|^2$ by dividing with the non-full spin content  $|\calZ_2^{\text{dec}}|^2$ that corresponds to degrees of freedom that decouple from the 5D theory. 
\begin{align}
\label{eq:indexT2}
\calZ_2^{S^4\times S^1}=&       
\left|\left[\calM\Big(\frac{\tilde{M}_1\tilde{N}_1}{\tilde{L}_1}\sqrt{\frac{\ft}{\fq}}\Big)\calM\Big(\frac{\tilde{M}_1\tilde{L}_1}{\tilde{N}_1}\sqrt{\frac{\ft}{\fq}}\Big)\calM\Big(\frac{\tilde{N}_1\tilde{L}_1}{\tilde{M}_1}\sqrt{\frac{\ft}{\fq}}\Big)\calM\Big(\tilde{M}_1\tilde{N}_1\tilde{L}_1\sqrt{\frac{\ft}{\fq}}\Big)\right]^{-1}\right|^2\nonumber\\
=&\Bigg[\Lambda^{4}(1-q)^{4(m_1^2+n_1^2+l_1^2)}\Up\left(\frac{Q}{2}+m_1+n_1+l_1\right)\Up\left(\frac{Q}{2}+m_1+n_1-l_1\right)\nonumber\\&\times \Up\left(\frac{Q}{2}+m_1-n_1+l_1\right)\Up\left(\frac{Q}{2}-m_1+n_1+l_1\right)\Bigg]^{-1}
\nonumber\\
=&\Bigg[\Lambda^{4}(1-q)^{\sum_{i=1}^4 u_i^2} \prod_{i=1}^4\Up\left(\frac{Q}{2}+u_i\right)\Bigg]^{-1}
.
\end{align}
In particular, we find by comparing with \eqref{eq:qweylinvariantT2structureconstants}
\beq
\fC_q(\balpha_1,\balpha_2,\balpha_3)=\Lambda^{4}\Up'(0)(1-q)^{-\chi_2}\calZ_2^{S^4\times S^1},
\eeq
as promised in \eqref{eq:mainequality}.
The index \eqref{eq:indexT2} of the $T_2$ theory can be expanded in powers of $x=\sqrt{\frac{\fq}{\ft}}$ with coefficients that can be interpreted either as sums of characters of SU$(2)\times$SU$(2)\times$SU$(2)$ or of SU$(4)$. Specifically, we find:
\begin{align}
\calZ_2^{S^4\times S^1}=&1+\chi_{({\bf 2,2,2})}x+\Big[\chi_{({\bf 3,1,1})}+\chi_{({\bf 1,3,1})}+\chi_{({\bf 1,1,3})}+\chi_{({\bf 3,3,3})}+\chi_{\bf 2}(y)\chi_{({\bf 2,2,2})}\Big]x^2+\calO(x^3)\nonumber\\
=&1+\left[\chi^{\text{SU}(4)}_{\bf  4}+\chi^{\text{SU}(4)}_{\bf  \bar{4}}\right]x+\Big[1+\chi^{\text{SU}(4)}_{\bf  10}+\chi^{\text{SU}(4)}_{\bf 15}+\chi^{\text{SU}(4)}_{\bf  \overline{10}}\\&+\chi_{\bf 2}(y)\big(\chi^{\text{SU}(4)}_{\bf  4}+\chi^{\text{SU}(4)}_{\bf  \bar{4}}\big)\Big]x^2+\calO(x^3),\nonumber
\end{align}
where we have the SU$(2)^3$ characters $\chi_{({\bf m_1,m_2,m_3})}\equiv \chi_{\bf m_1}(\tilde{M}_1)\chi_{\bf m_2}(\tilde{N}_1)\chi_{\bf m_3}(\tilde{L}_1)$ and the SU$(4)$ characters depend on the four variables $U_i$ with $\prod_{i=1}^4U_i=1$ that are given by
\beq
\tilde{M}_1=\sqrt{U_1U_3},\qquad \tilde{N}_1=\sqrt{U_2U_3},\qquad \tilde{L}_1=\sqrt{U_1U_2}.
\eeq
We thus also see by comparing with \eqref{eq:weylinvariantT2structureconstants} that the index $\calZ_2^{S^4\times S^1}$ has the same symmetry as the Weyl-invariant structure constants of the Liouville CFT \eqref{eq:weylinvariantT2structureconstants}, as was expected.

%%%%%%%%%%%%%%%%%%%%%%%%%%%%%%%%%%%%%%%%%%%%%%%%%%%%
\section{\texorpdfstring{\textbf{W}$_3$}{W3} from topological strings}
\label{sec:W3}
%%%%%%%%%%%%%%%%%%%%%%%%%%%%%%%%%%%%%%%%%%%%%%%%%%%%

In this section, we want to review our result for the 3-point structure constants of primaries for the case $N=3$ in its full glory and to investigate its symmetries and structure.

%%%%%%%%%%%%%%%%%%%%%%%%%%%%%%%%%%%%%%%%%%%%%%%%%%%%
\subsection{Slicing invariance}
%%%%%%%%%%%%%%%%%%%%%%%%%%%%%%%%%%%%%%%%%%%%%%%%%%%%

\begin{figure}[htbp]
  \centering
\includegraphics[height=5cm]{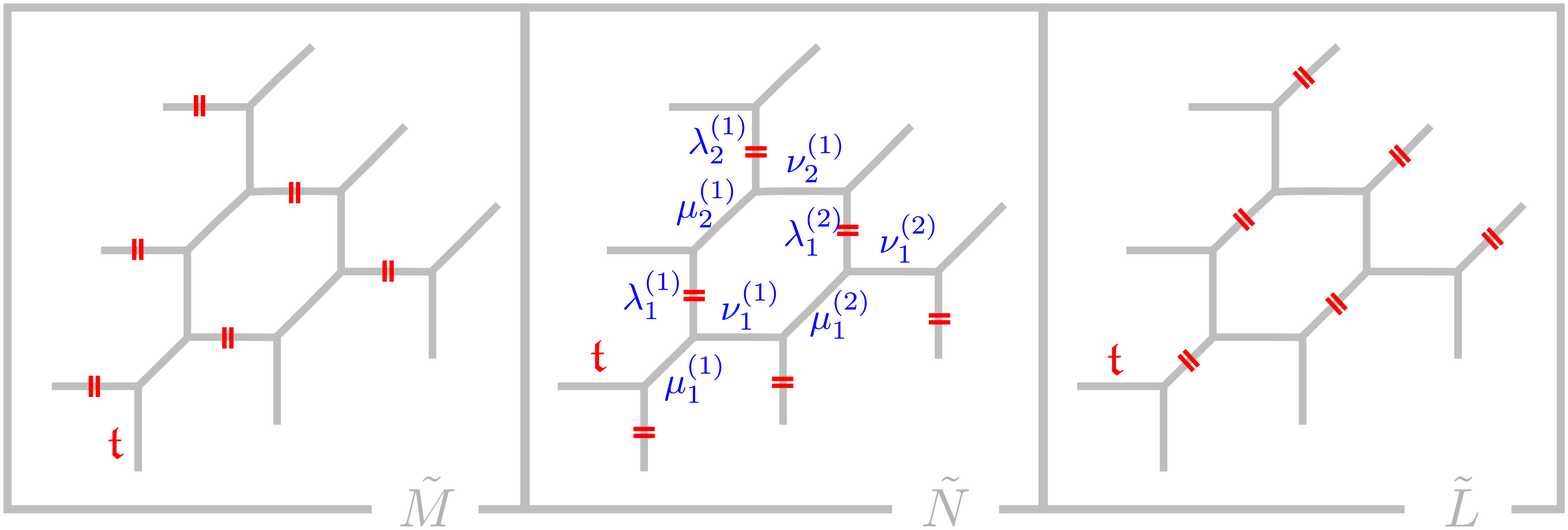}
  \caption{This figure shows the three different possible preferred directions for the $T_3$ junction. Each one is labeled by the K\"ahler moduli of the non-full spin content that is factorized. We also indicate the names of the partitions entering the instanton sums and to avoid clutter, we only do it for the middle one. }
  \label{fig:T3slices}
\end{figure}

We begin with slicing invariance. In figure~\ref{fig:T3slices}, we depict the three possible ways of choosing the preferred direction. Each one is labeled by the mass parameters that become prominent for that choice. The one we have used in section \ref{sec:TNpartitionfunction} for the determination of the strip partition functions is $\tilde{M}$.
For the choice $\tilde{M}$ of the preferred direction, we can compute the sum over the partitions $\lambda_i^{(j)}$ and  $\mu_i^{(j)}$, but not over  $\nu_i^{(j)}$. Similarly, for the choice $\tilde{N}$, we cannot perform the sum over the $\lambda_i^{(j)}$ and for the choice $\tilde{L}$ we cannot do it for the $\mu_i^{(j)}$. From equations \eqref{eq:ZTNperturbative} and \eqref{eq:ZTNsum}, we can read off the partition function for the $\tilde{M}$ choice. After some rearrangements, we find the cumbersome expression
\beqa
\label{eq:horribleT3amplitude}
\calZ_{3}^{\text{top}}&=& \frac{ \calM\left(\bA^2\tilde{N}_1^{-1} \tilde{L}_3\right) 
}{
\calM\left(\sqrt{\frac{\ft}{\fq}} \bA^{-1}\tilde{M}_1\tilde{N}_1\right)
\calM\left(\sqrt{\frac{\ft}{\fq}} \bA\tilde{M}_2^{-1}\tilde{N}_1^{-1}\right) 
\calM\left(\sqrt{\frac{\ft}{\fq}}\bA\tilde{M}_3^{-1}\tilde{N}_1^{-1}\right) 
 \calM\left(\sqrt{\frac{\ft}{\fq}}\bA\tilde{N}_3\tilde{L}_2^{-1}\right) }
\nonumber\\&&
\times  
 \frac{\calM\left(\frac{\ft}{\fq}\bA^2\tilde{N}_1^{-1} \tilde{L}_3\right) }{  \calM\left(\sqrt{\frac{\ft}{\fq}}\bA\tilde{M}_1\tilde{L}_3\right)\calM\left(\sqrt{\frac{\ft}{\fq}}\bA\tilde{M}_2 \tilde{L}_3\right)\calM\left(\sqrt{\frac{\ft}{\fq}}\bA^{-1}\tilde{M}_3^{-1}\tilde{L}_3^{-1}\right)
\calM\left(\sqrt{\frac{\ft}{\fq}}\bA\tilde{N}_2\tilde{L}_1^{-1}\right) }
\nonumber\\
&&\times \calM\left(\tilde{M}_1\tilde{M}_2^{-1}\right) \calM\left(\tilde{M}_1\tilde{M}_3^{-1}\right) \calM\left(\tilde{M}_2\tilde{M}_3^{-1}\right) \sum_{\boldsymbol{\nu}} \left(\frac{\tilde{N}_2\tilde{L}_1}{\tilde{N}_3\tilde{L}_2}\right)^{\frac{|\nu_1^{(2)}|}{2}}\left(\frac{\tilde{N}_1\tilde{L}_2}{\tilde{N}_2\tilde{L}_3}\right)^{\frac{|\nu_1^{(1)}|+|\nu_2^{(1)}|}{2}} \\
&&
\times 
\frac{\tN_{\nu_1^{(2)}\nu_2^{(1)}}\left(\ba+n_3-l_2-\nicefrac{\epsilon_+}{2}\right) \tN_{\nu_1^{(1)}\nu_1^{(2)}}\left(\ba+n_2-l_1-\nicefrac{\epsilon_+}{2}\right) }{\tN_{\nu_1^{(2)}\nu_1^{(2)}}\left(0\right)}
\nonumber\\
&&\times 
\frac{\prod_{k=1}^3\tN_{\nu_1^{(1)}\emptyset}\left(\ba-n_1-m_k-\nicefrac{\epsilon_+}{2}\right)\tN_{\emptyset\nu_2^{(1)}}\left(\ba+l_3+m_k-\nicefrac{\epsilon_+}{2}\right)}{\tN_{\nu_1^{(1)}\nu_1^{(1)}}\left(0\right)\tN_{\nu_2^{(1)}\nu_2^{(1)}}\left(0\right)\tN_{\nu_1^{(1)}\nu_2^{(1)}}\left(2\ba-n_1+l_3\right)\tN_{\nu_2^{(1)}\nu_1^{(1)}}\left(-2\ba+n_1-l_3\right)},\nonumber
\eeqa
where $\bA=e^{-\beta \ba}=\tilde{A}_1^{(1)}$ is the relabeled Coulomb modulus.

A direct computation using the topological vertex shows that the topological amplitude $\calZ_{3}^{\text{top}}$ for the  choice $\tilde{N}$ of preferred direction can be obtained from \eqref{eq:horribleT3amplitude} after the substitution
\begin{align}
\label{eq:slicingsymmetry1}
&m_k\rightarrow n_{4-k},&  &n_k\rightarrow m_{k},&  &l_k\rightarrow -l_{4-k},& 
\end{align}
 after exchanging $\ft\leftrightarrow \fq$. Furthermore, the amplitude $\calZ_{3}^{\text{top}}$ for the last remaining possible choice of preferred direction is obtained by setting in \eqref{eq:horribleT3amplitude}
\begin{align}
\label{eq:slicingsymmetry2}
&m_k\rightarrow -l_{4-k},&  &n_k\rightarrow -m_{4-k},&  &l_k\rightarrow n_{k},& 
\end{align}
without exchanging $\ft\leftrightarrow \fq$. Since it is thought and in some cases shown \cite{Iqbal:2007ii, Awata:2009yc} that the choice of preferred direction is irrelevant, the transformations \eqref{eq:slicingsymmetry1} and \eqref{eq:slicingsymmetry2} must be symmetries of the topological amplitude, {\it i.e.} we conjecture that \eqref{eq:horribleT3amplitude} is invariant under them:
\beq
\calZ_3^{\text{top}}(m_k,n_k,l_k)= \calZ_3^{\text{top}}(n_{4-k},m_k,-l_{4-k})=\calZ_3^{\text{top}}(-l_{4-k},-m_{4-k},n_{k}).
\eeq

%%%%%%%%%%%%%%%%%%%%%%%%%%%%%%%%%%%%%%%%%%%%%%%%%%%%
\subsection{Weyl invariance}
%%%%%%%%%%%%%%%%%%%%%%%%%%%%%%%%%%%%%%%%%%%%%%%%%%%%

The slicing invariance of the partition function can help us prove the Weyl covariance of the structure constants.
Using \eqref{eq:horribleT3amplitude} and the properties of the $\Up$ functions, the index
\beqa
\calZ_3^{S^4\times S^1}=\oint\frac{d\bA}{2\pi i \bA}\left|M(\ft,\fq)\right|^2\left|\frac{\calZ_{3}^{\text{top}}}{\calZ_{3}^{\text{dec}}}\right|^2
\eeqa
reads
\begin{align}
\label{eq:calIT3}
\calZ_3^{S^4\times S^1}&=\oint\frac{d\bA}{2\pi i \bA}\frac{(1-q)^{\frac{\chi'_3+\nicefrac{\epsilon_-^2}{4}}{\epsilon_1\epsilon_2}}}{\Lambda^8}\frac{\Up(\epsilon_1)\left[\prod_{i<j=1}^3\Up(n_i-n_j)\Up(l_j-l_i)\right]^{-1}}{\prod_{k=1}^3\Up(\ba-m_k-n_1+\nicefrac{\epsilon_+}{2})\Up(\ba+m_k+l_3+\nicefrac{\epsilon_+}{2})}\nonumber\\ &\times \frac{\Up(2\ba-n_1+l_3)\Up(-2\ba+n_1-l_3)}{\Up(\ba+n_3-l_2+\nicefrac{\epsilon_+}{2})\Up(\ba+n_2-l_1+\nicefrac{\epsilon_+}{2})} \Bigg|\sum_{\boldsymbol{\nu}} \left(\frac{\tilde{N}_2\tilde{L}_1}{\tilde{N}_3\tilde{L}_2}\right)^{\frac{|\nu_1^{(2)}|}{2}}\left(\frac{\tilde{N}_1\tilde{L}_2}{\tilde{N}_2\tilde{L}_3}\right)^{\frac{|\nu_1^{(1)}|+|\nu_2^{(1)}|}{2}} \\
&
\times 
\frac{\tN_{\nu_1^{(2)}\nu_2^{(1)}}\left(\ba+n_3-l_2-\nicefrac{\epsilon_+}{2}\right) \tN_{\nu_1^{(1)}\nu_1^{(2)}}\left(\ba+n_2-l_1-\nicefrac{\epsilon_+}{2}\right) }{\tN_{\nu_1^{(2)}\nu_1^{(2)}}\left(0\right)}
\nonumber\\
&\times 
\frac{\prod_{k=1}^3\tN_{\nu_1^{(1)}\emptyset}\left(\ba-n_1-m_k-\nicefrac{\epsilon_+}{2}\right)\tN_{\emptyset\nu_2^{(1)}}\left(\ba+l_3+m_k-\nicefrac{\epsilon_+}{2}\right)}{\tN_{\nu_1^{(1)}\nu_1^{(1)}}\left(0\right)\tN_{\nu_2^{(1)}\nu_2^{(1)}}\left(0\right)\tN_{\nu_1^{(1)}\nu_2^{(1)}}\left(2\ba-n_1+l_3\right)\tN_{\nu_2^{(1)}\nu_1^{(1)}}\left(-2\ba+n_1-l_3\right)}\Bigg|^2,\nonumber
\end{align}
where we the exponent of $(1-q)$ is 
\beq
\begin{split}
\chi'_3&=-2\sum_{i=1}^3m_i^2-\sum_{i<j=1}^{3}\left[\left(n_j-n_i+\frac{\epsilon_+}{2}\right)^2+\left(l_i-l_j+\frac{\epsilon_+}{2}\right)^2 \right]\\&-\frac{1}{N}\sum_{i<j=1}^3\left[(n_j-n_i)^2+(l_i-l_j)^2\right]-2\sum_{i=1}^3n_il_{4-i}+\frac{\epsilon_+^2}{2}.
\end{split}
\eeq
agreeing with  \eqref{eq:chiprimeN}. The additional factor of $\nicefrac{\epsilon_-^2}{4}$ in \eqref{eq:calIT3} comes from the factor of $\left|M(\ft,\fq)\right|^2$. 
In deriving expression \eqref{eq:calIT3}, we have used \eqref{eq:Mnormsquared} and \eqref{eq:McMahonUp}.

Now the invariance of $\calZ_{3}^{S^4\times S^1}$ under the Weyl reflections of SU$(3)^3$ is almost trivial to check. Affine Weyl transformations on the $\balpha_i$ act as usual Weyl transformations on the $m_i$, $n_i$ and $l_i$, {\it i.e.} they simply permute them. For the choice $\tilde{M}$ of preferred direction shown in \eqref{eq:calIT3}, we can easily see that the expression is invariant. However, while the invariance of  \eqref{eq:calIT3} under the  Weyl group of the first SU(3) is easy, the Weyl reflections of the remaining Weyl groups act non-trivially. At this point we need to use the fact that slicing invariance is a symmetry of the problem and by applying first \eqref{eq:slicingsymmetry1} or \eqref{eq:slicingsymmetry2} on \eqref{eq:calIT3} before acting with the Weyl reflection we can prove the complete invariance under Weyl reflections.

%%%%%%%%%%%%%%%%%%%%%%%%%%%%%%%%%%%%%%%%%%%%%%%%%%%%
\subsection{The index and \texorpdfstring{$E_6$}{E6} symmetry}
%%%%%%%%%%%%%%%%%%%%%%%%%%%%%%%%%%%%%%%%%%%%%%%%%%%%

While the invariance under Weyl reflections of the SU(3)'s are easy to see, the symmetry under $E_6$ transformations is not. For this, we expand in  $x= e^{-\frac{\beta \epsilon_+}{2}}$ before performing the integration. As shown in \cite{Bao:2013pwa, Hayashi:2013qwa} this leads to the index computed in \cite{Kim:2012gu}, that reads
\begin{align}
\label{eq:indexT3}
&\calZ_3^{S^4\times S^1}=1 + \chi_{\bf 78}^{E_6} x^2 + 
  \chi_{\bf 2}(y) (1+\chi_{\bf 78}^{E_6})x^3 + \left[1+\chi_{\bf 2430}^{E_6} +  \chi_{\bf 3}(y)(1+\chi_{\bf 78}^{E_6})\right] x^4\nonumber\\& + \left[\chi_{\bf 2}(y)\left(1+\chi_{\bf 78}^{E_6}+\chi_{\bf 2430}^{E_6} +\chi_{\bf 2928}^{E_6}) + \chi_{\bf 4}(y)(1+\chi_{\bf 78}^{E_6}\right)\right] x^5 + \left[2\chi_{\bf 78}^{E_6}+\chi_{\bf 2925}^{E_6}+\chi_{\bf 43758}^{E_6} \right. \nonumber\\&\left.+ \chi_{\bf 3}(y)\left(2+2\chi_{\bf 78}^{E_6}+\chi_{\bf 650}^{E_6}+2\chi_{\bf 2430}^{E_6}+\chi_{\bf 2925}^{E_6}\right) +  \chi_{\bf 5}(y)\left(1+\chi_{\bf 78}^{E_6}\right)\right] x^6+\calO(x^7).
 \end{align}
The fugacities $\tilde{M}$, $\tilde{N}$ and $\tilde{L}$ enter the $E_6$ characters as follows. We have an embedding SU$(3)^3\subset E_6$ and with the fundamental ${\bf 3}$ representation of the first SU(3) having the character $\chi_{\bf 3}=\tilde{M}_1 +\tilde{M}_2+\tilde{M}_3$ with similar expressions for the other SU(3) factors. The character of the 78-dimensional adjoint representation of $E_6$ then decomposes as 
\beq
\chi_{\bf 78}^{E_6}=\chi_{({\bf 8,1,1})}+\chi_{({\bf 1,8,1})}+\chi_{({\bf 1,1,8})}+\chi_{({\bf 3,\bar{3},3})}+\chi_{({\bf \bar{3},3,\bar{3}})},
\eeq
where $\chi_{({\bf j_1,j_2,j_3})}\colonequals \prod_{k=1}^3\chi_{\bf j_k}^{\text{SU}(3)}$. The other characters can be decomposed in a similar fashion, see appendix C of \cite{Bao:2013pwa} for more details.

Since we wish to identify the index as the $q$-deformed Weyl invariant structure constants and since we showed in section \ref{subsec:enchancedsymmetry} that the Weyl invariant structure constants have an $E_6$ symmetry, we have an additional piece of evidence in our favor. Furthermore, we can use the fact that \eqref{eq:polesofW3} captures all the poles of the Weyl invariant structure constants and that the position of the poles does not change under $q$-deformation to write another formula for the index\footnote{To be more precise, as we discussed in section \ref{subsec:qdefToda} and can be seen in equation   \eqref{eq:zeroesofUpsilon} of the appendix \ref{app:qUpsilon},
after the $q$-deformed versions of the functions have more poles. For for each single pole of the undeformed function, they have a whole tower of poles.}. Specifically, we make a guess for the $q$-deformation of \eqref{eq:polesofW3} and write 
\beq
\calZ_3^{S^4\times S^1}=\frac{\mathcal{F}_3}{\fZ_q(1)^3\prod_{i,j,k=1}^3\fZ_q(\tilde{M}_i\tilde{N}_j^{-1}\tilde{L}_k)\prod_{i<j=1}^3
\fZ_q(\tilde{M}_i\tilde{M}_j^{-1})\fZ_q(\tilde{N}_i\tilde{N}_j^{-1})\fZ_q(\tilde{L}_i\tilde{L}_j^{-1})},
\eeq
where 
\beq
\fZ_q(u;\ft,\fq)\colonequals \prod_{i,j=0}^{\infty}(1-u \ft^{-i-1}\fq^{j+1})(1-u^{-1}\ft^{-i-1}\fq^{j+1})
\eeq
is up to a constant the $q$-deformation of $\fZ$ and the compensating factor $\mathcal{F}_3$ is an unknown entire\footnote{That the  function $\mathcal{F}_3$ is entire follows from the facts that 1) the  function $\mathfrak{F}$ in (5.5) of \cite{Fateev:2008bm} is entire and 2) the Weyl covariant part has no poles.} function with the expansion in $x$ given by
\beq
\begin{split}
\mathcal{F}_3&=1+\chi_{\bf 2}(y)x^3+\left[\chi_{\bf 3}(y)-\chi_{\bf 650}^{E_6}\right]x^4+\left[\chi_{\bf 4}(y)-\left(\chi_{\bf 78}^{E_6}+\chi_{\bf 650}^{E_6}\right)\chi_{\bf 2}(y)\right]x^5\\
&+\left[\chi_{\bf 5}(y)-\left(\chi_{\bf 78}^{E_6}+\chi_{\bf 650}^{E_6}\right)\chi_{\bf 3}(y)+\left(\chi_{\bf 5284}^{E_6}+\chi_{\bf \overline{5284}}^{E_6}+\chi_{\bf 650}^{E_6}\right)\right]x^6+\calO(x^7).
\end{split}
\eeq
So far, we have no closed expression for the function $\mathcal{F}_3$. 

We end this section with one last remark. Our claim \eqref{eq:mainequality} states that
\beq
\fC(\balpha_1,\balpha_2,\balpha_3)=\text{const}\times \lim_{\beta\rightarrow 0} \beta^{-\chi_3}\calZ_3^{S^4\times S^1}.
\eeq
We see in \eqref{eq:indexT3} that $\calZ_3^{S^4\times S^1}$ is invariant under an $E_6$ symmetry and we saw in section \ref{subsec:enchancedsymmetry} that $\fC$ is invariant under that symmetry as well. A direct computation shows that $\chi_3$ given by \eqref{eq:chiNgeneral}
\beq
\chi_3=-\sum_{i<j=1}^3\left[(m_i-m_j)^2+(n_j-n_i)^2+(l_i-l_j)^2\right]
\eeq
is invariant under the $E_6$ Weyl tranformations \eqref{eq:E6Weyltransformations} as well.

%%%%%%%%%%%%%%%%%%%%%%%%%%%%%%%%%%%%%%%%%%%%%%%%%%%%
\section{Conclusions and Outlook}
%%%%%%%%%%%%%%%%%%%%%%%%%%%%%%%%%%%%%%%%%%%%%%%%%%%%

 In \cite{Bao:2013pwa} we calculated the 5D partition function of the non-Lagrangian $T_N$ theories on $S^4\times S^1$ using topological strings.
In this paper we take the next very important step and argue that it is possible to take the 4D limit ($\beta\rightarrow 0$ {\it i.e.} $q = e^{-\beta} \rightarrow 1$), thus obtaining the partition function of the 4D  non-Lagrangian $T_N$ theories on $S^4$.
Taking the 4D limit is {\it not} as simple as one might  naively think and it is definitely not as easy as for theories with a Lagrangian description. 

The first step in overcoming this difficulty was realizing that one can bring formula \eqref{eq:toppartitionfunction} into the  form \eqref{eq:finalexpressionforZTN} in which the individual building blocks are only the $\Upsilon_q$ functions and the ``Nekrasov functions'' $\tN_{\mu \nu}$ for which the 4D limit is well defined as individual functions \eqref{Y4Dlimit} \eqref{eq:defttN}. Our formula for the partition function \eqref{eq:defindex} is then written as a product of the factors\footnote{As we already stress in the main text, using the worlds  ``perturbative'' and ``instanton''   is an abuse of terminology.} $\left|\nicefrac{\calZ_N^{\text{pert}}}{\calZ_N^{\text{dec}}}\right|^2$ and $\left|\calZ_N^{\text{inst}}\right|^2$.
The first factor  $\left|\nicefrac{\calZ_N^{\text{pert}}}{\calZ_N^{\text{dec}}}\right|^2$ could be explicitly brought into a form that only includes products of the $\Upsilon_q$ functions times a divergent factor of $(1-q)^{\nicefrac{\chi_N'}{\epsilon_1\epsilon_2}}$. Thus, taking the limit is  straightforward after we divide by  $(1-q)^{\nicefrac{\chi_N'}{\epsilon_1\epsilon_2}}$.  However, for the $\left|\calZ_N^{\text{inst}}\right|^2$ piece we have a further obstacle to overcome. 
The sums that contain the ``Nekrasov functions'' $\tN_{\mu \nu}$ diverge if one naively takes the 4D limit,  in contrast with the usual sums in theories with Lagrangian description.  Schematically,  instead of having a coupling constant $q_{UV}=e^{2\pi i \tau_{UV}}$ as for theories with a Lagrangian description, where one can commute the limit with the sum, as for example in 
\beq
\label{eq:commutesumlimit1}
\sum_{\mu} \left(  q_{UV}^{5D} \right)^{|\mu|}\frac{\tN_{\mu \nu_1}(a_1) \cdots  \tN_{\mu \nu_L}(a_L)}{\tN_{\mu \lambda_1}(b_1) \cdots  \tN_{\mu \lambda_L}(b_L)}
\stackrel{\beta\rightarrow 0}{\longrightarrow} \sum_{\mu}  \left(  q_{UV}^{4D} \right)^{|\mu|} \frac{\ttN_{\mu \nu_1}(a_1) \cdots  \ttN_{\mu \nu_L}(a_L)}{\ttN_{\mu \lambda_1}(b_1) \cdots  \ttN_{\mu \lambda_L}(b_L)} ,
\eeq
for the case of the $T_N$ theories (that are isolated non trivial fixed points)  there is no of $q_{UV}$ but rather a combination $e^{-\beta x}$ of the mass parameters ($M=e^{-\beta m}$) and  instead of \eqref{eq:commutesumlimit1} we have
\beq
\label{eq:commutesumlimit2}
\sum_{\mu} \left(e^{-\beta x} \right)^{|\mu|}\frac{\tN_{\mu \nu_1}(a_1) \cdots  \tN_{\mu \nu_L}(a_L)}{\tN_{\mu \lambda_1}(b_1) \cdots  \tN_{\mu \lambda_L}(b_L)}
\stackrel{\beta\rightarrow 0}{\longrightarrow}
\beta^{\text{power}} \times \text{finite}
\eeq
that makes the sum diverge as $\beta^{\text{power}}$. 
Explicitly obtaining this power would require performing the sums in \eqref{eq:ZTNsum}. In this paper we do not do that, except for the $N=2$ case where the sum is given by \eqref{eq:Masatostrick2}.
However, by carefully studying the symmetry properties of the 3-point functions for general $N$, the properties of the $\Upsilon_q$ functions and the known $N=2$ case we manage to obtain this power of the divergence \eqref{eq:indexT2}. 
Our result is tested against the $q$-deformed version of the Fateev-Litvinov formula  with one semi-degenerate insertion \eqref{eq:qdefFLTodacorr}. Combining everything, we propose that the 4D limit of the superconformal index \eqref{eq:limit}, multiplied with $\beta$ raised to the appropriate power $-\chi_N$, will be finite and equal to the partition function of the $T_N$ theory on $S^4$.
Moreover, we explicitly computed in \eqref{eq:defnonfullspincontentv2}  the decoupled part $\left|\calZ_N^{\text{dec}}\right|^2$ and it is finite after extracting a divergent factor of  $\beta$ raised to the power $2\epsilon_+\sum_{k}\form{\balpha_k}{\rho}-\chi_N$. Finally, the full topological string partition function itself is finite after the divergent factor of $\beta$ to the power $2\epsilon_+\sum_{k}\form{\balpha_k}{\rho}$ has been removed.

Via the AGT-W correspondence, we translate our formula for the $T_N$ partition function to the 3-point structure constants  of three  generic primaries of the Toda field theories, both for the undeformed (4D AGT-W) as well as for the deformed  (5D  AGT-W) \textbf{W}$_N$ algebra. We give explicitly the parameter identification from the topological string  parameters to the gauge theory ones in appendix \ref{app:notation} and then to the 2D Toda parameters in equations \eqref{eq:identificationparameters}. We identify the 3-point structure constants of the Toda CFT with the topological string partition function in \eqref{eq:3pointfunctionsastopologicalstrings}.
A very nice byproduct of our work is our ability to give the exact definition of the $q$-deformed $\Up$ functions together the all their factors, which to our knowledge do not appear in the literature. This discussion appears in appendix \ref{app:qUpsilon}.

Moreover, we identified in \eqref{eq:main5Dequality} $\mathfrak{C}_q$, the Weyl invariant part of the $q$-deformed 3-point structure constants,  with the 5D superconformal index $\calZ^{S^4\times S^1}$, a powerful gauge theory object\footnote{The  superconformal index in any dimension is  the partition function of protected operators and is independent of the coupling constants of the theory, implying that it remains invariant under S-duality.
In 4D the superconformal index $S^3\times S^1$ was proven to be equivalent to a 2D TQFT  \cite{Gadde:2009kb,Gadde:2010te,Gadde:2011ik,Gadde:2011uv,Tachikawa:2012wi}. It is very possible that something very similar will also be proven for the 5D superconformal index   $S^4\times S^1$  (see \cite{Fukuda:2012jr} for some progress in this direction) and thus the Weyl invariant part of the $q$-deformed 3-point function could be discovered to obey special properties not visible from the CFT point of view, but realized only once one is using the superconformal index interpretation.}. This identification allows us to predict that the Weyl invariant part of the  $q$-deformed 3-point structure constants should have not just SU(N)$^3$ symmetry but also an  extended symmetry as  predicted by \cite{Seiberg:1996bd,Morrison:1996xf,Intriligator:1997pq} due to the existence of non-trivial UV fixed points for the 5D gauge theories.
We have explicitly checked in the Liouville case that the Weyl invariant part of the DOZZ formula \eqref{eq:weylinvariantT2structureconstants} enjoys SU(4) enhanced symmetry,  and that for the N=3 case the Weyl invariant structure constants have $E_6$ enhanced symmetry. Checking that the Weyl invariant  3-point structure constants for higher $N$ enjoy some, other than just the Weyl group of $SU(N)^3$, enhanced symmetry is an important future direction\footnote{The authors of  \cite{Hayashi:2013qwa}  were able to discover that some specialization (called ``Higgsing'' in the gauge theory jargon) of the parameters in the $T_4$ theory leads to an $E_7$ symmetry, while in \cite{Hayashi:2014wfa} a similar specialization of the parameters in $T_6$ leads to an $E_8$ symmetry. These specializations change the $T_N$ geometry significantly and in particular reduce the number of Coulomb moduli to one. It would be very interesting to see the meaning of this on the CFT side.}

The formula we give for the 3-point functions at this point is very implicit, since there are still integrals and sums that need to be performed. In a separate publication \cite{Misha}, we will show how at least some of the sums can be performed. In so doing, we will be able to explicitly calculate our degree of divergence $\chi_N$  that we conjectured in \eqref{eq:chiNgeneral}.
Moreover, beginning with our formulas and specializing some of the $\balpha$'s we should be able to obtain the formulas of \cite{Fateev:2005gs,Fateev:2007ab,Fateev:2008bm} for the cases of  degenerate or semi-degenerate primaries and for the semiclassical limit $b\rightarrow 0$. Conversely, our formulas predict highly nontrivial relations for the sums of ``Nekrasov functions'' $\tN_{\lambda \mu}$ by for example requiring that our formulas reproduces \eqref{eq:chiNgeneral} in the semi-degenerate case.

In this paper we give the Toda 3-point functions with three primaries, which however is not enough to solve Toda. 
To achieve that, we need to also compute the correlation functions of descendants, which as we discussed in the introduction is not as immediate as in the Liouville case. However, it is straightforward to see from the point of view of the topological strings what needs to be done in order to compute them. Specifically, we need to take the $T_N$ web diagrams from figure~\ref{fig:strip} and evaluate them with the refined topological vertex without putting empty Young diagram to the external legs. This will provide the general Ding-Iohara algebra interwiners.
The Ding-Iohara algebra \cite{Ding:1996mq} in the free boson representation (with $N$ free bosons) is known to become 
\beq
\mathcal{A} = \textbf{W}_N \otimes  \textbf{H}
\eeq
where  $\textbf{H}$ is the Heisenberg
algebra which is exactly the algebra that is needed to describe what is obtained from AGT-W \cite{Alba:2010qc,Fateev:2011hq}.
In particular, it is quite easy to obtain the 3-point function of two primaries and one descendant  and in fact the answer is just \eqref{eq:finalexpressionforZTN} without putting empty Young diagrams for $\boldsymbol{\nu}^{(0)}$. Such 3-point functions are already going to give us, via bootstrapping many higher point functions. Solving  this problem is work in progress \cite{FutoshiMasato}.

We would like to finish by remarking that for many reasons it seems to be much more advantageous to study the $q$-deformed version of the Toda field theories instead of the undeformed ones. For example, the functions $\Up$ behave in a sense a  bit better than the $\Upsilon$ ones, since for example the product formula \eqref{eq:defUp} is much simpler that \eqref{eq:defGamma2product} and \eqref{eq:defUpsilonproduct}. Furthermore, in the $q$-deformed case, we can use the topological string formalism to compute the partition functions, tools that are not directly available in the undeformed theory.

%%%%%%%%%%%%%%%%%%%%%%%%%%%%%%%%%%%%%%%%%%%%%%%%%%%%%%%%%%%%%%
\section*{Acknowledgments}
We would  like to thank first our collaborators on closely related projects Mikhail Isachenkov,  Masato Taki and Futoshi Yagi. In addition, we are thankful to Sara Pasquetti,  Paulina  Suchanek and J\"org Teschner for insightful comments and discussions. Finally, we are very grateful to Sylvain Ribault for making several useful comments on our draft.

We thank CERN, the CERN-Korea Theory Collaboration
funded by the National Research Foundation (Korea) and the C.N.Yang Institute for Theoretical Physics (Stony Brook) for their hospitality during the finishing stage of this work.

%APPENDIX
%%%%%%%%%%%%%%%%%%%%%%%%%%%%%%%%%%%%%%%%%%%%%%%%%%%%
\appendix
%%%%%%%%%%%%%%%%%%%%%%%%%%%%%%%%%%%%%%%%%%%%%%%%%%%%

%%%%%%%%%%%%%%%%%%%%%%%%%%%%%%%%%%%%%%%%%%%%%%%%%%%
\section{Parametrization of the \texorpdfstring{$T_N$}{TN} junction}
\label{app:notation}

\begin{figure}[htbp]
  \centering
\includegraphics[height=12cm]{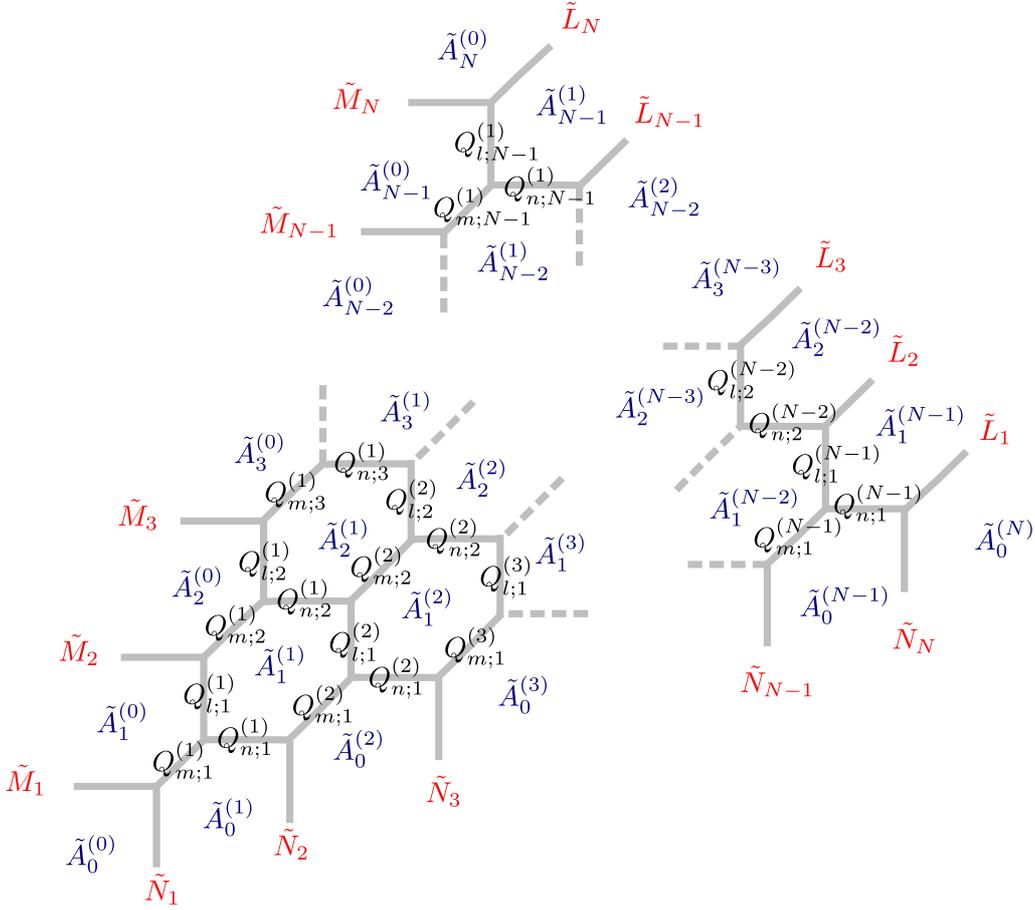}
  \caption{Parametrization for $T_N$. We denote the K\"ahler moduli parameters corresponding to the horizontal lines as $Q_{n;i}^{(j)}$, to the vertical lines as $Q_{l;i}^{(j)}$, and to tilted lines as $Q_{m;i}^{(j)}$. We denote the breathing modes as $\tilde{A}^{(j)}_i$. The index $j$ labels the strips in which the diagram can be decomposed. }
  \label{fig:TNparam}
\end{figure}

We gather in this appendix all necessary formulas for the parametrizations of the K\"ahler moduli. 
First, the ``interior'' Coulomb moduli $\tilde{A}_j^{(i)}=e^{-\beta a_i^{(j)}}$ are independent, while the  ``border'' ones are given by
\beq
\label{eq:borderAdefMNL}
\tilde{A}_{i}^{(0)} = \prod_{k=1}^i \tilde{M}_k,
\qquad
\tilde{A}_{0}^{(j)} = \prod_{k=1}^j \tilde{N}_k,
\qquad
\tilde{A}_{i}^{(N-i)}  = \prod_{k=1}^i \tilde{L}_k.
\eeq
The parameters labeling the positions of the flavors branes obey the relations
\beq
\label{eq:externalparametersconstraints}
\prod_{k=1}^N\tilde{M}_k=\prod_{k=1}^N\tilde{N}_k=\prod_{k=1}^N\tilde{L}_k=1\Longleftrightarrow \sum_{k=1}^Nm_k=\sum_{k=1}^Nn_k=\sum_{k=1}^Nl_k=0.
\eeq
Therefore, $\tilde{A}_0^{(0)}=\tilde{A}_N^{(0)}=\tilde{A}_0^{(N)}=1$ and we can invert  relation \eqref{eq:borderAdefMNL} as
\beq
\label{eq:AfunctionMNL}
\tilde{M}_i=\frac{\tilde{A}_i^{(0)}}{\tilde{A}_{i-1}^{(0)}},\qquad \tilde{N}_i=\frac{\tilde{A}_0^{(i)}}{\tilde{A}_{0}^{(i-1)}},\qquad \tilde{L}_i=\frac{\tilde{A}_i^{(N-i)}}{\tilde{A}_{i-1}^{(N-i+1)}}.
\eeq
All placements are illustrated in figure~\ref{fig:TNparam}.

The   K\"ahler parameters associated to the edges of the $T_N$ junction are related to the $\tilde{A}_i^{(j)}$ as follows
\begin{align}
\label{eq:PQR}
Q_{n;i}^{(j)}
 = \frac{\tilde{A}_{i}^{(j)} \tilde{A}_{i-1}^{(j)}}%
    {\tilde{A}_{i}^{(j-1)} \tilde{A}_{i-1}^{(j+1)}},
\qquad
Q_{l;i}^{(j)}
 = \frac{\tilde{A}_{i}^{(j)} \tilde{A}_{i}^{(j-1)}}%
    {\tilde{A}_{i-1}^{(j)} \tilde{A}_{i+1}^{(j-1)}},
\qquad
Q_{m;i}^{(j)}
 = \frac{\tilde{A}_{i}^{(j-1)} \tilde{A}_{i-1}^{(j)}}%
    {\tilde{A}_{i}^{(j)} \tilde{A}_{i-1}^{(j-1)}}.
\end{align}
For each inner hexagon of \eqref{fig:TNparam}, the following two constraints are satisfied
\begin{equation}
Q_{l;i}^{(j)} Q_{m;i+1}^{(j)} = Q_{m;i}^{(j+1)} Q_{l;i}^{(j+1)},
\qquad
Q_{n;i}^{(j)} Q_{m;i}^{(j+1)} = Q_{m;i+1}^{(j)} Q_{n;i+1}^{(j)} .
\end{equation}
Furthermore, we find the following identities relating them to the masses: 
\begin{align}
\label{eq:quotientMNL}
Q_{m;i}^{(1)} Q_{l;i}^{(1)} = \frac{\tilde{M}_i}{\tilde{M}_{i+1}},
\qquad
Q_{m;1}^{(i)} Q_{n;1}^{(i)} = \frac{\tilde{N}_i}{\tilde{N}_{i+1}},
\qquad
Q_{n;i}^{(N-i)} Q_{l;i}^{(N-i)} = \frac{\tilde{L}_i}{\tilde{L}_{i+1}}.
\end{align}
Using the above, we find the following expressions for the products appearing in the $T_N$ partition function:
\beq
\begin{split}
\prod_{k=i}^jQ_{l;k}^{(r)}Q_{m;k+1}^{(r)}=&\prod_{k=i}^j\frac{\Big(\tilde{A}_k^{(r)}\Big)^2}{\tilde{A}_{k-1}^{(r)}\tilde{A}_{k+1}^{(r)}}=\frac{\tilde{A}_i^{(r)}\tilde{A}_j^{(r)}}{\tilde{A}_{i-1}^{(r)}\tilde{A}_{j+1}^{(r)}},\\
\prod_{k=i}^jQ_{l;k}^{(r)}Q_{m;k}^{(r)}=&\prod_{k=i}^j\frac{\Big(\tilde{A}_k^{(r-1)}\Big)^2}{\tilde{A}_{k-1}^{(r-1)}\tilde{A}_{k+1}^{(r-1)}}=\frac{\tilde{A}_i^{(r-1)}\tilde{A}_j^{(r-1)}}{\tilde{A}_{i-1}^{(r-1)}\tilde{A}_{j+1}^{(r-1)}},\\Q_{m;j}^r\prod_{k=i}^{j-1}Q_{l;k}^{(r)}Q_{m;k}^{(r)}=&\frac{\tilde{A}_j^{(r-1)}\tilde{A}_{j-1}^{(r)}}{\tilde{A}_{j}^{(r)}\tilde{A}_{j-1}^{(r-1)}}\frac{\tilde{A}_i^{(r-1)}\tilde{A}_{j-1}^{(r-1)}}{\tilde{A}_{i-1}^{(r-1)}\tilde{A}_{j}^{(r-1)}}=\frac{\tilde{A}_i^{(r-1)}\tilde{A}_{j-1}^{(r)}}{\tilde{A}_{i-1}^{(r-1)}\tilde{A}_{j}^{(r)}},\\Q_{l;i}^r\prod_{k=i+1}^{j}Q_{l;k}^{(r)}Q_{m;k}^{(r)}=&\frac{\tilde{A}_i^{(r)}\tilde{A}_i^{(r-1)}}{\tilde{A}_{i-1}^{(r)}\tilde{A}_{i+1}^{(r-1)}}\frac{\tilde{A}_{i+1}^{(r-1)}\tilde{A}_j^{(r-1)}}{\tilde{A}_{i}^{(r-1)}\tilde{A}_{j+1}^{(r-1)}}=\frac{\tilde{A}_i^{(r)}\tilde{A}_j^{(r-1)}}{\tilde{A}_{i-1}^{(r)}\tilde{A}_{j+1}^{(r-1)}}.
\end{split}
\eeq
Furthermore, the following two follow directly from \eqref{eq:quotientMNL} and are used in the derivation of the ``perturbative'' part of the topological string partition function \eqref{eq:ZTNperturbative}
\beq
\label{eq:prodQmQl}
\prod_{j=1}^iQ_{m;j}^{(r)}=\frac{\tilde{A}_0^{(r)}\tilde{A}_i^{(r-1)}}{\tilde{A}_0^{(r-1)}\tilde{A}_i^{(r)}}, \qquad \prod_{k=i}^{N-r}Q_{l;k}^{(r)}=\frac{\tilde{A}_i^{(r-1)}\tilde{A}_{N-r}^{(r)}}{\tilde{A}_{i-1}^{(r)}\tilde{A}_{N-r+1}^{(r-1)}}.
\eeq

%%%%%%%%%%%%%%%%%%%%%%%%%%%%%%%%%%%%%%%%%%%%%%%
\section{Conventions and notations for SU$(N)$}
\label{app:sln}
%%%%%%%%%%%%%%%%%%%%%%%%%%%%%%%%%%%%%%%%%%%%%%%

The purpose of this appendix is to summarize our SU$(N)$ conventions. The weights of the fundamental representation of SU$(N)$ are $h_i$ with $\sum_{i=1}^N h_i=0$. We remind that the scalar product is defined via $\form{h_i}{h_j}=\delta_{ij}-\frac{1}{N}$.
The simple roots are 
\beq
e_k\colonequals h_k-h_{k+1 }, \qquad k=1,\ldots, N-1,
\eeq
and the positive roots $e>0$ are contained in the set
\beq
\Delta^+\colonequals \{h_i-h_j\}_{i<j=1}^N=\{e_i\}_{i=1}^{N-1}\cup\{e_i+e_{i+1}\}_{i=1}^{N-2}\cup\cdots \cup \{e_1+\cdots+e_{N-1}\}.
\eeq
The Weyl vector $\rho$ for SU$(N)$ is given by 
\beq
\label{eq:defWeylvector}
\rho\colonequals \frac{1}{2}\sum_{e>0}e=\frac{1}{2}\sum_{i<j=1}^N(h_i-h_j)=\sum_{i=1}^N\frac{N+1-2i}{2}h_i=\omega_1+\cdots+\omega_{N-1},
\eeq
and it obeys $\form{\rho}{e_i}=1$ for all $i$. 
The $N-1$ fundamental weights $\omega_i$ of SU$(N)$ are given by
\beq
\omega_i=\sum_{k=1}^i h_k, \qquad i=1,\dots, N-1
\eeq
and the corresponding finite dimensional representations are the $i$-fold antisymmetric tensor product of the fundamental representation. They obey the scalar products $\form{e_i}{\omega_j}=\delta_{ij}$, {\it i.e.} they are a dual basis. Furthermore, we find the following scalar products useful
\beq 
\label{eq:SUNscalarproducts1}
 \form{\rho}{h_j}=\frac{N+1}{2}-j,\qquad \form{\rho}{\omega_i}=\frac{i(N-i)}{2}, \qquad (h_j,\omega_i)=\left\{\begin{array}{ll}1-\frac{i}{N}& j\leq i\\-\frac{i}{N} & j>i \end{array}\right. ,
\eeq
as well as
\beq
\label{eq:SUNscalarproducts2}
\form{\omega_i}{\omega_j}=\frac{\text{min}(i,j)\left(N-\text{max}(i,j)\right)}{N},\qquad \form{\rho}{\rho}=\frac{N(N^2-1)}{12}.
\eeq
After some work, one can  prove using the scalar products \eqref{eq:SUNscalarproducts1} and \eqref{eq:SUNscalarproducts2} that
\beq
\label{eq:SUNscalarproducts3}
\frac{1}{N}\sum_{e>0}\form{\balpha_1}{e}\form{\balpha_2}{e}=\form{\balpha_1}{\balpha_2},
\eeq
for any two weights $\balpha_i$.

The Weyl group of SU$(N)$ is isomorphic to $S_N$ and is generated by the $N-1$ Weyl reflections associated to the simple roots. If $\balpha$ is a weight, we define the  Weyl reflections with respect to the simple root $e_i$
\beq
\label{eq:defWeyltransformations}
\fw_i\cdot \balpha \colonequals  \balpha-2\frac{\form{e_i}{\balpha}}{\form{e_i}{e_i}}e_i=\balpha-\form{e_i}{\balpha}e_i.
\eeq
Furthermore, we define the affine Weyl reflections with respect to $e_i$ as follows
\beq
\label{eq:defaffineWeyltransformations}
\fw_i\circ \balpha\colonequals \fQ+\fw_i\cdot(\balpha-\fQ)=\fw_i\cdot \balpha+Qe_i=\balpha-\form{\balpha-\fQ}{e_i}e_i,
\eeq
where $\fQ\colonequals Q\rho=(b+b^{-1})\rho$.

%%%%%%%%%%%%%%%%%%%%%%%%%%%%%%%%%%%%%%%%%%%%%%%
\section{Special functions}
\label{app:special}
%%%%%%%%%%%%%%%%%%%%%%%%%%%%%%%%%%%%%%%%%%%%%%%

For the reader's convenience, we gather here the definitions and properties of all special functions used in the main text.

%%%%%%%%%%%%%%%%%%%%%%%%%%%%%%%%%%%%%%%%%%%%%%%%%%%%%
\subsection{The \texorpdfstring{$\Upsilon$}{Y} function.}
\label{app:Upsilon}
%%%%%%%%%%%%%%%%%%%%%%%%%%%%%%%%%%%%%%%%%%%%%%%%%%%%%

The purpose of this part of the appendix is to summarize the known identities for the functions used in the undeformed Liouville and Toda CFT. 
We begin with the function $\Upsilon(x)$ which is defined for $0<\Re(x)<Q=b+b^{-1}$ as the integral
\beq
\log \Upsilon(x)\colonequals  \int_{0}^{\infty}\frac{dt}{t}\left[\left(\frac{Q}{2}-x\right)^2e^{-t}-\frac{\sinh^2\left[\left(\frac{Q}{2}-x\right)\frac{t}{2}\right]}{\sinh \frac{b t}{2}\sinh \frac{t}{2b}}\right].
 \eeq 
It is clear from the definition that 
\beq
\Upsilon(x)=\Upsilon(Q-x), \qquad  \Upsilon\left(\frac{Q}{2}\right)=1. 
\eeq
One can show from the alternative definition below that the following shift identities are obeyed
\beq
\label{eq:shiftUpsilon4D}
\Upsilon(x+b)=\gamma(x b)b^{1-2bx}\Upsilon(x),\qquad \Upsilon(x+b^{-1})=\gamma(x b^{-1})b^{2xb^{-1}-1}\Upsilon(x).
\eeq
where $\gamma(x)\colonequals \frac{\Gamma(x)}{\Gamma(1-x)}$.  An useful implication is 
\beq
\label{eq:extrashiftsUpsilon4D}
\Upsilon(x+Q)=b^{2(b^{-1}-b)x}\frac{\Gamma\big(1+bx\big)\Gamma\big(b^{-1}x\big)}{\Gamma\big(1-bx\big)\Gamma\big(-b^{-1}x\big)}\Upsilon(x),
\eeq
which is used in the derivation of the reflection amplitude \eqref{eq:defreflectionamplitude}.
It follows from \eqref{eq:shiftUpsilon4D} that $\Upsilon$ is an entire function with zeroes at
\beq
\label{eq:zeroesofUpsilon}
x=-n_1 b -n_2 b^{-1},\quad \text { or } \quad x=(n_1+1) b +(n_2+1) b^{-1},
\eeq
where $n_i\in \mathbb{N}_0$. 

The function $\Upsilon$ can be connected to the Barnes Double Gamma function $\Gamma_2(x|\omega,\omega_2)$.
First, we define $\Gamma_2(x|\omega_1,\omega_2)$ via the \textit{analytic continuation} (the sum is only well-defined if $\Re(t)>2$) of 
\beq
\log \Gamma_2(s|\omega_1,\omega_2)=\left[\frac{\partial}{\partial t}\sum_{n_1,n_2=0}^{\infty}(s+n_1\omega_1+n_2\omega_2)^{-t}\right]_{t=0}.
\eeq
From this definition, one can prove (see A.54 of \cite{Nakayama:2004vk}) the \textit{difference property}
\beq
\frac{\Gamma_2(s+\omega_1|\omega_1,\omega_2)}{\Gamma_2(s|\omega_1,\omega_2)}=\frac{\sqrt{2\pi}}{\omega_2^{\frac{s}{\omega_2}-\frac{1}{2}}\Gamma\Big(\frac{s}{\omega_2}\Big)},\qquad \frac{\Gamma_2(s+\omega_2|\omega_1,\omega_2)}{\Gamma_2(s|\omega_1,\omega_2)}=\frac{\sqrt{2\pi}}{\omega_1^{\frac{s}{\omega_1}-\frac{1}{2}}\Gamma\Big(\frac{s}{\omega_1}\Big)}.
\eeq
In order to express the $\Upsilon$ function using the Barnes double Gamma function, we have to first define  the \textit{normalized} function
\beq
\label{eq:defGammabviaBarnes}
 \Gamma_b(x)\colonequals \frac{ \Gamma_2(x|b,b^{-1})}{ \Gamma_2(\frac{Q}{2}|b,b^{-1})}.
\eeq
The log of the function $\Gamma_b(x)$ has an integral representation as
\beq
\log\Gamma_b(x)=\int_{0}^{\infty}\frac{dt}{t}\left(\frac{e^{-x t}-e^{-\frac{Q t}{2}}}{(1-e^{-t b })(1-e^{-t b^{-1}})}-\frac{\left(\frac{Q}{2}-x\right)^2}{2}e^{-t}-\frac{\frac{Q}{2}-x}{t}\right).
\eeq
Then, using \eqref{eq:defGammabviaBarnes} we can express the $\Upsilon(x)$ as 
\beq
\label{eq:defUpsilonproduct}
\Upsilon(x)= \frac{1}{\Gamma_b(x)\Gamma_b(Q-x)}.
\eeq
This, together with the difference properties of $\Gamma_2$ proves the shift identities \eqref{eq:shiftUpsilon4D}. 
Also of  interest is the function  $\textbf{G}(x)$ introduced  in \cite{Fateev:2008bm} with  the  properties
\beq
\label{eq:propertiesG}
\textbf{G}(x+b)=\frac{b^{\nicefrac{1}{2}-bx}}{\sqrt{2\pi}}\Gamma(bx)\textbf{G}(x),\qquad \textbf{G}(x+b^{-1})=\frac{b^{b^{-1}x-\nicefrac{1}{2}}}{\sqrt{2\pi}}\Gamma(b^{-1}x)\textbf{G}(x).
\eeq
The zeroes of this function are $x=-mb-nb^{-1}$ for $m,\,n\in \mathbb{N}_0$. 
If we normalize it by setting $\textbf{G}(\frac{Q}{2})=1$, then we have $\textbf{G}(x)=\frac{1}{\Gamma_b(x)}$. Furthermore, \cite{Fateev:2008bm} also introduce the function $\fZ$ as
\beq
\label{eq:deffrakZ}
\fZ(x)=\textbf{G}(Q+x)\textbf{G}(Q-x)=\frac{b^{b^{-1}x-bx}}{2\pi}x\Gamma(bx)\Gamma(b^{-1}x)\Upsilon(x).
\eeq

One very often encounters a product formula for the function $\Gamma_2=\prod_{n_1,n_2}(x+\omega_1n_1+\omega_2n_2)^{-1}$ that is unfortunately not quite correct. To get the product formula  for $\Gamma_2(x)$ working, one has to use (A.62) of \cite{Nakayama:2004vk}. Specifically, we set for $\Re(s)>2$
\beq
\chi(s|\omega_1,\omega_2)\colonequals \sum'_{n_1,n_2\geq 0}\frac{1}{(\omega_1n_1+\omega_2n_2)^s},
\eeq
where the prime removes the value $(n_1,n_2)=(0,0)$ from the sum. The function $\chi(s|\omega_1,\omega_2)$ can be analytically continued for all $s\in \mathbb{C}$ except for $s=1$ and $s=2$ where there are poles. We have the residues
\beq
\text{Res}(\chi(s|\omega_1,\omega_2),s=1)=\frac{1}{2}\left(\frac{1}{\omega_1}+\frac{1}{\omega_2}\right),\qquad \text{Res}(\chi(s|\omega_1,\omega_2),s=2)=\frac{1}{\omega_1\omega_2}
\eeq
and the finite parts
\beqa
\text{Res}\Big(\frac{\chi(s|\omega_1,\omega_2)}{s-1},s=1\Big)&=&-\frac{\log \omega_1}{\omega_1}+\frac{1}{2}\left(\frac{1}{\omega_1}-\frac{1}{\omega_2}\right)\log \omega_2+\frac{\gamma}{\omega_1}+\frac{\gamma}{2\omega_2}-\frac{1}{2\omega_1}\log2\pi\nonumber\\&&-\frac{i}{b}\int_{0}^{\infty}\frac{\psi(i\frac{\omega_1}{\omega_2}y+1)-\psi(-i\frac{\omega_1}{\omega_2}y+1)}{e^{2\pi y}-1}dy\nonumber\\
\text{Res}\Big(\frac{\chi(s|\omega_1,\omega_2)}{s-2},s=2\Big)&=&\frac{\zeta(2) }{\omega_1^2}+\frac{\zeta(2)}{2\omega_2^2}+\frac{1}{\omega_1\omega_2}\left(\gamma-1-\log \omega_2\right)\nonumber\\&&-\frac{i}{\omega_2}\int_{0}^{\infty}\frac{\zeta_H(2,i\frac{\omega_1}{\omega_2}y+1)-\zeta_H(2,-i\frac{\omega_1}{\omega_2}y+1)}{e^{2\pi y}-1}dy,
\eeqa
where $\psi$ is the digamma function, $\gamma$ is the Euler - Mascheroni constant and $\zeta_H(s,q)$ is the Hurwitz-$\zeta$ function with ($\Re(s)>1$ and $\Re(q)>0$)
\beq
\zeta_H(s,q)\colonequals \sum_{n=0}^{\infty}\frac{1}{(q+n)^s}.
\eeq
Finally, using the shorthands $\alpha\colonequals \text{Res}(\frac{\chi(s|\omega_1,\omega_2)}{s-1},s=1)$ and $\beta\colonequals \text{Res}(\frac{\chi(s|\omega_1,\omega_2)}{s-2},s=2)+\text{Res}(\chi(s|\omega_1,\omega_2),s=2)$ we obtain
\beq
\label{eq:defGamma2product}
\Gamma_2(x|\omega_1,\omega_2)=\frac{e^{-\alpha x +\frac{\beta x^2}{2}}}{x} \prod'_{n_1,n_2\geq 0}\frac{e^{\frac{x}{\omega_1n_1+\omega_2n_2}-\frac{x^2}{2(\omega_1n_1+\omega_2n_2)^2}}}{1+\frac{x}{\omega_1n_1+\omega_2n_2}}.
\eeq

%%%%%%%%%%%%%%%%%%%%%%%%%%%%%%%%%%%%%%%%%%%%%%%%%%%%%
\subsection{The \texorpdfstring{$q$}{q}-deformed \texorpdfstring{$\Upsilon$}{Y} function.}
\label{app:qUpsilon}
%%%%%%%%%%%%%%%%%%%%%%%%%%%%%%%%%%%%%%%%%%%%%%%%%%%%%

In this subsection, we wish to summarize some results involving the $q$-deformed $\Upsilon$ functions. First we begin by defining the shifted factorials\footnote{A good source for the properties of the shifted factorials is \cite{Nishizawa:2001}.} (we require for convergence that $|q_i|<1$ for all $i$)
\beq
(x;q_1,\ldots, q_r)_{\infty}\colonequals \prod_{i_1=0,\ldots, i_r=0}^{\infty}(1-x q_1^{i_1}\cdots q_r^{i_r}). 
\eeq
We can extend the definition of the shifted factorial for all values of $q_i$ by imposing the relations
\beq
\label{eq:shiftedfactorialinversion}
(x;q_1,\ldots,q_i^{-1},\ldots, q_r)_{\infty}=\frac{1}{(xq_i;q_1,\ldots, q_r)_{\infty}}.
\eeq
Furthermore, they obey the following shifting properties
\beq
\label{eq:shiftingfactorials}
(q_jx;q_1,\ldots, q_r)_{\infty}=\frac{(x;q_1,\ldots, q_r)_{\infty}}{(x;q_1,\ldots,q_{j-1},q_{j+1},\ldots, q_r)_{\infty}}.
\eeq
We then define the function $\calM(u;\ft,\fq)$ as
\beq
\label{eq:defcalMeverywhere}
\calM(u;\ft,\fq)\colonequals (u \fq ;\ft,\fq)_{\infty}^{-1}=\left\{\begin{array}{ll}
\prod_{i,j=1}^{\infty}(1-u\ft^{i-1}\fq^j)^{-1} & \text{ for } |\ft|<1, |\fq|<1\\
\prod_{i,j=1}^{\infty}(1-u\ft^{i-1}\fq^{1-j}) & \text{ for } |\ft|<1, |\fq|>1\\
\prod_{i,j=1}^{\infty}(1-u\ft^{-i}\fq^j) & \text{ for } |\ft|>1, |\fq|<1\\
\prod_{i,j=1}^{\infty}(1-u\ft^{-i}\fq^{1-j})^{-1} & \text{ for } |\ft|>1, |\fq|>1\\ \end{array}\right.,
\eeq
converging for all $u$. This function can be written as a plethystic exponential 
\beq
\label{eq:defMpexp}
\calM(u;\ft,\fq)=\exp\left[\sum_{m=1}^{\infty} \frac{u^m}{m}\frac{\fq^m}{(1-\ft^m)(1-\fq^m)}\right],
\eeq
which converges for all $\ft$ and all $\fq$ provided that $|u|<\fq^{-1+\theta(|\fq|-1)}\ft^{\theta(|\ft|-1)}$. Here and elsewhere $\theta(x)=1$ if $x>0$ and is zero otherwise. The following identity is obvious from the definition
\beq
\label{eq:exchangerelationMN}
\calM(u;\fq,\ft)=\calM(u\nicefrac{\ft}{\fq};\ft,\fq).
\eeq
From the analytic properties of the shifted factorials \eqref{eq:shiftedfactorialinversion}, we read the identities
\beq
\label{eq:inversionidentities}
\calM(u;\ft^{-1},\fq)=\frac{1}{\calM(u\ft;\ft,\fq)}, \qquad \calM(u;\ft,\fq^{-1})=\frac{1}{\calM(u\fq^{-1};\ft,\fq)},
\eeq
while from \eqref{eq:shiftingfactorials} we take the following shifting identities
\beq
\label{eq:calMshift}
\calM(u\ft;\ft,\fq)=(u \fq; \fq)_{\infty} \calM(u;\ft,\fq),\qquad \calM(u\fq;\ft,\fq)=(u \fq; \ft)_{\infty} \calM(u;\ft,\fq).
\eeq
We define the $q$-deformed $\Upsilon$ function as
\beq
\label{eq:defUp}
\begin{split}
\Up(x|\epsilon_1,\epsilon_2)=&(1-q)^{-\frac{1}{\epsilon_1\epsilon_2}\left(x-\frac{\epsilon_+}{2}\right)^2}\prod_{n_1,n_2=0}^{\infty}\frac{(1-q^{x+n_1\epsilon_1+n_2\epsilon_2})(1-q^{\epsilon_+-x+n_1\epsilon_1+n_2\epsilon_2})}{(1-q^{\nicefrac{\epsilon_+}{2}+n_1\epsilon_1+n_2\epsilon_2})}\\
=&(1-q)^{-\frac{1}{\epsilon_1\epsilon_2}\left(x-\frac{\epsilon_+}{2}\right)^2}\left|\frac{\calM(q^{-x};\ft,\fq)}{\calM(\sqrt{\frac{\ft}{\fq}};\ft,\fq)}\right|^2,
\end{split}
\eeq
where we have used the definition \eqref{eq:normsquared} for the norm squared.  If follows from the definition that $\Up(\nicefrac{\epsilon_+}{2}|\epsilon_1,\epsilon_2)=1$, that $\Up(x|\epsilon_1,\epsilon_2)=\Up(\epsilon_+-x|\epsilon_1,\epsilon_2)$ and  that $\Up(x|\epsilon_1,\epsilon_2)=\Up(x|\epsilon_2,\epsilon_1)$. Furthermore, from the shifting identities for $\calM$, we can easily prove that 
\beq
\label{eq:shiftidentitiesqUpsilon}
\Up(x+\epsilon_1|\epsilon_1,\epsilon_2)=\left(\frac{1-q}{1-q^{\epsilon_2}}\right)^{1-2\epsilon_2^{-1}x}\gamma_{q^{\epsilon_2}}(x\epsilon_2^{-1})\Up(x|\epsilon_1,\epsilon_2),
\eeq
together with a similar equation for the shift with $\epsilon_2$. Here, we have used the definition of the $q$-deformed $\Gamma$ and $\gamma$ functions
\beq
\label{eq:defGammaq}
\Gamma_q(x)\colonequals (1-q)^{1-x}\frac{(q;q)_{\infty}}{(q^x;q)_{\infty}},\qquad \gamma_q(x)\colonequals\frac{\Gamma_q(x)}{\Gamma_q(1-x)}=(1-q)^{1-2x}\frac{(q^{1-x};q)_{\infty}}{(q^{x};q)_{\infty}},
\eeq
valid for $|q|<1$. They obey  $\Gamma_q(x+1)=\frac{1-q^x}{1-q}\Gamma_q(x)$, implying  $\gamma_q(x+1)=\frac{(1-q^x)(1-q^{-x})}{(1-q)^2}\gamma_q(x)$.  Because of the normalization of $\Up(x|\epsilon_1,\epsilon_2)$ and since the factors of the right hand side of \eqref{eq:shiftidentitiesqUpsilon} have a well defined limit for $q\rightarrow 1$, we find by comparing functional identities that\footnote{The $q\rightarrow 1$ limit has also been checked numerically for the case $b=\epsilon_1=\epsilon_2^{-1}$. }
\beq
\label{Y4Dlimit}
\Up(x+\epsilon_1|\epsilon_1,\epsilon_2)\stackrel{q\rightarrow 1}{\longrightarrow } \Upsilon(x|\epsilon_1,\epsilon_2)\colonequals \frac{\Gamma_2\big(\frac{\epsilon_+}{2}|\epsilon_1,\epsilon_2\big)^2}{\Gamma_2\big(x|\epsilon_1,\epsilon_2\big)\Gamma_2\big(\epsilon_+-x|\epsilon_1,\epsilon_2\big)}.
\eeq
In particular, the function $\Upsilon(x)$ defined in subsection \ref{app:Upsilon} is equal to $\Upsilon(x|b,b^{-1})$. We shall often just write $\Up(x)$ instead of $\Up(x|\epsilon_1,\epsilon_2)$ and indicate in the text whether the $\epsilon_i$ parameters are arbitrary or whether $b=\epsilon_1=\epsilon_2^{-1}$. 

For the rest of the section, we set $b=\epsilon_1=\epsilon_2^{-1}$. One very useful implication of \eqref{eq:shiftidentitiesqUpsilon} for the derivation of the reflection amplitude \eqref{eq:qreflection} 
\beq
\label{eq:specialshiftidentityqUpsilon}
\Up(x+Q)=\left[\frac{\big(1-q^{b^{-1}}\big)^{b}\big(1-q^{b}\big)^{b^{-1}}}{(1-q)^{Q}}\right]^{2x}\frac{\Gamma_{q^{b^{-1}}}(1+bx)\Gamma_{q^{b}}(b^{-1}x)}{\Gamma_{q^{b^{-1}}}(1-bx)\Gamma_{q^{b}}(-b^{-1}x)}\Up(x),
\eeq
which reduces to \eqref{eq:extrashiftsUpsilon4D} in the limit $q\rightarrow 1$.  
We finish this part of the appendix with two small remarks. 
First, the zeroes of $\Up$ are given by
\beq
\label{eq:zeroesofqUpsilon}
x=-n_1\epsilon_1-n_2\epsilon_2+\frac{2\pi i }{\log q}m,\qquad  x=(n_1+1)\epsilon_1+(n_2+1)\epsilon_2+\frac{2\pi i }{\log q}m',
\eeq
where $n_i\in \mathbb{N}_0$ and $m$ and $m'$ are integer. We thus see by comparing with \eqref{eq:zeroesofUpsilon} that for each zero of $\Upsilon$ we have a whole tower, Kaluza-Klein like, of zeroes of $\Up$. The new zeroes are $q$-dependent, but the ones that are also zeroes of $\Upsilon$, \textit{i.e.} those with $m=0$ are $q$-independent. 
The tower of zeros is obtained by beginning with the  $q$-independent $m=0$ zero and shifting by multiples of 
$\frac{2\pi i }{\log q} = - \frac{2\pi i }{\beta}$.
Second, we will need to evaluate the derivative of $\Up(x)$ at $x=0$. Since the zero of $\Up(x)$ at $x=0$ is due to the factor $(1-q^x)$ in the numerator of \eqref{eq:defUp}, we find that the only piece of the derivative that survives is 
\beq
\label{eq:derivativeofUp}
\Up'(0)=\frac{\beta}{1-q}\Up(b). 
\eeq

%%%%%%%%%%%%%%%%%%%%%%%%%%%%%%%%%%%%%%%%%%%%%%%
\subsection{The finite product functions}
\label{app:finiteproduct}
%%%%%%%%%%%%%%%%%%%%%%%%%%%%%%%%%%%%%%%%%%%%%%%

In this subsection $\epsilon_1$ and $\epsilon_2$ are general. 
In the definition of the topological string amplitudes, we often need to use the following two functions given by finite products
\beq
\begin{split}
\tilde{Z}_{\nu}(\ft,\fq)\colonequals & \prod_{i=1}^{\ell(\nu)}\prod_{j=1}^{\nu_i}\left(1-\ft^{\nu^t_j-i+1}\fq^{\nu_i-j}\right)^{-1},\\
\calN_{\lambda\mu}(Q;\ft,\fq)\colonequals &\prod_{i,j=1}^{\infty}\frac{1-Q\ft^{i-1-\lambda_j^t}\fq^{j-\mu_i}}{1-Q\ft^{i-1}\fq^j}\\=&\prod_{(i,j)\in\lambda}(1-Q\ft^{\mu_j^t-i}\fq^{\lambda_i-j+1})\prod_{(i,j)\in\mu}(1-Q\ft^{-\lambda_j^t+i-1}\fq^{-\mu_i+j}).
\end{split}
\eeq
We shall use in the following $|\lambda|\colonequals \sum_{i=1}^{\ell(\lambda)}\lambda_i$ and $||\lambda||^2\colonequals \sum_{i=1}^{\ell(\lambda)}\lambda_i^2$, where $\ell(\lambda)$ is the number of rows of the partition $\lambda$.
We observe that in some cases the function $\calN_{\lambda\mu}$ behaves like a delta function, for instance $\calN_{\lambda\emptyset}(\frac{\ft}{\fq})=\calN_{\emptyset\lambda}(1)=\delta_{\lambda\emptyset}$. Furthermore, we find a relation allowing us to express the product of two $\tilde{Z}_\mu$ through
\beq
\label{eq:NasproductoftwoZ}
\calN_{\mu\mu}(1;\ft,\fq)
=\left(-\sqrt{\frac{\fq}{\ft}}\right)^{|\mu|}\ft^{-\frac{||\mu^t||^2}{2}}\fq^{-\frac{||\mu||^2}{2}}\left(\tilde{Z}_{\mu}(\ft,\fq)\tilde{Z}_{\mu^t}(\fq,\ft)\right)^{-1}.
\eeq
Using the identities 
\beq
\sum_{(i,j)\in \lambda}i=\frac{1}{2}\left(||\lambda^t||^2+|\lambda|\right),\qquad \sum_{(i,j)\in \lambda}\mu_i=\sum_{i=1}^{\min\{\ell(\lambda),\ell(\mu)\}}\lambda_i\mu_i,
\eeq
we find the exchange identities
\beq
\begin{split}
 \calN_{\lambda\mu}(Q;\fq,\ft)&=\calN_{\mu^t\lambda^t}(Q\frac{\ft}{\fq};\ft,\fq),\\
 \calN_{\lambda\mu}(Q^{-1};\ft,\fq)&=\left(-Q^{-1}\sqrt{\frac{\fq}{\ft}}\right)^{|\lambda|+|\mu|}\ft^{\frac{-||\lambda^t||^2+||\mu^t||^2}{2}}\fq^{\frac{||\lambda||^2-||\mu||^2}{2}}\calN_{\mu\lambda}(Q\frac{\ft}{\fq};\ft,\fq).
\end{split}
\eeq
From \cite{Kozcaz:2010af, Bao:2013pwa} we take the following summation formula 
\beq
\label{eq:Masatostrick1}
\sum_{\mu}\left(\sqrt{\frac{\fq}{\ft}}Q_3\right)^{|\mu|}\frac{\calN_{\mu\emptyset}\Big(\sqrt{\frac{\ft}{\fq}}Q_1\Big)\calN_{\emptyset\mu}\Big(\sqrt{\frac{\ft}{\fq}}Q_2\Big)}{\calN_{\mu\mu}(1)}=\frac{\calM\Big(Q_1Q_3\Big)\calM\Big(\frac{\ft}{\fq} Q_2 Q_3\Big)}{\calM\Big(\sqrt{\frac{\ft}{\fq}}  Q_3\Big)\calM\Big(\sqrt{\frac{\ft}{\fq}}  Q_1 Q_2 Q_3\Big)}.
\eeq
In the 4D limit, it is often useful to use the rescaled $\calN$  functions that we refer to as ``Nekrasov'' functions ($Q=e^{-\beta m}$)
\beq
\label{eq:newNoldN}
\calN_{\lambda\mu}(Q;\ft,\fq)=\left(Q\sqrt{\frac{\fq}{\ft}}\right)^{\frac{|\lambda|+|\mu|}{2}}\ft^{\frac{||\mu^t||^2-||\lambda^t||^2}{4}}\fq^{\frac{||\lambda||^2-||\mu||^2}{4}}\tN_{\lambda\mu}(m;\epsilon_1,\epsilon_2),
\eeq
where, using the parametrization \eqref{eq:deftq}, the new functions are given by
\beq
\label{eq:deftN}
\begin{split}
\tN_{\lambda\mu}(m;\epsilon_1,\epsilon_2)&=\prod_{(i,j)\in \lambda}2\sinh\frac{\beta}{2}\left[m+\epsilon_1(\lambda_i-j+1)+\epsilon_2(i-\mu^t_j)\right]\nonumber\\&\times \prod_{(i,j)\in \mu}2\sinh\frac{\beta}{2}\left[m+\epsilon_1(j-\mu_i)+\epsilon_2(\lambda^t_j-i+1)\right].
\end{split}
\eeq
The new function obeys the simpler exchange identities
\begin{align}
\label{eq:propertiestN}
\tN_{\lambda\mu}(m;-\epsilon_2,-\epsilon_1)&=\tN_{\mu^t\lambda^t}(m-\epsilon_1-\epsilon_2;\epsilon_1,\epsilon_2),\nonumber\\
\tN_{\lambda\mu}(-m;\epsilon_1,\epsilon_2)&=(-1)^{|\lambda|+|\mu|}\tN_{\mu\lambda}(m-\epsilon_1-\epsilon_2;\epsilon_1,\epsilon_2),\\
\tN_{\lambda\mu}(m;\epsilon_2,\epsilon_1)&=\tN_{\lambda^t\mu^t}(m;\epsilon_1,\epsilon_2),\nonumber
\end{align}
as well as the summation formula
\begin{multline}
\label{eq:Masatostrick2}
\sum_{\mu}e^{-\beta(\frac{m_1}{2}+\frac{m_2}{2}+m_3)|\mu|}\frac{\tN_{\mu\emptyset}\Big(m_1-\frac{\epsilon_+}{2}\Big)\tN_{\emptyset\mu}\Big(m_2-\frac{\epsilon_+}{2}\Big)}{\tN_{\mu\mu}(0)}=\\=\frac{\calM\Big(e^{-\beta(m_1+m_3)}\Big)\calM\Big(e^{-\beta(m_2+m_3-\epsilon_+)}\Big)}{\calM\Big(e^{-\beta(m_3-\frac{\epsilon_+}{2})}\Big)\calM\Big(e^{-\beta(m_1+m_2+m_3-\frac{\epsilon_+}{2})}\Big)}.
\end{multline}
We finish this section by remarking that in the limit $\beta\rightarrow 0$, the functions $\tN_{\lambda\mu}$ behave as 
\beq
\tN_{\lambda\mu}\stackrel{\beta\rightarrow 0}{\longrightarrow }\beta^{|\lambda|+|\mu|}\ttN_{\lambda\mu},
\eeq
 where we  have defined
\beq
\label{eq:defttN}
\begin{split}
\ttN_{\lambda\mu}(m;\epsilon_1,\epsilon_2)&=\prod_{(i,j)\in \lambda}\left[m+\epsilon_1(\lambda_i-j+1)+\epsilon_2(i-\mu^t_j)\right]\\&\times \prod_{(i,j)\in \mu}\left[m+\epsilon_1(j-\mu_i)+\epsilon_2(\lambda^t_j-i+1)\right].
\end{split}
\eeq
Thus for instance ratios of $\tN_{\lambda\mu}$ that are balanced in the sense that the same partitions appear in the numerator and in the denominator have proper limits for $\beta\rightarrow 0$. 

%%%%%%%%%%%%%%%%%%%%%%%%%%%%%%%%%%%%%%%%%%%%%%%%%%%%
\section{Computation of the $T_{N}$ partition function}
\label{app:TNpartitionfunction}
%%%%%%%%%%%%%%%%%%%%%%%%%%%%%%%%%%%%%%%%%%%%%%%%%%%%

In this part of the appendix, we wish to put together the computations that bring us from equations \eqref{eq:strippartitionfunction1} and \eqref{eq:toppartitionfunction} to \eqref{eq:finalexpressionforZTN}, \eqref{eq:ZTNperturbative} and \eqref{eq:ZTNsum}.
We define the function
\begin{equation}
\label{eq:defcalR}
\calR_{\lambda\mu}(Q;\ft,\fq)\colonequals \prod_{i,j=1}^{\infty}\left(1-Q \ft^{i-\frac{1}{2}-\lambda_j}\fq^{j-\frac{1}{2}-\mu_i}\right)=\calM(Q\sqrt{\frac{\ft}{\fq}};\ft,\fq)^{-1}\calN_{\lambda^t\mu}(Q\sqrt{\frac{\ft}{\fq}};\ft,\fq),
\end{equation}
and, after using some Cauchy identities, we rewrite \eqref{eq:strippartitionfunction1} as equation (4.67) of \cite{Bao:2013pwa}:
\begin{align}
\label{eq:strippartitionfunction2}
\calZ_{\boldsymbol{\nu}\boldsymbol{\tau}}^{\text{strip}}(\boldsymbol{Q}_m, \boldsymbol{Q}_l;\ft,\fq)=&\prod_{j=1}^{L+1}t^{\frac{||\nu_j^t||^2}{2}}\tilde{Z}_{\nu_j^t}(\fq,\ft)\prod_{j=1}^L\fq^{\frac{||\tau_j||^2}{2}}\tilde{Z}_{\tau_j}(\ft,\fq)\nonumber\\&\times\prod_{i\leq j=1}^L\frac{\calR_{\nu_i^t\tau_j}\Big(Q_{m;j}\prod_{k=i}^{j-1}Q_{m;k}Q_{l;k}\Big)
\calR_{\tau_i^t\nu_{j+1}}\Big(Q_{l;i}\prod_{k=i+1}^jQ_{m;k}Q_{l;k}\Big)}{\calR_{\nu_i^t\nu_{j+1}}\Big(\prod_{k=i}^jQ_{l;k}Q_{m;k}\sqrt{\frac{\fq}{\ft}}\Big)}\nonumber\\&\times \prod_{i\leq j=1}^{L-1}\calR_{\tau_i^t\tau_{j+1}}\Big(\prod_{k=i}^jQ_{l;k}Q_{m;k+1}\sqrt{\frac{\ft}{\fq}}\Big)^{-1}.
\end{align}
The complete $T_{N}$ diagram is made out of $N$ such strips, as depicted in figure~\ref{fig:strip} and written down in equation \eqref{eq:toppartitionfunction}. We remind that $\nu_{j}^{(0)}=\emptyset$ for all $j$, see the right part of figure~\ref{fig:strip}. We can redefine the strip partition function without affecting the topological string partition function \eqref{eq:toppartitionfunction}, by moving half of the $\tilde{Z}$ from one strip to the another. Specifically, we move the $\tilde{Z}_{\nu_i^t}(\fq,\ft)\ft^{\frac{||\nu_i^t||^2}{2}}$ of the left lines to the right ones, so that they become $\tilde{Z}_{\tau_i^t}(\fq,\ft)\ft^{\frac{||\tau_i^t||^2}{2}}$ for the strip on the left. This redefinition doesn't change $\calZ_N$, since the partitions to the extreme left of the $T_N$ diagram are all empty.
Putting it all together, we get a new strip partition function,
\begin{align}
\label{eq:strippartitionfunction3}
\calZ_{\boldsymbol{\nu}\boldsymbol{\tau}}^{\text{strip}\,\prime}(\boldsymbol{Q}_m, \boldsymbol{Q}_l;\ft,\fq)=&\prod_{j=1}^L\ft^{\frac{||\tau_j^t||^2}{2}}\fq^{\frac{||\tau_j||^2}{2}}\tilde{Z}_{\tau_j}(\ft,\fq)\tilde{Z}_{\tau_j^t}(\fq,\ft)\nonumber\\&\times\prod_{i\leq j=1}^L\frac{\calR_{\nu_i^t\tau_j}\Big(Q_{m;j}\prod_{k=i}^{j-1}Q_{m;k}Q_{l;k}\Big)
\calR_{\tau_i^t\nu_{j+1}}\Big(Q_{l;i}\prod_{k=i+1}^jQ_{m;k}Q_{l;k}\Big)}{\calR_{\nu_i^t\nu_{j+1}}\Big(\prod_{k=i}^jQ_{l;k}Q_{m;k}\sqrt{\frac{\fq}{\ft}}\Big)}\nonumber\\&\times \prod_{i\leq j=1}^{L-1}\calR_{\tau_i^t\tau_{j+1}}\Big(\prod_{k=i}^jQ_{l;k}Q_{m;k+1}\sqrt{\frac{\ft}{\fq}}\Big)^{-1}.
\end{align}
We can get rid of the $\tilde{Z}$ functions using \eqref{eq:NasproductoftwoZ}. Putting things together in the products and replacing the $\calR$ functions by using \eqref{eq:defcalR}, we get
\begin{align}
\label{eq:Znewstrip}
&\calZ_{\boldsymbol{\nu}\boldsymbol{\tau}}^{\text{strip}\,\prime}(\boldsymbol{Q}_m, \boldsymbol{Q}_l;\ft,\fq)=\prod_{i\leq j=1}^{L-1}\calM\Big(\frac{\ft}{\fq}\prod_{k=i}^jQ_{l;k}Q_{m;k+1}\Big)\nonumber\\&\times \prod_{i\leq j=1}^L\frac{\calM\Big(\prod_{k=i}^jQ_{l;k}Q_{m;k}\Big)}{\calM\Big(\sqrt{\frac{\ft}{\fq}}Q_{m;j}\prod_{k=i}^{j-1}Q_{m;k}Q_{l;k}\Big)\calM\Big(\sqrt{\frac{\ft}{\fq}}Q_{l;i}\prod_{k=i+1}^{j}Q_{m;k}Q_{l;k}\Big)}
\prod_{k=1}^L\left(-\sqrt{\frac{\ft}{\fq}}\right)^{|\tau_k|}\nonumber\\
&\times\prod_{i\leq j=1}^L\frac{\calN_{\nu_i\tau_j}\Big(\sqrt{\frac{\ft}{\fq}}Q_{m;j}\prod_{k=i}^{j-1}Q_{m;k}Q_{l;k}\Big)
\calN_{\tau_i\nu_{j+1}}\Big(\sqrt{\frac{\ft}{\fq}}Q_{l;i}\prod_{k=i+1}^jQ_{m;k}Q_{l;k}\Big)}{\calN_{\nu_i\nu_{j+1}}\Big(\prod_{k=i}^jQ_{l;k}Q_{m;k}\Big)\calN_{\tau_i\tau_{j}}\Big(\frac{\ft}{\fq}\prod_{k=i}^{j-1}Q_{l;k}Q_{m;k+1}\Big)}.
\end{align}
We can straightforwardly obtain \eqref{eq:ZTNperturbative} by taking only the $\calM$ dependent terms of the strip partition functions, plugging them in \eqref{eq:toppartitionfunction} and replacing the K\"ahler parameters $Q_m$, and $Q_l$ by the $\tilde{A}$'s using the formulas \eqref{eq:prodQmQl} of appendix \ref{app:notation}. Thus, we get the  ``perturbative part'' of the topological string $T_N$ partition \eqref{eq:ZTNperturbative}, {\it i.e.} the part that is independent of the partitions entering the sum.

Using the functions $\tN$ defined in \eqref{eq:deftN}, the relations \eqref{eq:newNoldN} and performing a shift of the factors from one strip to the one standing on its left, which implies the following change:
\beq
\prod_{i\leq j=1}^L\left(\frac{\ft}{\fq}\right)^{\frac{|\nu_i|+|\nu_{j+1}|}{4}}\longrightarrow
\prod_{i\leq j=1}^{L-1}\left(\frac{\ft}{\fq}\right)^{\frac{|\tau_i|+|\tau_{j+1}|}{4}},
\eeq
 we can write the ``instanton'' part of the redefined strip as 
\begin{multline}
\label{eq:newstrip2}
\calZ^{\text{strip, inst}\,\prime\prime}_{\boldsymbol{\nu}\boldsymbol{\tau}}(\boldsymbol{Q}_m, \boldsymbol{Q}_l;\ft,\fq)=\prod_{k=1}^L(-1)^{|\tau_k|}\prod_{i\leq j=1}^LQ_{m;i}^{\frac{|\tau_j|-|\nu_{j+1}|}{2}}Q_{l;j}^{\frac{|\tau_i|-|\nu_i|}{2}}\\\times \frac{\tN_{\nu_i\tau_j}(\sum_{k=i}^{j-1}q_{l;k}+\sum_{k=i}^{j}q_{m;k}-\frac{\epsilon_+}{2})\tN_{\tau_i\nu_{j+1}}(\sum_{k=i}^{j}q_{l;k}+\sum_{k=i+1}^{j}q_{m;k}-\frac{\epsilon_+}{2})}{\tN_{\nu_i\nu_{j+1}}(\sum_{k=i}^j(q_{l;k}+q_{m;k}))\tN_{\tau_i\tau_{j}}(\sum_{k=i}^{j-1}q_{l;k}+\sum_{k=i+1}^{j}q_{m;k}-\epsilon_+)},
\end{multline}
where we have  used $Q_{m;j}^{(i)}=e^{-\beta q_{m;j}^{(i)}}$, $Q_{l;j}^{(i)}=e^{-\beta q_{l;j}^{(i)}}$ and $Q_{n;j}^{(i)}=e^{-\beta q_{n;j}^{(i)}}$.  Before we move on, let us remark that 
\beq
\label{eq:newstripinstantonQpart}
\prod_{i\leq j=1}^LQ_{m;i}^{\frac{|\tau_j|-|\nu_{j+1}|}{2}}Q_{l;j}^{\frac{|\tau_i|-|\nu_i|}{2}}=\prod_{i=1}^L\left(\prod_{j=1}^iQ_{m;j}\prod_{k=i}^LQ_{l;k}\right)^{\frac{|\tau_i|}{2}}\prod_{i=1}^{L+1}\left(\prod_{j=1}^{i-1}Q_{m;j}\prod_{k=i}^LQ_{l;k}\right)^{-\frac{|\nu_i|}{2}}.
\eeq
Armed with \eqref{eq:newstrip2}, we can compute \eqref{eq:ZTNsum}.
We have 
\beq
\label{eq:TNpartitionfunctioninstantonnew}
\calZ_{N}^{\text{inst}}=\prod_{r=1}^{N}\Big(-\boldsymbol{Q}_n^{(r)}\Big)^{|\boldsymbol{\nu}^{(r)}|}Z^{\text{strip, inst}\,\prime\prime}_{\boldsymbol{\nu}^{(r-1)}\boldsymbol{\nu}^{(r)}}(\boldsymbol{Q}_m^{(r)}, \boldsymbol{Q}_l^{(r)};\ft,\fq).
\eeq
First, we consider the part of the sum of $\calZ_{N}^{\text{inst}}$ that doesn't involve the $\tN$ functions. It consists solely of the $\prod_{r=1}^{N}\Big(-\boldsymbol{Q}_n^{(r)}\Big)^{|\boldsymbol{\nu}^{(r)}|}$ term (The minus sign will be canceled by the $\prod_{k=1}^L(-1)^{|\tau_k|}$ part of  \eqref{eq:newstrip2}) of \eqref{eq:TNpartitionfunctioninstantonnew} and of the product of \eqref{eq:newstripinstantonQpart} over all the strips, where $\boldsymbol{\nu}$ is to be replaced by $\boldsymbol{\nu}^{(r-1)}$ and $\boldsymbol{\tau}$ by $\boldsymbol{\nu}^{(r)}$. Explicitly, this part of the summand has the form (the length $L$ of the strip is given by $N-r$, where $r$ numbers the strips from left to right)
\beqa
\label{eq:lengthyQcomputation}
&&\prod_{r=1}^N\prod_{i=1}^{N-r}\Big(Q_{n;i}^{(r)}\Big)^{|\nu_i^{(r)}|}\prod_{i=1}^{N-r}\left(\prod_{j=1}^iQ_{m;j}^{(r)}\prod_{k=i}^{N-r}Q_{l;k}^{(r)}\right)^{\frac{|\nu_i^{(r)}|}{2}}\prod_{i=1}^{N-r+1}\left(\prod_{j=1}^{i-1}Q^{(r)}_{m;j}\prod_{k=i}^{N-r}Q_{l;k}^{(r)}\right)^{-\frac{|\nu_i^{(r-1)}|}{2}}\nonumber\\
&&=\prod_{r=1}^N\prod_{i=1}^{N-r}\Big(Q_{n;i}^{(r)}\Big)^{|\nu_i^{(r)}|}\left(\prod_{j=1}^iQ_{m;j}^{(r)}\prod_{k=i}^{N-r}Q_{l;k}^{(r)}\right)^{\frac{|\nu_i^{(r)}|}{2}}\left(\prod_{j=1}^{i-1}Q^{(r+1)}_{m;j}\prod_{k=i}^{N-r-1}Q_{l;k}^{(r+1)}\right)^{-\frac{|\nu_i^{(r)}|}{2}}\nonumber\\
&&=\prod_{r=1}^N\prod_{i=1}^{N-r}\left[\left(Q_{n;i}^{(r)}\right)^2\frac{\prod_{j=1}^iQ_{m;j}^{(r)}\prod_{k=i}^{N-r}Q_{l;k}^{(r)}}{\prod_{j=1}^{i-1}Q^{(r+1)}_{m;j}\prod_{k=i}^{N-r-1}Q_{l;k}^{(r+1)}}\right]^{\frac{|\nu_i^{(r)}|}{2}}\nonumber\\&&=\prod_{r=1}^N\prod_{i=1}^{N-r}\left[\frac{\Big(\tilde{A}_0^{(r)}\tilde{A}_{N-r}^{(r)}\Big)^2}{\tilde{A}_0^{(r-1)}\tilde{A}_0^{(r+1)}\tilde{A}_{N-r+1}^{(r-1)}\tilde{A}_{N-r-1}^{(r+1)}}\right]^{\frac{|\nu_i^{(r)}|}{2}}=\prod_{r=1}^N\prod_{i=1}^{N-r}\left[\frac{\tilde{N}_r\tilde{L}_{N-r}}{\tilde{N}_{r+1}\tilde{L}_{N-r+1}}\right]^{\frac{|\nu_i^{(r)}|}{2}}
\eeqa
where in the second line we have used the fact that $\boldsymbol{\nu}^{(0)}$ consists entirely of empty partitions, in the fourth we have used the very useful formulas \eqref{eq:prodQmQl}
 and in the last equation have used \eqref{eq:borderAdefMNL}.

Taking now  the last line of \eqref{eq:lengthyQcomputation} and adding the remaining $\tN$ parts, we get the full instanton part of the $T_N$ partition function that we wrote in \eqref{eq:ZTNsum}.

%%%%%%%%%%%%%%%%%%%%%%%%%%%%%%%%%%%%%%%%%%
%Bibliography

%\bibliographystyle{JHEP}
%\bibliography{Biblio}

\begin{thebibliography}{10}

\bibitem{Bao:2013pwa}
L.~Bao, V.~Mitev, E.~Pomoni, M.~Taki, and F.~Yagi, {\it {Non-Lagrangian
  Theories from Brane Junctions}},  {\em JHEP} {\bf 1401} (2014) 175,
  [\href{http://arxiv.org/abs/1310.3841}{{\tt arXiv:1310.3841}}].

\bibitem{Alday:2009aq}
L.~F. Alday, D.~Gaiotto, and Y.~Tachikawa, {\it {Liouville Correlation
  Functions from Four-dimensional Gauge Theories}},  {\em Lett.Math.Phys.} {\bf
  91} (2010) 167--197, [\href{http://arxiv.org/abs/0906.3219}{{\tt
  arXiv:0906.3219}}].

\bibitem{Wyllard:2009hg}
N.~Wyllard, {\it {A(N-1) conformal Toda field theory correlation functions from
  conformal N = 2 SU(N) quiver gauge theories}},  {\em JHEP} {\bf 0911} (2009)
  002, [\href{http://arxiv.org/abs/0907.2189}{{\tt arXiv:0907.2189}}].

\bibitem{Gaiotto:2009ma}
D.~Gaiotto, {\it {Asymptotically free N=2 theories and irregular conformal
  blocks}},  \href{http://arxiv.org/abs/0908.0307}{{\tt arXiv:0908.0307}}.

\bibitem{Hayashi:2013qwa}
H.~Hayashi, H.-C. Kim, and T.~Nishinaka, {\it {Topological strings and 5d $T_N$
  partition functions}},  \href{http://arxiv.org/abs/1310.3854}{{\tt
  arXiv:1310.3854}}.

\bibitem{Dorn:1994xn}
H.~Dorn and H.~Otto, {\it {Two and three point functions in Liouville theory}},
   {\em Nucl.Phys.} {\bf B429} (1994) 375--388,
  [\href{http://arxiv.org/abs/hep-th/9403141}{{\tt hep-th/9403141}}].

\bibitem{Zamolodchikov:1995aa}
A.~B. Zamolodchikov and A.~B. Zamolodchikov, {\it {Structure constants and
  conformal bootstrap in Liouville field theory}},  {\em Nucl.Phys.} {\bf B477}
  (1996) 577--605, [\href{http://arxiv.org/abs/hep-th/9506136}{{\tt
  hep-th/9506136}}].

\bibitem{Fateev:2009aw}
V.~Fateev and A.~Litvinov, {\it {On AGT conjecture}},  {\em JHEP} {\bf 1002}
  (2010) 014, [\href{http://arxiv.org/abs/0912.0504}{{\tt arXiv:0912.0504}}].

\bibitem{Mironov:2009qn}
A.~Mironov and A.~Morozov, {\it {Proving AGT relations in the large-c limit}},
  {\em Phys.Lett.} {\bf B682} (2009) 118--124,
  [\href{http://arxiv.org/abs/0909.3531}{{\tt arXiv:0909.3531}}].

\bibitem{Hadasz:2010xp}
L.~Hadasz, Z.~Jaskolski, and P.~Suchanek, {\it {Proving the AGT relation for
  $N_f = 0,1,2$ antifundamentals}},  {\em JHEP} {\bf 1006} (2010) 046,
  [\href{http://arxiv.org/abs/1004.1841}{{\tt arXiv:1004.1841}}].

\bibitem{Alba:2010qc}
V.~A. Alba, V.~A. Fateev, A.~V. Litvinov, and G.~M. Tarnopolskiy, {\it {On
  combinatorial expansion of the conformal blocks arising from AGT
  conjecture}},  {\em Lett.Math.Phys.} {\bf 98} (2011) 33--64,
  [\href{http://arxiv.org/abs/1012.1312}{{\tt arXiv:1012.1312}}].

\bibitem{Mironov:2010qe}
A.~Mironov, A.~Morozov, and S.~Shakirov, {\it {Towards a proof of AGT
  conjecture by methods of matrix models}},  {\em Int.J.Mod.Phys.} {\bf A27}
  (2012) 1230001, [\href{http://arxiv.org/abs/1011.5629}{{\tt
  arXiv:1011.5629}}].

\bibitem{Mironov:2010pi}
A.~Mironov, A.~Morozov, and S.~Shakirov, {\it {A direct proof of AGT conjecture
  at beta = 1}},  {\em JHEP} {\bf 1102} (2011) 067,
  [\href{http://arxiv.org/abs/1012.3137}{{\tt arXiv:1012.3137}}].

\bibitem{Fateev:2011hq}
V.~Fateev and A.~Litvinov, {\it {Integrable structure, W-symmetry and AGT
  relation}},  {\em JHEP} {\bf 1201} (2012) 051,
  [\href{http://arxiv.org/abs/1109.4042}{{\tt arXiv:1109.4042}}].

\bibitem{Kanno:2013vi}
H.~Kanno, K.~Maruyoshi, S.~Shiba, and M.~Taki, {\it {$W_3$ irregular states and
  isolated N=2 superconformal field theories}},  {\em JHEP} {\bf 1303} (2013)
  147, [\href{http://arxiv.org/abs/1301.0721}{{\tt arXiv:1301.0721}}].

\bibitem{Mironov:2013oaa}
S.~Mironov, A.~Morozov, and Y.~Zenkevich, {\it {Generalized Jack polynomials
  and the AGT relations for the $SU(3)$ group}},  {\em JETP Lett.} {\bf 99}
  (2014) 109--113, [\href{http://arxiv.org/abs/1312.5732}{{\tt
  arXiv:1312.5732}}].

\bibitem{Schiappa:2009cc}
R.~Schiappa and N.~Wyllard, {\it {An A(r) threesome: Matrix models, 2d CFTs and
  4d N=2 gauge theories}},  {\em J.Math.Phys.} {\bf 51} (2010) 082304,
  [\href{http://arxiv.org/abs/0911.5337}{{\tt arXiv:0911.5337}}].

\bibitem{Awata:2009ur}
H.~Awata and Y.~Yamada, {\it {Five-dimensional AGT Conjecture and the Deformed
  Virasoro Algebra}},  {\em JHEP} {\bf 1001} (2010) 125,
  [\href{http://arxiv.org/abs/0910.4431}{{\tt arXiv:0910.4431}}].

\bibitem{Awata:2010yy}
H.~Awata and Y.~Yamada, {\it {Five-dimensional AGT Relation and the Deformed
  beta-ensemble}},  {\em Prog.Theor.Phys.} {\bf 124} (2010) 227--262,
  [\href{http://arxiv.org/abs/1004.5122}{{\tt arXiv:1004.5122}}].

\bibitem{Mironov:2011dk}
A.~Mironov, A.~Morozov, S.~Shakirov, and A.~Smirnov, {\it {Proving AGT
  conjecture as HS duality: extension to five dimensions}},  {\em Nucl.Phys.}
  {\bf B855} (2012) 128--151, [\href{http://arxiv.org/abs/1105.0948}{{\tt
  arXiv:1105.0948}}].

\bibitem{Itoyama:2013mca}
H.~Itoyama, T.~Oota, and R.~Yoshioka, {\it {2d-4d Connection between
  $q$-Virasoro/W Block at Root of Unity Limit and Instanton Partition Function
  on ALE Space}},  {\em Nucl.Phys.} {\bf B877} (2013) 506--537,
  [\href{http://arxiv.org/abs/1308.2068}{{\tt arXiv:1308.2068}}].

\bibitem{Bao:2011rc}
L.~Bao, E.~Pomoni, M.~Taki, and F.~Yagi, {\it {M5-Branes, Toric Diagrams and
  Gauge Theory Duality}},  {\em JHEP} {\bf 1204} (2012) 105,
  [\href{http://arxiv.org/abs/1112.5228}{{\tt arXiv:1112.5228}}].

\bibitem{Nieri:2013yra}
F.~Nieri, S.~Pasquetti, and F.~Passerini, {\it {3d \& 5d gauge theory partition
  functions as q-deformed CFT correlators}},
  \href{http://arxiv.org/abs/1303.2626}{{\tt arXiv:1303.2626}}.

\bibitem{Nieri:2013vba}
F.~Nieri, S.~Pasquetti, F.~Passerini, and A.~Torrielli, {\it {5D partition
  functions, q-Virasoro systems and integrable spin-chains}},
  \href{http://arxiv.org/abs/1312.1294}{{\tt arXiv:1312.1294}}.

\bibitem{Aganagic:2013tta}
M.~Aganagic, N.~Haouzi, C.~Kozcaz, and S.~Shakirov, {\it {Gauge/Liouville
  Triality}},  \href{http://arxiv.org/abs/1309.1687}{{\tt arXiv:1309.1687}}.

\bibitem{Aganagic:2014oia}
M.~Aganagic, N.~Haouzi, and S.~Shakirov, {\it {$A_n$-Triality}},
  \href{http://arxiv.org/abs/1403.3657}{{\tt arXiv:1403.3657}}.

\bibitem{Taki:2014fva}
M.~Taki, {\it {On AGT-W Conjecture and q-Deformed W-Algebra}},
  \href{http://arxiv.org/abs/1403.7016}{{\tt arXiv:1403.7016}}.

\bibitem{Iqbal:2012xm}
A.~Iqbal and C.~Vafa, {\it {BPS Degeneracies and Superconformal Index in
  Diverse Dimensions}},  \href{http://arxiv.org/abs/1210.3605}{{\tt
  arXiv:1210.3605}}.

\bibitem{Fateev:2005gs}
V.~Fateev and A.~Litvinov, {\it {On differential equation on four-point
  correlation function in the Conformal Toda Field Theory}},  {\em JETP Lett.}
  {\bf 81} (2005) 594--598, [\href{http://arxiv.org/abs/hep-th/0505120}{{\tt
  hep-th/0505120}}].

\bibitem{Fateev:2007ab}
V.~Fateev and A.~Litvinov, {\it {Correlation functions in conformal Toda field
  theory. I.}},  {\em JHEP} {\bf 0711} (2007) 002,
  [\href{http://arxiv.org/abs/0709.3806}{{\tt arXiv:0709.3806}}].

\bibitem{Fateev:2008bm}
V.~Fateev and A.~Litvinov, {\it {Correlation functions in conformal Toda field
  theory II}},  {\em JHEP} {\bf 0901} (2009) 033,
  [\href{http://arxiv.org/abs/0810.3020}{{\tt arXiv:0810.3020}}].

\bibitem{Dotsenko:1984ad}
V.~Dotsenko and V.~Fateev, {\it {Four Point Correlation Functions and the
  Operator Algebra in the Two-Dimensional Conformal Invariant Theories with the
  Central Charge $c < 1$}},  {\em Nucl.Phys.} {\bf B251} (1985) 691.

\bibitem{Benini:2009gi}
F.~Benini, S.~Benvenuti, and Y.~Tachikawa, {\it {Webs of five-branes and N=2
  superconformal field theories}},  {\em JHEP} {\bf 0909} (2009) 052,
  [\href{http://arxiv.org/abs/0906.0359}{{\tt arXiv:0906.0359}}].

\bibitem{Awata:2005fa}
H.~Awata and H.~Kanno, {\it {Instanton counting, Macdonald functions and the
  moduli space of D-branes}},  {\em JHEP} {\bf 0505} (2005) 039,
  [\href{http://arxiv.org/abs/hep-th/0502061}{{\tt hep-th/0502061}}].

\bibitem{Iqbal:2007ii}
A.~Iqbal, C.~Kozcaz, and C.~Vafa, {\it {The Refined topological vertex}},  {\em
  JHEP} {\bf 0910} (2009) 069, [\href{http://arxiv.org/abs/hep-th/0701156}{{\tt
  hep-th/0701156}}].

\bibitem{Kozcaz:2010af}
C.~Kozcaz, S.~Pasquetti, and N.~Wyllard, {\it {$A \& B$ model approaches to
  surface operators and Toda theories}},  {\em JHEP} {\bf 1008} (2010) 042,
  [\href{http://arxiv.org/abs/1004.2025}{{\tt arXiv:1004.2025}}].

\bibitem{Kim:2012gu}
H.-C. Kim, S.-S. Kim, and K.~Lee, {\it {5-dim Superconformal Index with
  Enhanced En Global Symmetry}},  {\em JHEP} {\bf 1210} (2012) 142,
  [\href{http://arxiv.org/abs/1206.6781}{{\tt arXiv:1206.6781}}].

\bibitem{Hayashi:2014wfa}
H.~Hayashi and G.~Zoccarato, {\it {Exact partition functions of Higgsed 5d
  $T_N$ theories}},  \href{http://arxiv.org/abs/1409.0571}{{\tt
  arXiv:1409.0571}}.

\bibitem{Misha}
M.~Isachenkov, V.~Mitev, and E.~Pomoni, {\it {Work in progress}}, .

\bibitem{Belavin:1984vu}
A.~Belavin, A.~M. Polyakov, and A.~Zamolodchikov, {\it {Infinite Conformal
  Symmetry in Two-Dimensional Quantum Field Theory}},  {\em Nucl.Phys.} {\bf
  B241} (1984) 333--380.

\bibitem{FutoshiMasato}
V.~Mitev, E.~Pomoni, M.~Taki, and F.~Yagi, {\it {Work in progress}}, .

\bibitem{Teschner:1995yf}
J.~Teschner, {\it {On the Liouville three point function}},  {\em Phys.Lett.}
  {\bf B363} (1995) 65--70, [\href{http://arxiv.org/abs/hep-th/9507109}{{\tt
  hep-th/9507109}}].

\bibitem{Teschner:2001rv}
J.~Teschner, {\it {Liouville theory revisited}},  {\em Class.Quant.Grav.} {\bf
  18} (2001) R153--R222, [\href{http://arxiv.org/abs/hep-th/0104158}{{\tt
  hep-th/0104158}}].

\bibitem{Aharony:1997ju}
O.~Aharony and A.~Hanany, {\it {Branes, superpotentials and superconformal
  fixed points}},  {\em Nucl.Phys.} {\bf B504} (1997) 239--271,
  [\href{http://arxiv.org/abs/hep-th/9704170}{{\tt hep-th/9704170}}].

\bibitem{Aharony:1997bh}
O.~Aharony, A.~Hanany, and B.~Kol, {\it {Webs of (p,q) five-branes,
  five-dimensional field theories and grid diagrams}},  {\em JHEP} {\bf 9801}
  (1998) 002, [\href{http://arxiv.org/abs/hep-th/9710116}{{\tt
  hep-th/9710116}}].

\bibitem{Brandhuber:1997ua}
A.~Brandhuber, N.~Itzhaki, J.~Sonnenschein, S.~Theisen, and S.~Yankielowicz,
  {\it {On the M theory approach to (compactified) 5-D field theories}},  {\em
  Phys.Lett.} {\bf B415} (1997) 127--134,
  [\href{http://arxiv.org/abs/hep-th/9709010}{{\tt hep-th/9709010}}].

\bibitem{Katz:1996fh}
S.~H. Katz, A.~Klemm, and C.~Vafa, {\it {Geometric engineering of quantum field
  theories}},  {\em Nucl.Phys.} {\bf B497} (1997) 173--195,
  [\href{http://arxiv.org/abs/hep-th/9609239}{{\tt hep-th/9609239}}].

\bibitem{Katz:1997eq}
S.~Katz, P.~Mayr, and C.~Vafa, {\it {Mirror symmetry and exact solution of 4-D
  N=2 gauge theories: 1.}},  {\em Adv.Theor.Math.Phys.} {\bf 1} (1998) 53--114,
  [\href{http://arxiv.org/abs/hep-th/9706110}{{\tt hep-th/9706110}}].

\bibitem{Assi:2014exa}
A.~Z. Assi, {\it {Topological Amplitudes and the String Effective Action}},
  \href{http://arxiv.org/abs/1402.2428}{{\tt arXiv:1402.2428}}.

\bibitem{Kashani-Poor:2014lxa}
A.-K. Kashani-Poor, {\it {Computing $Z_{\textbf{top}}$}},
  \href{http://arxiv.org/abs/1408.1240}{{\tt arXiv:1408.1240}}.

\bibitem{Leung:1997tw}
N.~C. Leung and C.~Vafa, {\it {Branes and toric geometry}},  {\em
  Adv.Theor.Math.Phys.} {\bf 2} (1998) 91--118,
  [\href{http://arxiv.org/abs/hep-th/9711013}{{\tt hep-th/9711013}}].

\bibitem{Awata:2009yc}
H.~Awata and H.~Kanno, {\it {Changing the preferred direction of the refined
  topological vertex}},  {\em J.Geom.Phys.} {\bf 64} (2013) 91--110,
  [\href{http://arxiv.org/abs/0903.5383}{{\tt arXiv:0903.5383}}].

\bibitem{Iqbal:2012mt}
A.~Iqbal and C.~Kozcaz, {\it {Refined Topological Strings and Toric Calabi-Yau
  Threefolds}},  \href{http://arxiv.org/abs/1210.3016}{{\tt arXiv:1210.3016}}.

\bibitem{Gopakumar:1998ii}
R.~Gopakumar and C.~Vafa, {\it {M theory and topological strings. 1.}},
  \href{http://arxiv.org/abs/hep-th/9809187}{{\tt hep-th/9809187}}.

\bibitem{Gopakumar:1998jq}
R.~Gopakumar and C.~Vafa, {\it {M theory and topological strings. 2.}},
  \href{http://arxiv.org/abs/hep-th/9812127}{{\tt hep-th/9812127}}.

\bibitem{Gadde:2009kb}
A.~Gadde, E.~Pomoni, L.~Rastelli, and S.~S. Razamat, {\it {S-duality and 2d
  Topological QFT}},  {\em JHEP} {\bf 1003} (2010) 032,
  [\href{http://arxiv.org/abs/0910.2225}{{\tt arXiv:0910.2225}}].

\bibitem{Gadde:2010te}
A.~Gadde, L.~Rastelli, S.~S. Razamat, and W.~Yan, {\it {The Superconformal
  Index of the $E_6$ SCFT}},  {\em JHEP} {\bf 1008} (2010) 107,
  [\href{http://arxiv.org/abs/1003.4244}{{\tt arXiv:1003.4244}}].

\bibitem{Gadde:2011ik}
A.~Gadde, L.~Rastelli, S.~S. Razamat, and W.~Yan, {\it {The 4d Superconformal
  Index from q-deformed 2d Yang-Mills}},  {\em Phys.Rev.Lett.} {\bf 106} (2011)
  241602, [\href{http://arxiv.org/abs/1104.3850}{{\tt arXiv:1104.3850}}].

\bibitem{Gadde:2011uv}
A.~Gadde, L.~Rastelli, S.~S. Razamat, and W.~Yan, {\it {Gauge Theories and
  Macdonald Polynomials}},  {\em Commun.Math.Phys.} {\bf 319} (2013) 147--193,
  [\href{http://arxiv.org/abs/1110.3740}{{\tt arXiv:1110.3740}}].

\bibitem{Tachikawa:2012wi}
Y.~Tachikawa, {\it {4d partition function on $S^1 x S^3$ and 2d Yang-Mills with
  nonzero area}},  {\em PTEP} {\bf 2013} (2013) 013B01,
  [\href{http://arxiv.org/abs/1207.3497}{{\tt arXiv:1207.3497}}].

\bibitem{Fukuda:2012jr}
Y.~Fukuda, T.~Kawano, and N.~Matsumiya, {\it {5D SYM and 2D q-Deformed YM}},
  {\em Nucl.Phys.} {\bf B869} (2013) 493--522,
  [\href{http://arxiv.org/abs/1210.2855}{{\tt arXiv:1210.2855}}].

\bibitem{Seiberg:1996bd}
N.~Seiberg, {\it {Five-dimensional SUSY field theories, nontrivial fixed points
  and string dynamics}},  {\em Phys.Lett.} {\bf B388} (1996) 753--760,
  [\href{http://arxiv.org/abs/hep-th/9608111}{{\tt hep-th/9608111}}].

\bibitem{Morrison:1996xf}
D.~R. Morrison and N.~Seiberg, {\it {Extremal transitions and five-dimensional
  supersymmetric field theories}},  {\em Nucl.Phys.} {\bf B483} (1997)
  229--247, [\href{http://arxiv.org/abs/hep-th/9609070}{{\tt hep-th/9609070}}].

\bibitem{Intriligator:1997pq}
K.~A. Intriligator, D.~R. Morrison, and N.~Seiberg, {\it {Five-dimensional
  supersymmetric gauge theories and degenerations of Calabi-Yau spaces}},  {\em
  Nucl.Phys.} {\bf B497} (1997) 56--100,
  [\href{http://arxiv.org/abs/hep-th/9702198}{{\tt hep-th/9702198}}].

\bibitem{Ding:1996mq}
J.-t. Ding and K.~Iohara, {\it {Generalization and deformation of Drinfeld
  quantum affine algebras}},  {\em Lett.Math.Phys.} {\bf 41} (1997) 181--193.

\bibitem{Nakayama:2004vk}
Y.~Nakayama, {\it {Liouville field theory: A Decade after the revolution}},
  {\em Int.J.Mod.Phys.} {\bf A19} (2004) 2771--2930,
  [\href{http://arxiv.org/abs/hep-th/0402009}{{\tt hep-th/0402009}}].

\bibitem{Nishizawa:2001}
M.~Nishizawa, {\it {An elliptic analogue of the multiple gamma function}},
  {\em J. Phys. A: Math. Gen.} {\bf 34} (2001) 7411.

\end{thebibliography}
\providecommand{\href}[2]{#2}\begingroup\raggedright\endgroup

\end{document}